\documentclass[floats,floatfix,amssymb,prd,twocolumn,superscriptaddress,nofootinbib,preprintnumbers]{revtex4-1}
\usepackage{float}
\usepackage{placeins}
\usepackage{subcaption}
\usepackage{ragged2e}

\DeclareCaptionJustification{justified}{\justifying}
\makeatletter
\newcommand{\subsetsim}{\mathrel{\mathpalette\subset@sim\relax}}
\newcommand{\subset@sim}[2]{%
  \vtop{\offinterlineskip\m@th
    \ialign{\hfil##\cr
      $#1\subset$\cr\noalign{\kern0.5pt}\scalebox{0.9}{$#1\sim$}\cr
    }%
  }%
}
\makeatother

\usepackage{amssymb,amsmath,verbatim,mathtools,needspace,enumitem,etoolbox,graphicx,microtype,afterpage,bm}

\usepackage[dvipsnames, usenames]{xcolor}
\definecolor{linkcolor}{rgb}{0.0,0.3,0.5}

\usepackage{booktabs}
\usepackage[unicode, colorlinks=true, linkcolor=linkcolor, citecolor=linkcolor, filecolor=linkcolor,urlcolor=linkcolor, pdfusetitle]{hyperref}
\usepackage{cleveref}

\usepackage[all]{hypcap}
\usepackage[T1]{fontenc}
\usepackage[utf8]{inputenc}
\usepackage{tabularx}
\usepackage{float}
\usepackage{multirow}
\usepackage{pifont}
\usepackage{lmodern}

\allowdisplaybreaks
\usepackage{tikz}
\usetikzlibrary{shapes.misc, positioning, arrows.meta}
\usepackage{framed}
\hypersetup{colorlinks, citecolor=bluscuro, linkcolor=black, urlcolor=bluscuro}
\definecolor{rossos}{cmyk}{0,1,1,0.55}
\definecolor{bluscuro}{rgb}{0.15, 0.2, .85}
\definecolor{bluchiaro}{cmyk}{1,.3,0.,0.1}
\definecolor{ForestGreen}{rgb}{0.13, 0.55, 0.13}
\definecolor{TLGreen}{RGB}{50, 164, 49}
\definecolor{TLOrange}{RGB}{231,180,22}
\definecolor{TLRed}{RGB}{204,50,50}

\usepackage[normalem]{ulem}

\newcommand{\TLBullet}[1]{\raisebox{-5pt}{\scalebox{0.23}{\begin{tikzpicture}\shadedraw[rounded corners=15pt, top color=gray!84!black,bottom color=black, line width=.6pt] (0,0) rectangle ++(6,2); \ifthenelse{#1=1}{\draw[fill=green,line width=1.pt]  (1,1) circle(.75cm);}{\draw[fill=green!35!black,line width=1.pt]  (1,1) circle(.75cm);}\ifthenelse{#1=2}{\draw[fill=yellow,line width=1.pt]  (3,1) circle(.75cm);}{\draw[fill=yellow!60!black,line width=1.pt]  (3,1) circle(.75cm);}\ifthenelse{#1=3}{\draw[fill=red,line width=1.pt]  (5,1) circle(.75cm);}{\draw[fill=red!50!black,line width=1.pt]  (5,1) circle(.75cm);}\end{tikzpicture}}}}

\newcommand{\bs}{\begin{subequations}}
\newcommand{\es}{\end{subequations}}

\newcommand{\be}{\begin{equation}}
\newcommand{\ee}{\end{equation}}

\def\lsim{\mathrel{\rlap{\lower4pt\hbox{\hskip0.5pt$\sim$}}
    \raise1pt\hbox{$<$}}}         
\def\gsim{\mathrel{\rlap{\lower4pt\hbox{\hskip0.5pt$\sim$}}
    \raise1pt\hbox{$>$}}}         

\usepackage{siunitx}
\DeclareSIUnit \parsec {pc}
\DeclareSIUnit \arcsecondfull {arcsec}
\DeclareSIUnit \year{yr}
\DeclareSIUnit \day{day}
\DeclareSIUnit \hour{hr}
\DeclareSIUnit \radiant{rad}
\DeclareSIUnit \degfull{deg}
\DeclareSIUnit \erg {erg}
\DeclareSIUnit \Lsun {L_\odot}
\DeclareSIUnit \Msun {M_\odot}
\DeclareSIUnit \AstroUnit {au}

\usepackage{nicefrac}

\newcommand{\sapienza}{Dipartimento di Fisica, Sapienza Università 
	di Roma, Piazzale Aldo Moro 5, 00185, Roma, Italy}
\newcommand{\infn}{INFN, Sezione di Roma, Piazzale Aldo Moro 2, 00185, Roma, Italy}

\newcommand{\ifae}{
Institut de Fisica d’Altes Energies (IFAE), The Barcelona Institute}

\begin{document}

\title{Accuracy of ringdown models calibrated to numerical relativity simulations}

\author{Francesco Crescimbeni}
\email{francesco.crescimbeni@uniroma1.it}
\affiliation{\sapienza}
\affiliation{\infn}

\author{Gregorio Carullo}
\email{g.carullo@bham.ac.uk}
\affiliation{School of Physics and Astronomy \& Institute for Gravitational Wave Astronomy, University of Birmingham,
Birmingham, B15 2TT, United Kingdom} 

\author{Emanuele Berti}
\email{berti@jhu.edu}
\affiliation{William H. Miller III Department of Physics and Astronomy, Johns Hopkins
University, 3400 North Charles Street, Baltimore, Maryland, 21218, USA} 

\author{Giada Caneva Santoro}
\email{ gcaneva@ifae.es}
\affiliation{\ifae}

\author{Mark Ho-Yeuk Cheung}
\email{mcheung@ias.edu}
\affiliation{William H. Miller III Department of Physics and Astronomy, Johns Hopkins
University, 3400 North Charles Street, Baltimore, Maryland, 21218, USA}
\affiliation{School of Natural Sciences, Institute for Advanced Study, Einstein Drive, Princeton, New Jersey, 08540, USA}

\author{Paolo Pani}
\email{paolo.pani@uniroma1.it}
\affiliation{\sapienza}
\affiliation{\infn}

\begin{abstract}
\noindent
The ``ringdown'' stage of gravitational-wave signals from binary black hole mergers, mainly consisting of a superposition of quasinormal modes emitted by the merger remnant, is a key tool to test fundamental physics and to probe black hole dynamics.
However, ringdown models are known to be accurate only in the late-time, stationary regime.
A key open problem in the field is to understand if these models are robust when extrapolated to earlier times, and if they can faithfully recover a larger portion of the signal.
We address this question through a systematic time-domain calculation of the mismatch between non-precessing, quasi-circular ringdown models parameterised by the progenitor binary's degrees of freedom and full numerical relativity inspiral-merger-ringdown waveforms from the Simulating eXtreme Spacetimes (SXS) simulation catalog.
For the best-performing models, the mismatch is typically in the range $[10^{-6}, 10^{-4}]$ for the $(\ell,|m|)= (2,2)$ harmonic, and $[10^{-4}, 10^{-2}]$ for higher-order modes.
Our findings inform ongoing observational searches for quasinormal modes, and underscore the need for improved modeling of higher-order modes to meet the sensitivity requirements of future gravitational-wave detectors.
\end{abstract}

\maketitle

\section{Introduction}\label{intro}
The first observation of a gravitational-wave (GW) signal emitted by the coalescence of two black holes (BHs)~\cite{Abbott:2016blz} has revolutionized our understanding of the Universe, with important implications in astrophysics, fundamental physics, and cosmology~\cite{Abbott:2023aa,Abbott:2021xx,Abbott:2023yy,Abbott:2023zz}.
While the early inspiral signal provides information on the dynamics of the two objects, the merger and ringdown stages offer a unique observational window into the strong-field dynamical regime of general relativity (GR) and the nature of the remnant object~\cite{Abbott:2019prv,Abbott:2020jks,Berti:2015itd,Berti:2018vdi,Cardoso:2019rvt}.

Within black hole perturbation theory~\cite{Regge:1957td,Zerilli:1970se, Teukolsky:1972my, Teukolsky:1973ha}, it is possible to describe part of the signal as a superposition of damped sinusoids, or quasinormal modes (QNMs)~\cite{Vishveshwara:1970zz,Press:1971wr,Leaver:1986gd}. 
As a consequence of BH uniqueness theorems, the QNM frequencies and damping times of uncharged BHs belong to an infinite set of discrete values that solely depend on the final mass and final spin of the remnant~\cite{Israel:1967wq,Hawking:1972qk,Carter:1971zc,Robinson:1975bv}.
The research program aimed at measuring these frequencies and damping times is known as ``black hole spectroscopy''~\cite{Detweiler:1980gk,Dreyer:2003bv,Berti:2005ys,Berti:2007zu,Meidam:2014jpa} (see~\cite{Berti:2025hly} for a recent review). A recent breakthrough was the detection by the LIGO-Virgo-Kagra (LVK) collaboration of GW250114, a loud event for which at least two QNMs (the fundamental mode and the first overtone) were confidently identified~\cite{LIGOScientific:2025obp, LIGOScientific:2025rid}.
The degree of excitation of different QNMs in a merger remnant depends on the parameters of the binary progenitors. 
The complex amplitudes of the modes grow dynamically in time during the plunge-merger stage~\cite{DeAmicis:2025xuh}.
They later saturate to constant values after a timescale of $\sim 10-20 \, GM/c^3$ past the peak of the radiation, where $M$ denotes the merger remnant mass, $G$ the gravitational constant, and $c$ the speed of light (here and in the rest of the paper we use geometrical units, $G=c=1$).
Through numerical relativity (NR) and analytical work, one can model these constant excitation amplitudes, using them to infer the binary properties for massive signals for which the inspiral is unobservable~\cite{Kamaretsos:2011um,Kamaretsos:2012bs, Barausse:2014oca, Meidam:2014jpa, Miller:2025eak, Hadar:2009ip, Zhang:2013ksa, Kuchler:2025hwx, DeAmicis:2025xuh}. 
Further, accurately modeling the QNM amplitudes and phases reduces systematic biases in parameter estimation and boosts searches of new physics~\cite{Maggio:2022hre,Forteza:2023abc,Toubiana:2023cwr,Toubiana:2024abc,Gupta:2024def,Yi:2025pxe,Pompili:2025cdc,Hu:2025sea}. This is particularly important for future, high signal-to-noise ratio (SNR) observations with the space interferometer LISA~\cite{LISA:2024hlh} and with third-generation ground-based detectors, like the Einstein Telescope (ET)~\cite{Abac:2025saz} and Cosmic Explorer (CE)~\cite{Evans:2023euw}.

Large efforts have been devoted to modeling the complex QNM excitation amplitudes via calibration to numerical simulations~\cite{Berti:2025hly}.
In this work, we focus on two such models: the post-Newtonian inspired closed-form expressions for the QNMs excitation amplitudes of Ref.~\cite{London:2018nxs} (henceforth \texttt{London}) and the model of Ref.~\cite{Cheung:2023vki} (henceforth \texttt{Cheung}), which extended those predictions through an efficient and robust fitting algorithm capable of handling a larger number of modes, including quadratic QNMs~\cite{Gleiser:1996yc,London:2014cma,Cheung:2022rbm,Mitman:2022qdl,Baibhav:2023clw,Bucciotti:2023ets, Perrone:2023jzq, Redondo-Yuste:2023seq, Ma:2024qcv,Bucciotti:2024zyp,Bourg:2024jme,Zhu:2024rej}.
Both models consider quasi-circular, spin-aligned binaries, although these assumptions have recently been relaxed by including either precession~\cite{Nobili:2025ydt} or eccentricity~\cite{Carullo:2024smg}.

The precise time at which the transition between such a dynamical phase and a stationary, constant-amplitude QNM phase takes place depends on the binary parameters and is only partially known.
Earlier times, during merger and early post-merger phases, are contaminated by the prompt response and dynamical growth~\cite{Leaver:1986gd,Zhang:2013ksa,Price:2015gia,Lagos:2022otp,Kuchler:2025hwx,DeAmicis:2025xuh, Oshita:2025qmn,Lu:2025vol} or non-modal effects beyond the linear approximation~\cite{Sberna:2021eui, Lagos:2022otp, Sberna:2021eui, Redondo-Yuste:2023ipg,May:2024rrg, Capuano:2024qhv, Zhu:2024rej}.
This implies an uncertainty on the starting time (relative to the waveform peak amplitude) at which it makes sense to start the QNM analysis.
Starting at early times can induce systematics in the measurement of QNMs, since a pure damped sinusoid model is invalid; conversely, starting too late leads to an exponential reduction in the ringdown SNR.
Ringdown models that go beyond the assumption of purely QNM-based signals have been developed within the effective-one-body framework to address this issue~\cite{Buonanno:1999cu,Buonanno:2000ef,Damour:2001tt}. 
In this approach, the entire post-merger portion is modeled via phenomenological ans\"atze for the time-dependent QNM amplitudes that retain information about the progenitor's parameters, and effectively incorporate transient and nonlinear effects in the early post-merger regime~\cite{Damour:2014yha,DelPozzo:2017rka,Bohe:2017bkm,Cotesta:2018fzk,Nagar:2020eul,Nagar:2020xsk,Estelles:2020twz,Pompili:2023tna}.
These models have the advantage of capturing the entire post-merger signal starting from the peak, but their parameterization does not reflect our first-principles understanding of the underlying physics, and hence they are less useful for the interpretation of possible signatures of new physics.
Moreover, the numerical ans\"atze with multiple free parameters makes it significantly more complicated to easily identify physical quantities of interest.
Out of these models, in this paper we will focus on the non-precessing, quasi-circular \texttt{TEOBPM} template~\cite{Damour:2014yha,DelPozzo:2017rka,Nagar:2020eul,Nagar:2020xsk}.

In this work, we aim to determine the starting time for which these three numerically-informed ringdown models (\texttt{London}, \texttt{Cheung}, and \texttt{TEOBPM}) provide an accurate description of the signal by directly comparing them to NR simulations.
To this end, we perform time-domain mismatch computations~\cite{Isi:2021iql,Siegel:2024jqd}.
These mismatch comparisons do not suffer from overfitting issues, since there is no ``fitting'' involved in our work: the physical modal content has already been determined in~\cite{London:2018gaq, Cheung:2023vki, Nagar:2020xsk,Pompili:2023tna} and is not varied here.

When comparing \texttt{London} and \texttt{Cheung} to numerical relativity simulations, we find that both models fit the waveform well at late times, with mismatches $\sim 10^{-4}$ at times $\sim 25 $M after the merger peak.
For most $(\ell, m)$ waveform harmonics, the \texttt{Cheung} model performs relatively better at earlier times because it includes overtones, it extracts the QNM amplitudes in a time window where they are most stable, and it is built from the same simulation catalog used in this work to validate the models.
When we also include \texttt{TEOBPM} in the comparison, we find that the mismatch between the most accurate ringdown models and numerical simulations typically reaches values between $[10^{-6}, 10^{-4}]$ for the $(\ell,|m|)=$(2,2) harmonic, and $[10^{-4}, 10^{-2}]$ for higher-order modes.
While this level of mismatch is typically acceptable for current detections, it may be insufficient to prevent systematic biases for the high-SNR events anticipated with future detectors~\cite{Baibhav:2018rfk,Baibhav:2020tma,Toubiana:2023cwr,Capuano:2025kkl}.
Our calculations provide a practical criterion to identify the optimal starting time for a given NR-informed ringdown model, and they can be readily applied to current QNM searches through a public repository implementing ready-to-use starting-time interpolants for a given required accuracy~\cite{crescimbeni2025interpolating}.

The remainder of the paper is organized as follows. In Sec.~\ref{sec:inference_framework}, we detail the ringdown modal content, the time-domain mismatch calculation, and its application to the ringdown regime. The main results are presented in Sec.~\ref{sec:results}, and our conclusions are summarized in Sec.~\ref{sec:discussions_conclusions}.

\section{Comparing ringdown models and numerical relativity data}
\label{sec:inference_framework}

\subsection{Ringdown modeling}

We start by describing how we model ringdown signals. 
We will limit attention to quasi-circular, non-precessing binaries. 
Following the conventions of Ref.~\cite{Berti:2025hly}, we can express the ringdown waveform in GR as
\begin{equation}
    h(t) = \sum_{\ell, m, n, \pm}^{\infty} \mathcal{A}^{\pm}_{\ell mn}e^{i\phi^{\pm}_{\ell mn}} \, S^{\pm}_{\ell mn}(\iota, \varphi) \, e^{i \omega^{\pm}_{\ell mn}(t - t^{\pm}_{\ell mn})}\,
    \label{spheroidal_dec}
\end{equation}
where $(\ell,m,n)$ are the angular, azimuthal, and overtone indices, respectively; 
$+$ and $-$ denote the prograde and retrograde modes (see Sec.5.1.6 of~\cite{Berti:2025hly}).
When discussing contributions of a given QNM index, we use the compact notation $(\ell,m,n)^+$ and $(\ell,m,n)^-$.
$_{-2}S_{\ell mn}(\iota, \varphi, \omega^{\pm}_{\ell mn})$ denotes the spin-$2$ spheroidal harmonics;
$\omega^{\pm}_{\ell mn} = 2\pi f^{\pm}_{\ell mn} - i/\tau^{\pm}_{\ell mn}$ are the complex QNM frequencies of the remnant BH, with mass M and spin $\chi$;
$t^{\pm}_{\ell mn}$ is a conventional reference starting time;
$\mathcal{A}_{\ell mn}^{\pm}$ and $\phi_{\ell mn}^{\pm}$ are the amplitudes and phases corresponding to the $(\ell ,m,n)^{\pm}$ modes, which depend on the progenitor parameters $\theta=(\eta,\chi_1,\chi_2)$, where the symmetric mass ratio is $\eta=\frac{m_1m_2}{(m_1+m_2)^2}$ in terms of the binary component masses $m_1$, $m_2$, and $\chi_1,\chi_2$ are the dimensionless progenitor spin components parallel to the orbital angular momentum.
The GW strain in NR simulations is usually expanded in spin-$2$ spherical harmonics ${}_{-2}Y_{\ell m} \left(\iota, \varphi\right)$:
\begin{equation}
    h(t) = h_+(t) - i h_\times(t) = \sum_{\ell,m} h_{\ell m}(t) \; {}_{-2} Y_{\ell m} \left(\iota, \varphi\right) \,.
    \label{spherical_dec}
\end{equation}
In general, Eqs.~\eqref{spheroidal_dec} and~\eqref{spherical_dec} are equivalent upon expanding the spheroidal harmonics in terms of spherical ones
\begin{equation}
    _{-2}S_{\ell mn}(\iota, \varphi, \omega^{\pm}_{\ell mn}) = \sum_{\ell'} \mu_{m\ell'\ell n}^{\pm} \, {}_{-2} Y_{\ell' m}(\theta, \varphi)\,,
\end{equation}
where $\mu^{\pm}_{m\ell\ell'n}$ are the complex spherical-spheroidal mixing factors~\cite{Press:1973zz,Berti:2005gp}, see Ref.~\cite{Berti:2014fga}. Hence, we can write the $h_{\ell m}(t)$ components of the ringdown strain as 
\begin{align}
    h_{\ell m}(t) = \sum_{\ell'=2}^{\infty} \sum_{n=0}^{\infty} \sum_{\pm} \mu_{m\ell\ell'n}^{\pm} \mathcal{A}^{\pm}_{\ell'mn} e^{i\phi_{\ell' mn}^{\pm}} e^{i \omega^{\pm}_{\ell'mn} (t - t^{\pm}_{\ell'm n})}.
    \label{h_l|m|}
\end{align}
Then, each spherical component $(\ell,m)$ corresponds to a superposition of spheroidal QNMs with the same value of $m$ but different values of $\ell$.
Since we focus on non-precessing systems, negative-$m$ modes are related to positive-$m$ modes by symmetry (see e.g.~\cite{Berti:2007fi,Blanchet:2013haa,London:2018gaq}):
\begin{equation}
 h_{\ell,-m} = (-1)^{\ell} h^{*}_{\ell,m}.   
\end{equation}

\subsection{Time-domain formalism and construction of the autocorrelation function}

Here, we describe the mismatch computation between two signals in the time domain following Refs.~\cite{Isi:2021iql, Siegel:2024jqd}.
The starting point is the detector's noise power spectral density in the frequency domain, which we convert into an acyclic estimate of the auto-covariance function (ACF).
We model the instrumental noise as a discrete random process, denoted by $n_i = n(t_i)$ for $i = 1, \ldots, N$, where the time samples are $t_i \in \{0, \Delta t, 2\Delta t, \ldots, (N-1)\Delta t\}$, and the total signal duration is $T = N \Delta t$.
We assume the noise to be Gaussian, and as such fully characterized by its mean and covariance matrix. Without loss of generality, we can set the mean of the noise to zero:
\begin{equation}
    \mu = E[n_i] = 0\,,
\end{equation}
where $E[\cdot]$ denotes the expectation value, and we denote the covariance matrix by
$C_{ij} = E[n_i n_j]$\,.

We also assume the noise to be \textit{stationary}, so the covariance matrix does not depend on the individual times $t_i,\,t_j$, but only on their difference $|t_i-t_j|$:
\begin{equation}
    C_{ij} = \rho(|i - j|)\,,
    \label{cov_m}
\end{equation}
where $\rho(k)$ is known as the ACF. If, in addition to stationarity, we impose periodic boundary conditions, then $\rho(k)=\rho(N-k)$ is said to be cyclic. This function can be estimated by performing the inverse Fourier transform of the power spectral density (PSD) $S(f)$:
\begin{equation}\label{eq:acf_from_psd}
    \rho(k) = \frac{1}{2T} \sum_{j=0}^{N-1} S(|f_j|) e^{2\pi i jk / N}\,,
\end{equation}
where $f_i \in \{0, \Delta f, \ldots, f_{\text{max}}\}$, and
\begin{equation}
    \Delta f = \frac{f_s}{N} = \frac{1}{T}\,,
\end{equation}
with sampling frequency $f_s = 2 f_{\text{max}}$.
In this work we use the LIGO design sensitivity curve provided in Ref.~\cite{LIGOT1800044}.

In practice, since the PSD is a discrete frequency series, computing the ACF requires Fourier transform algorithms (e.g.,~\cite{ifft}). 
Artifacts at high frequencies may be present due to aliasing when the sampling frequency $f_s$ is smaller than the Nyquist frequency, i.e., $f_s < 2f_{\rm max}$; also, care must be taken at lower frequencies to avoid contributions from signal portions below the nominal detector sensitivity.
To prevent these effects, we apply a window function to the edges of the PSD prior when computing its inverse Fourier transform (see Appendix~\ref{sec:window_PSD} for details). 

Given Eq.~\eqref{cov_m}, we can define the scalar product between two signals $x(t)$ and $y(t)$ in the time domain as
\begin{equation}
    \langle x|y\rangle = \sum_{i,j=0}^{N} x_i C_{ij}^{-1} y_j\,.
    \label{sc_prod_full}
\end{equation}

The formalism described above can be applied to GW data analysis of full inspiral-merger-ringdown~(IMR) signals. A ``windowing'' function can be applied to set the IMR signal to zero at the boundary, when the emission has fallen below a detectable level.
This smooth tapering additionally suppresses spectral leakage when performing the Fourier transform~\cite{LIGOScientific:2019hgc}. 
However, the data for ringdown-only analyses are nonzero on the left boundary (being preceded by the merger).
In this case, a standard windowing procedure can still be used, at the price of losing part of the signal or introducing contamination from the merger~\cite{Carullo:2018sfu,Isi:2021iql}. 

To avoid these issues, we work in the time domain without assuming cyclic boundary conditions. 
We restrict our analysis to the time window $\underline t_{\rm trunc}$, defined as 
\begin{equation}
    \underline t_{\rm trunc}=[t_{\rm start},...,t_{\rm end}]
\end{equation}
where $t_{\rm start}$ and $t_{\rm end}$ can be associated to a discretized pair $t_{\rm s}$, $t_{\rm e}$ such that:
\begin{equation}
\begin{cases}
t_{\rm s} \equiv \min(t_i | t_{\rm start} \leq t_i)\,,\\
t_{\rm e} \equiv \max(t_i | t_{\rm end} \geq t_i)\,. 
\end{cases}
\end{equation}

\begin{figure}
    \centering
    \includegraphics[width=1.02\linewidth]{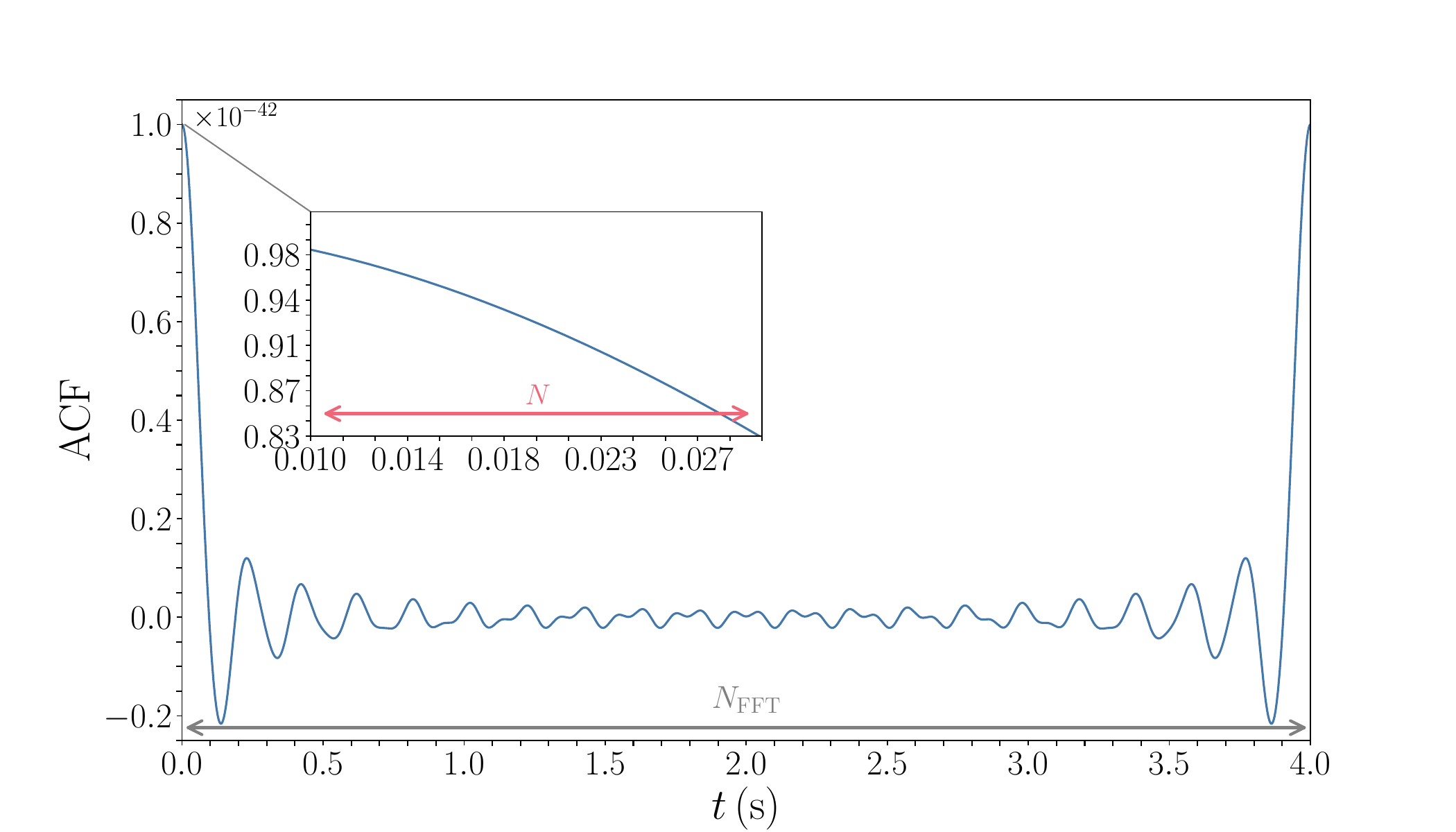}
    \caption{ACF as a function of time in the range $t=[0,~4]\,{\rm s}$, where $T$ is the observation time. The inset shows the truncated ACF in the time range $[t_{\rm s},t_{\rm e}]$, of length $N$.}
    \label{fig:acf_trunc}
\end{figure}

We associate the index $i=0$ to $t=0$, and $i=N_{\text{trunc}}$ to $t_{\rm e}-t_{\rm s}$, where $N_{\text{trunc}}=\text{dim}(\underline t_{\text{trunc}})$.
We construct the autocovariance matrix using an acyclic estimate of the ACF, taking the ACF constructed from Eq.~\eqref{eq:acf_from_psd} (of length $N_{\rm FFT}$) and truncate it at the desired ringdown length $N_{\rm trunc}$, with $N_{\rm FFT}\gg N_{\rm trunc}$.
The resulting $\rho^A(k)$ is illustrated in Fig.~\ref{fig:acf_trunc}, and the corresponding autocovariance matrix is $C^A_{ij}=\rho^A(|i-j|)$. 
Given two ringdown-like signals $x(t)$, $y(t)$, we define a scalar product in the range $t\in [t_{\rm s}, ..., t_{\rm e}]$ as follows:

\begin{equation}
    \langle x|y\rangle_{\rm tr} = \sum_{i,j=0}^{N_{\text{trunc}}} x_i (C^A_{ij})^{-1} y_j\,.
    \label{sc_prod}
\end{equation}%

We are interested in comparing a given ringdown model $h^{\rm rd}_{\ell m} (t)$, where ``rd'' stands for ringdown, against an NR waveform $h^{\rm NR}_{\ell m} (t)$. We define the mismatch between the two waveforms as

\begin{equation}
    \mathcal{M}^{(\ell,m)}=1-\frac{\langle h^{\rm rd}_{\ell m}|h^{\rm NR}_{\ell m}\rangle_{\rm tr}}{\sqrt{\langle h^{\rm rd}_{\ell m}|h^{\rm rd}_{\ell m}\rangle_{\rm tr}\langle h^{\rm NR}_{\ell m}|h^{\rm NR}_{\ell m}\rangle_{\rm tr}}}\,,
    \label{TD_mismatch_tr}
\end{equation}
where we recall that the scalar product of Eq.~\eqref{sc_prod} is defined in the time window 
\begin{equation}
t \in \left[t_{\rm start}^{(\ell,m)},\, t_{\rm end}^{(\ell,m)}\right]\,.
\end{equation}

\section{Mismatch for quasi-circular, non-precessing binaries}
\label{sec:results}

\begin{figure*}[h]
\centering
\includegraphics[width=0.49\textwidth]{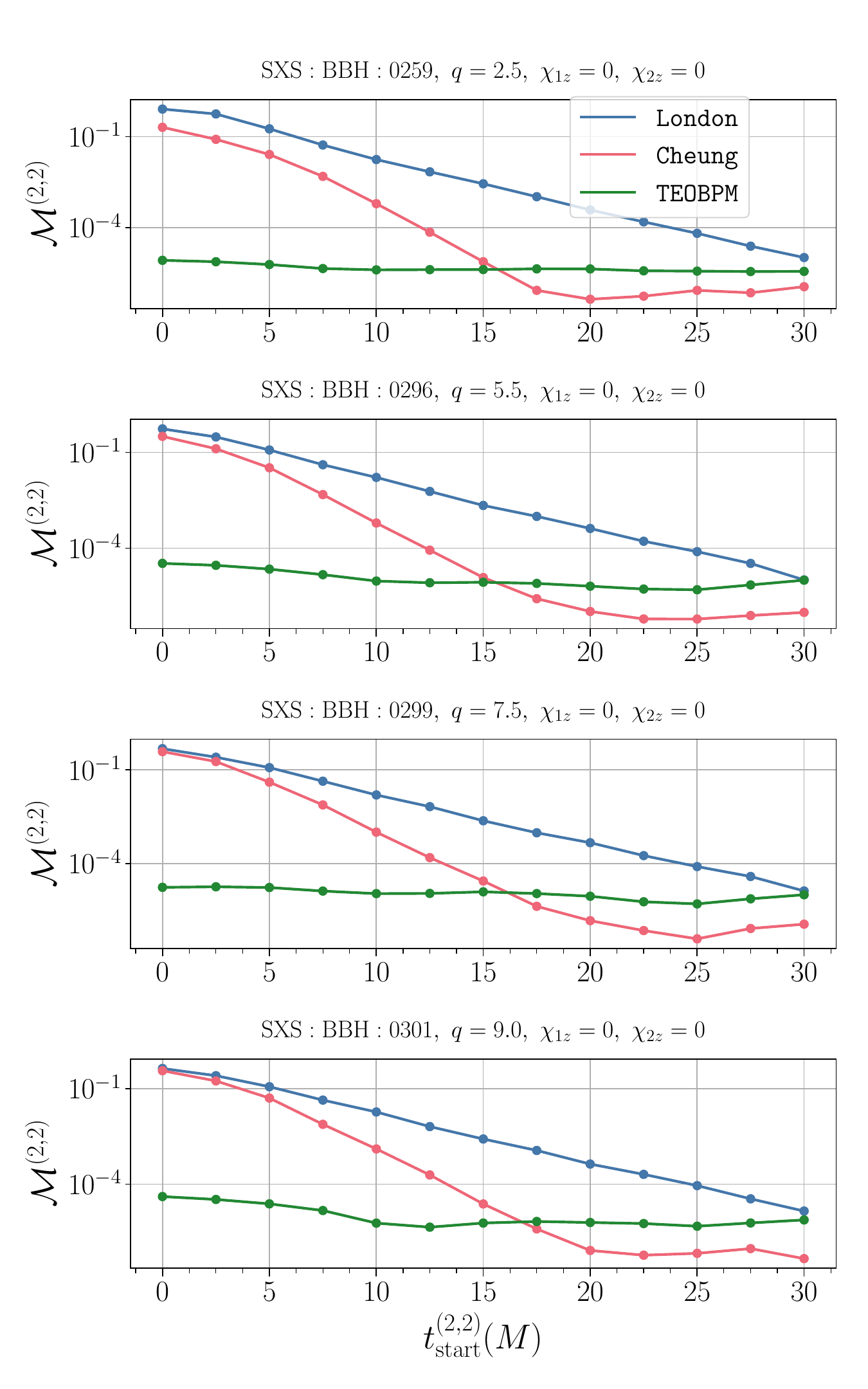}
\includegraphics[width=0.49\textwidth]{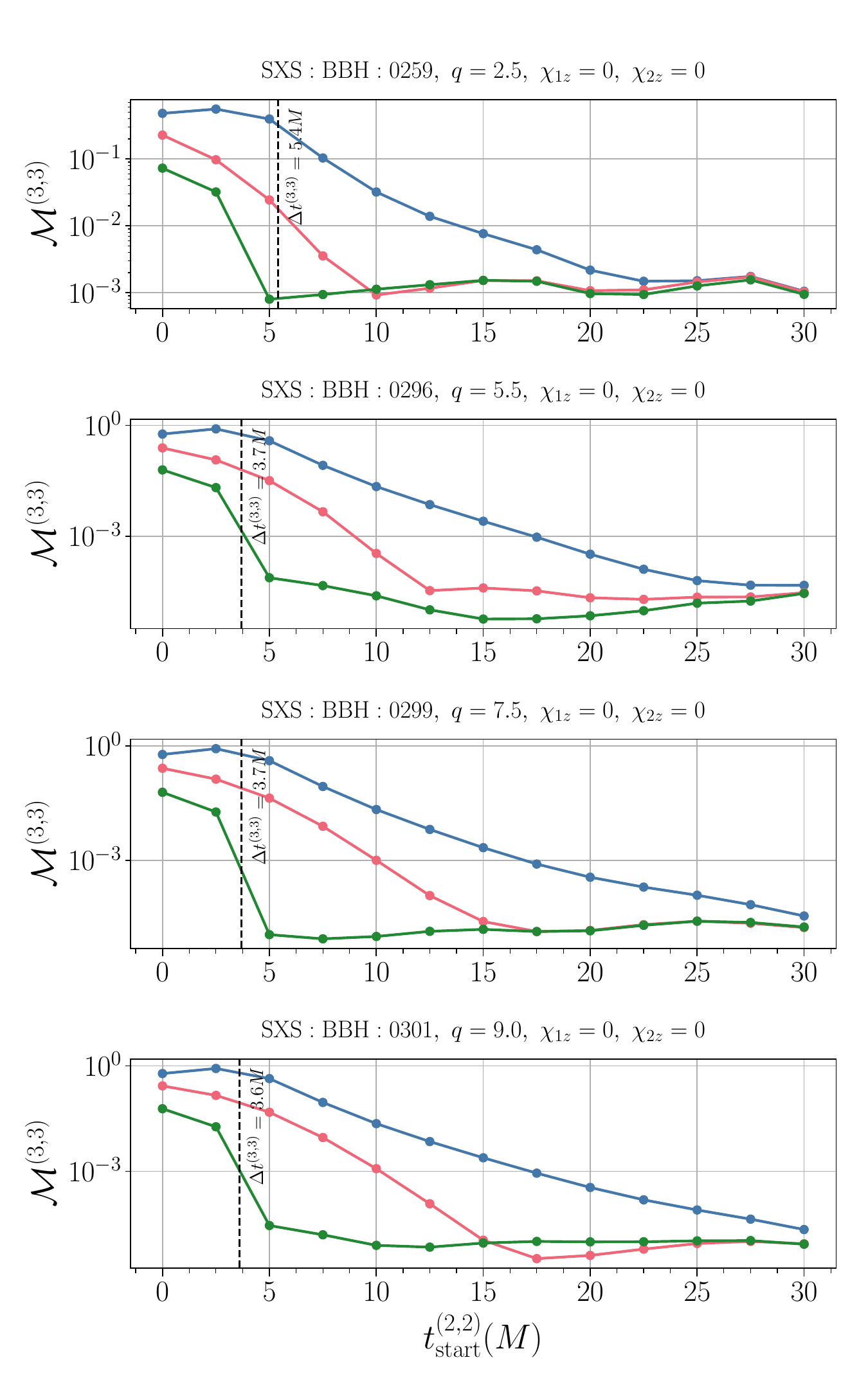}
\caption{Mismatches between the NR waveforms and the \texttt{London} (blue), \texttt{Cheung} (red), and \texttt{TEOBPM} (green) models as a function of the starting time $t^{(2,2)}_{\rm peak}$, in units of the remnant mass. The left column refers to $(\ell,|m|)=(2,2)$, and the right column to $(\ell,|m|)=(3,3)$. In both cases, the origin of the starting time is taken to be the peak of the $(2,2)$ waveform, $t^{(2,2)}_{\rm peak}$. Different rows refer to representative non-spinning NR simulations with varying mass ratios. Dashed vertical black lines mark the time delay $\Delta t^{(\ell,|m|)}$ between the peak of the $(\ell,|m|)$ multipole and the peak of the $(2,2)$ multipole, as defined in Eq.~\eqref{delay}.}
\label{fig:KerrBinary_mm_h+_hx_zero_spin}
\end{figure*}

In this section, we present the main results of our work. We use the mismatch, computed as in Eq.~\eqref{TD_mismatch_tr}, to compare ringdown models against the SXS catalog of NR simulations~\cite{Boyle:2019kee, Scheel:2025jct}. Each of the simulations we considered is labeled \texttt{SXS:BBH:NNNN}, with \texttt{NNNN} denoting a unique simulation ID corresponding to a set of progenitor parameters. We restrict our analysis to non-precessing quasi-circular binaries with spin components in the $x$ and $y$ directions such that $|\chi_{i,x}|, |\chi_{i,y}| \lesssim 10^{-3}$, and with eccentricity $\lesssim10^{-3}$.
Throughout this work, we denote the resulting ensemble of simulations by ${\cal I}_{\rm tot}$.
In the following, we will focus on the progenitor parameters $\theta = (\eta, \chi_{1,z}, \chi_{2,z})$, where $\chi_{1,z}$ and $\chi_{2,z}$ denote the dimensionless spin components projected along the $z$-axis.
 
To access the $(\ell,|m|)$ harmonics of each NR waveform $h^{\rm NR}_{\ell m} (t,\theta)$, we make use of the \texttt{sxs} package~\cite{SXSPackage_v2025.0.15}.
The latter is interfaced through \texttt{bayRing}, a publicly available \texttt{python} package~\cite{carullo_gregorio_2023_8284026} that relies on \texttt{pyRing}~\cite{pyRing} to access ringdown waveform models, and on the \texttt{qnm}~\cite{Stein:2019mop} package to compute QNM frequencies. 
We quote results for GW150914-like values of the remnant mass, $M=62M_{\odot}$, and of the luminosity distance, $d_L=410\,{\rm Mpc}$~\cite{Abbott:2016blz}.

We compute ringdown waveforms $h^{\rm rd}_{\ell m} (t,\theta)$ using the set of parameters $\theta$ associated with the SXS simulation with the given ID.
The \texttt{London} and \texttt{Cheung} models are late-ringdown models composed of pure QNM superpositions with amplitudes and phases that depend on progenitor parameters, while the frequencies and damping times depend on $(M,\chi)$ as predicted by Kerr BH perturbation theory~\cite{Detweiler:1980gk,Berti:2005ys}.
The \texttt{London} amplitudes are calibrated in the parameter space $(\eta, \chi_+, \chi_-)$ at $t=20$M after the time $t^{(2,2)}_{\rm peak}$, at which the absolute value of the $(2,2)$ strain reaches its maximum. The calibration was performed using 101 non-precessing simulations from the Georgia Tech catalog, dated before 2018~\cite{Jani:2016wkt}.
In the \texttt{Cheung} model, the amplitudes and phases of the ringdown are fitted in the same parameter space using a subset of 188 non-precessing simulations from the SXS catalog (we denote this subset of simulations as $\mathcal{I}_{\rm C}$).
Amplitudes and phases are extracted at varying starting times, depending both on the specific simulation and on the stability of the mode under investigation. In practice, this means selecting a time region where the mode amplitudes have stabilized -- i.e., where the signal is well described by a QNM superposition. We will identify those models, using the same nomenclature as in Chapter 6 of Ref.~\cite{Berti:2025hly}, as belonging to the \texttt{KerrBinary} class.

Note that in this work we use simulations from the SXS catalog to quantify the accuracy of the models. This can favor the \texttt{Cheung} model, as it was built using waveforms belonging to the same catalog.
To mitigate it, in some sections of this work, we will only test the models on a subset of the SXS simulations that were not used to build the \texttt{Cheung} model.

The \texttt{TEOBPM} model belongs to the class of NR-informed models covering the entire post-peak regime, identified in Chapter 6 of Ref.~\cite{Berti:2025hly} as the \texttt{KerrPostmerger} class.
Beside modeling the $n=0$ QNM amplitude as a function of the progenitor parameters (just like the \texttt{KerrBinary} models), the model effectively includes non-modal and transient effects that produce time-dependent amplitudes and phases $\{A_{\ell mn}, \phi_{\ell mn}(t)\}$ close to the peak of the waveform amplitude. 
Further, the \texttt{TEOBPM} model depends on $(m_1,m_2,\chi_1,\chi_2)$.

For each simulation, and for each model, we vary the starting time $t_{\rm start}$ between $t^{(2,2)}_{\rm peak}$ and $t_{\rm end}=80$M, and compute the mismatches with respect to the plus and cross polarizations of the signal, denoted by $\mathcal{M}^{\ell, m}_+(t_{\rm start}|t_{\rm end},\theta)$ and $\mathcal{M}^{\ell, m}_\times(t_{\rm start}|t_{\rm end},\theta)$, respectively.
Since the two quantities are simply related by a phase shift, we will plot the \textit{averaged} mismatch, defined as
\begin{equation}
\mathcal{M}^{(\ell,|m|)}=\frac{\mathcal{M}^{(\ell,|m|)}_++\mathcal{M}^{(\ell,|m|)}_{\times}}{2}\,.
\end{equation}

In Fig.~\ref{fig:KerrBinary_mm_h+_hx_zero_spin} we show some examples of the mismatches $\mathcal{M}^{(2,2)}$ and $\mathcal{M}^{(3,3)}$ as a function of the starting time for non-spinning SXS simulations with different mass ratios.
%
Let us first focus on the $(2,2)$ (left) panels. The mismatches of the ringdown models \texttt{London} and \texttt{Cheung} generally decrease with the starting time, since QNM superpositions are a good description of the signal far from the peak.
On the contrary, the mismatch of the \texttt{TEOBPM} model is approximately flat, because the model is designed to also reproduce the merger and early post-merger parts of the signal.
The mismatches of the \texttt{Cheung} model are lower than those of the \texttt{London} model at large enough starting times, roughly by a factor ${\cal O}(10-100)$.
The general trends for the $(3,3)$ multipoles for the \texttt{KerrBinary} models (right panels) are similar to the $(2,2)$.
The \texttt{TEOBPM} mismatch with respect to the numerical $(3,3)$ multipole shows a rapid decrease at early starting times. This is because higher multipoles with $(\ell,|m|)\neq(2,2)$ typically peak after $t_{\rm start}=t^{(2,2)}_{\rm peak}$, with a time delay
\begin{equation}
\Delta t^{(\ell,|m|)}=t^{(\ell,|m|)}_{\rm peak}-t^{(2,2)}_{\rm peak}.
\label{delay}
\end{equation}
The observed drop in the \texttt{TEOBPM} mismatch corresponds to this delay time.

\subsection{$(\ell,|m|)$ mismatch accuracy for a given starting time}
\label{sec:res1}

\begin{figure*}[]
    \includegraphics[width=0.88\textwidth]{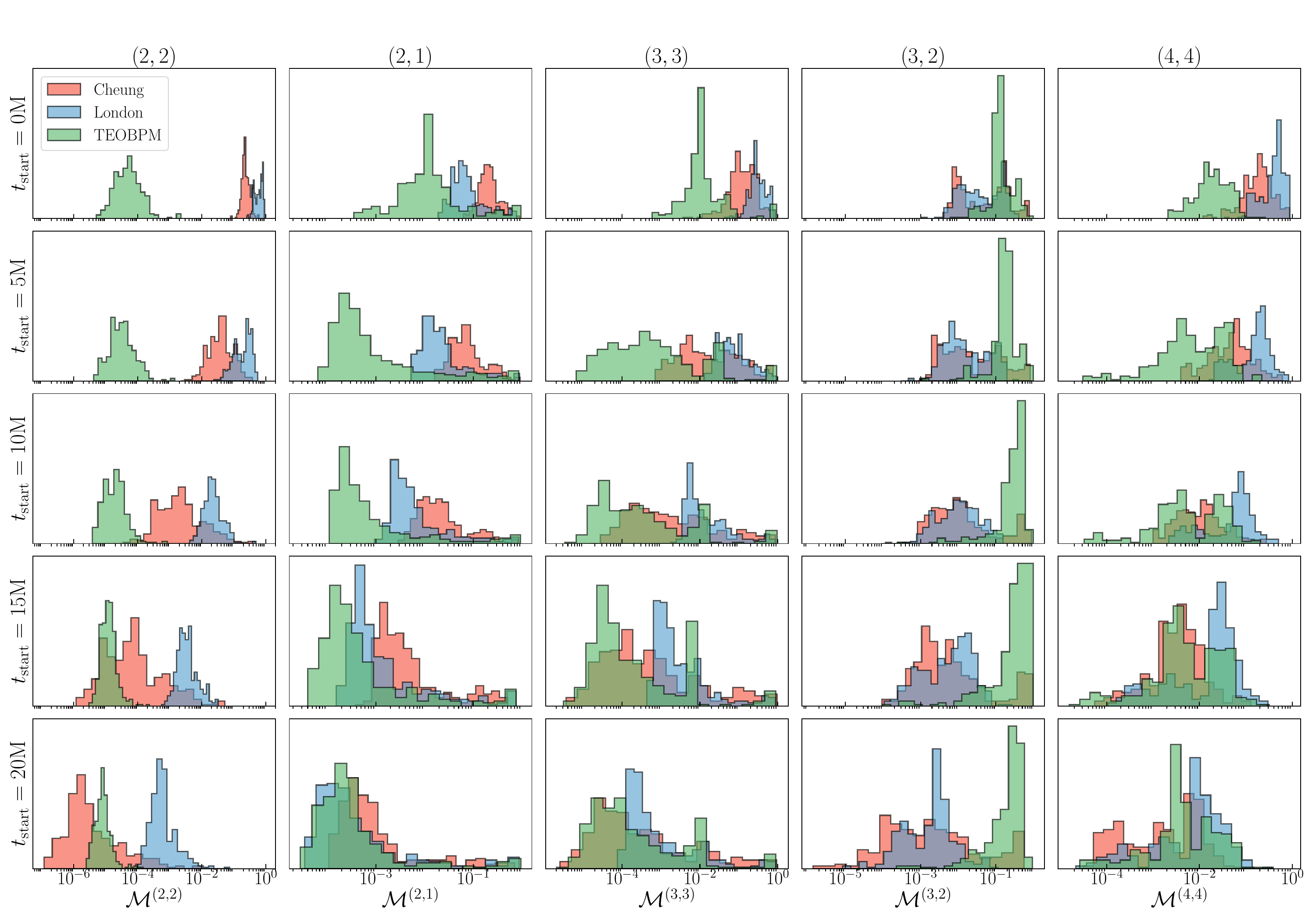}
    \caption{Histograms of mismatches, computed with the simulation set $\mathcal{I}_{\rm tot - C}$, of the \texttt{London} (blue), the \texttt{Cheung} (red), and \texttt{TEOBPM} (green) models at different $(\ell,|m|)$, reported on the different rows. On the columns instead, moving from left to right, are reported the different starting times, defined with repspect to the given $(\ell,|m|)$ mode, from $0$ to $20$M, at steps of $5$M.}
    \label{fig:mm_distrs_l_vs_c}
\end{figure*}

We now address the following question: 
\textit{given a ringdown model, what is the mismatch with respect to the $(\ell,|m|)$ multipole of the corresponding NR simulation catalog at some fixed starting time $t^{(\ell,|m|)}_{\rm start}$?} 
In particular, which of the available models yields the lowest mismatch?

Since the \texttt{Cheung} model was calibrated on a large fraction of SXS data, denoted as $\mathcal{I}_{\rm C}$, we perform mismatches for all models using the subset of simulations not employed for training the \texttt{Cheung} model,
\begin{equation}
    \mathcal{I}_{\rm tot-C} = \mathcal{I}_{\rm tot} \setminus \mathcal{I}_{\rm C}\, .
\end{equation}

As we show in Fig. \ref{mm_distrs_cheung_dataset} in Appendix~\ref{sec:additional_results}, the results do not change appreciably if we consider also the simulations in the set $\mathcal{I}_{\rm C}$: the mismatches of the \texttt{Cheung} model computed over the $\mathcal{I}_{\rm C}$ and $\mathcal{I}_{\rm tot-C}$ sets have similar distributions. 

We apply the procedure explained at the beginning of this section to the subset $\mathcal{I}_{\rm tot-C}$ for the dominant multipoles $(\ell,|m|)=\{(2,2), (2,1), (3,3), (3,2),(4,4)\}$. 
The mismatch histograms at selected starting times for the three ringdown models are plotted in Fig.~\ref{fig:mm_distrs_l_vs_c}. Since the \texttt{TEOBPM} model is calibrated from the peak of each $(\ell,|m|)$ multipole, in this plot we measure the starting time $t^{(\ell,|m|)}_{\rm start}$ with respect to the peak of the $(\ell,|m|)$ multipole, i.e., the starting times presented in Fig.~\ref{fig:mm_distrs_l_vs_c} are defined as
\begin{equation}
    t^{(\ell,|m|)}_{\rm start}=t_{\rm start}-\Delta t^{(\ell,|m|)}\,.
\end{equation}
For completeness, in Appendix~\ref{sec:additional_results} we show a version of this plot in which the mismatches are computed with respect to $t^{(2,2)}_{\rm peak}$ instead. 
The two plots address different questions: in Fig.~\ref{fig:mm_distrs_l_vs_c} we ask how accurate the ringdown waveform is for each mode, while by measuring the mismatches relative to $t^{(2,2)}_{\rm peak}$ (as in Appendix~\ref{sec:additional_results}) we can understand how the different models perform in a data-analysis setting, when \texttt{TEOBPM} is used to analyze the entire post-$t^{(2,2)}_{\rm peak}$ signal~\cite{Gennari:2023gmx}.

For a given $(\ell,|m|)$, the distribution of the \texttt{London} and \texttt{Cheung} mismatches is peaked at progressively lower values for higher starting times. This is particularly evident for the $(\ell,|m|)=(2,2)$ mode, and it occurs because the amplitude fits in the  \texttt{London} model were performed at $t_{\rm start}= 20$M after the peak: the larger the starting time, the closer we are to the fitting dataset. 
By contrast, as shown in Fig.~\ref{fig:KerrBinary_mm_h+_hx_zero_spin}, \texttt{TEOBPM} has roughly similar mismatch distributions at different starting times. 
The \texttt{London} model has a lowest mismatch of ${\cal O}(10^{-4})$ for $t_{\rm start}=20$M, while the \texttt{Cheung} model can have mismatches as low as $\lesssim10^{-7}$, performing better than \texttt{TEOBPM} at $\approx 20$M. 
Our computed \texttt{TEOBPM} mismatches at $t_{\rm start}=0$M are in good agreement with those computed in Ref.~\cite{Gennari:2023gmx}, which did not apply a treatment of the PSD boundaries in the mismatch definition.

For the $(\ell,|m|)=(3,3)$, and $(4,4)$ multipoles, \texttt{TEOBPM} is more accurate than the \texttt{London} and \texttt{Cheung} models at early times, and comparable to those models at late times.
The only exception is the $(4,4)$ multipole, for which the \texttt{Cheung} model mismatch histograms show a tail at low values.
This is likely due to the inclusion of quadratic modes, which are absent in the \texttt{TEOBPM} model.

The trend is reversed in the $(3,2)$ multipole, for which \texttt{TEOBPM} has the largest mismatches at all times.
As pointed out in Ref.~\cite{Gennari:2023gmx}, this is due to the lack of mode mixing, which is instead included in the \texttt{KerrBinary} set of models.

The $(\ell,|m|)=(2,1)$ multipole has an interesting behavior: the \texttt{Cheung} mismatches are worse than the \texttt{London} mismatches for most simulations.
This is because the \texttt{London} model includes the $(2,1,0)^+$ mode only, while the \texttt{Cheung} model includes also the $(2,1,0)^-$ and $(2,1,1)^+$ modes.
The $(2,1,0)^-$ mode does not contribute much compared to the other two modes. 
The $(2,1,1)^+$ mode does contribute, with an amplitude that can be up to one order of magnitude higher than the $(2,1,0)^+$ mode when extrapolated back to the waveform peak time.
Therefore, if the $(2,1,1)^+$ mode is not as precisely modeled as the $(2,1,0)^+$ mode, its inclusion in the model can actually hurt the model's accuracy for two main reasons.
First, the $(2,1,1)^+$ mode was not confidently found in a significant fraction of simulations in Ref.~\cite{Cheung:2023vki}, especially in the regions of the parameter space corresponding to large $\chi_{\rm eff}$ or $\eta \sim 1/4$ (see the figure hosted online at \url{https://mhycheung.github.io/jaxqualin/mode_md/2.1.1}).
The fact that there are not enough simulations with high $\chi_{\rm eff}$ to infer the $(2,1,1)^+$ mode amplitude means that the polynomial fit model is extrapolated, possibly giving unreliable results.
Secondly, the uncertainty of the extracted amplitudes for the $(2,1,1)^+$ mode is relatively high compared to all other modes listed in Ref.~\cite{Cheung:2023vki}, since the $(2,1)$ multipole has low amplitude and is more prone to being contaminated by mode mixing.
This reduces the accuracy of the model even when extrapolation is not necessary.
At the end of this section, we will discuss what happens to the mismatch distribution when we remove one mode at a time.

\subsection{Starting time of ringdown models from fixed mismatch threshold}
\label{sec:res2}

In this subsection, we address a complementary question regarding \texttt{KerrBinary} models: \textit{what is the earliest starting time $t_{\rm start}$ at which a target mismatch can be achieved?}
Our analysis will determine the time intervals over which a given model provides sufficient agreement with NR simulations. 
In practise, we will find that if the chosen mismatch threshold is too stringent, there may be regions of the parameter space $(\eta, \chi_+, \chi_-)$ where the model fails to meet the desired accuracy for {\em any} starting time.

\subsubsection{Starting time statistics for a given mismatch threshold}

To address the above question, we generate a histogram of the starting times $t^{(\ell,|m|)}_{\rm start}$ at which the mismatch reaches a threshold value of $\mathcal{M}_{\rm thr}=[10.0, 3.5, 0.5, 0.1]\%$ (if at all).
We only report results for the \texttt{KerrBinary} models: the need to introduce a starting time arises because these models lack accuracy at early times. 
To have a fair comparison, we compare the \texttt{London} and \texttt{Cheung} models over the dataset $\mathcal{I}_{\rm tot - C}$.
A summary of our results is presented in Table~\ref{tab:mismatch_crossings_times}.

Let us first focus on the $(2,2)$ mode. For a given threshold, the \texttt{Cheung} model typically achieves a lower mean starting time with respect to the \texttt{London} model. The reason is clear from Fig.~\ref{fig:KerrBinary_mm_h+_hx_zero_spin}: the mismatch as a function of the starting time decays faster for the \texttt{Cheung} model, and therefore $t_{\rm start}(\mathcal{M}=\mathcal{M}_{\rm thr})$ is lower. 

This is not always true for higher modes. The starting times corresponding to a given threshold mismatch are often comparable, and for $(\ell,|m|)=(2,1)$ case the \texttt{London} model actually performs better. This is consistent with the second column of Fig.~\ref{fig:mm_distrs_l_vs_c}: there the distribution of the mismatches for the \texttt{Cheung} model peaks at higher values with respect to the corresponding distribution for the \texttt{London} model.

\begin{table}[!t]
\centering
\scalebox{0.84}{%
\begin{tabular}{ccccccc}
\hline
$\mathcal{M}_{\rm thr}$ & Model & $t_{\rm start}^{(2,2)}\,[M]$ & $t_{\rm start}^{(2,1)}\,[M]$ & $t_{\rm start}^{(3,3)}\,[M]$ & $t_{\rm start}^{(3,2)}\,[M]$ & $t_{\rm start}^{(4,4)}\,[M]$ \\
\hline
10.0\% & \texttt{London} & $6.7 \pm 1.7$ & $-1.0 \pm 4.0$ & $4.5 \pm 3.1$ & $-1.6 \pm 6.4$ & $7.5 \pm 4.7$ \\
& \texttt{Cheung} & $3.5 \pm 1.9$ & $4.4 \pm 4.4$ & $1.9 \pm 4.5$ & $-1.3 \pm 6.8$ & $3.3 \pm 4.0$ \\
\hline
3.5\% & \texttt{London} & $9.2 \pm 1.7$ & $3.3 \pm 4.0$ & $7.4 \pm 3.5$ & $2.2 \pm 7.6$ & $12.0 \pm 5.3$ \\
& \texttt{Cheung} & $5.9 \pm 2.2$ & $8.5 \pm 4.5$ & $4.4 \pm 4.9$ & $0.7 \pm 8.3$ & $7.4 \pm 5.7$ \\
\hline
0.5\% & \texttt{London} & $14.4 \pm 2.4$ & $10.4 \pm 3.8$ & $12.0 \pm 2.5$ & $11.1 \pm 6.4$ & $17.5 \pm 6.0$ \\
& \texttt{Cheung} & $9.4 \pm 2.6$ & $13.9 \pm 3.5$ & $6.8 \pm 3.4$ & $9.2 \pm 5.6$ & $13.9 \pm 5.3$ \\
\hline
0.1\% & \texttt{London} & $18.9 \pm 2.6$ & $14.7 \pm 3.4$ & $16.4 \pm 2.4$ & $15.1 \pm 4.5$ & $16.6 \pm 4.8$ \\
& \texttt{Cheung} & $11.9 \pm 2.9$ & $17.7 \pm 2.3$ & $10.3 \pm 4.4$ & $15.7 \pm 3.9$ & $16.9 \pm 4.0$ \\
\hline
\end{tabular}}%
\caption{Mean $\pm$ 1-$\sigma$  values of the starting time $t_{\rm start}$ at which the \texttt{KerrBinary} models drop below a given mismatch threshold compared to the NR simulations.}
\label{tab:mismatch_crossings_times}
\end{table}

\subsubsection{Fraction of simulations that reach a given mismatch threshold}

\begin{figure}[!t]
    \centering
    \includegraphics[width=0.9\linewidth]{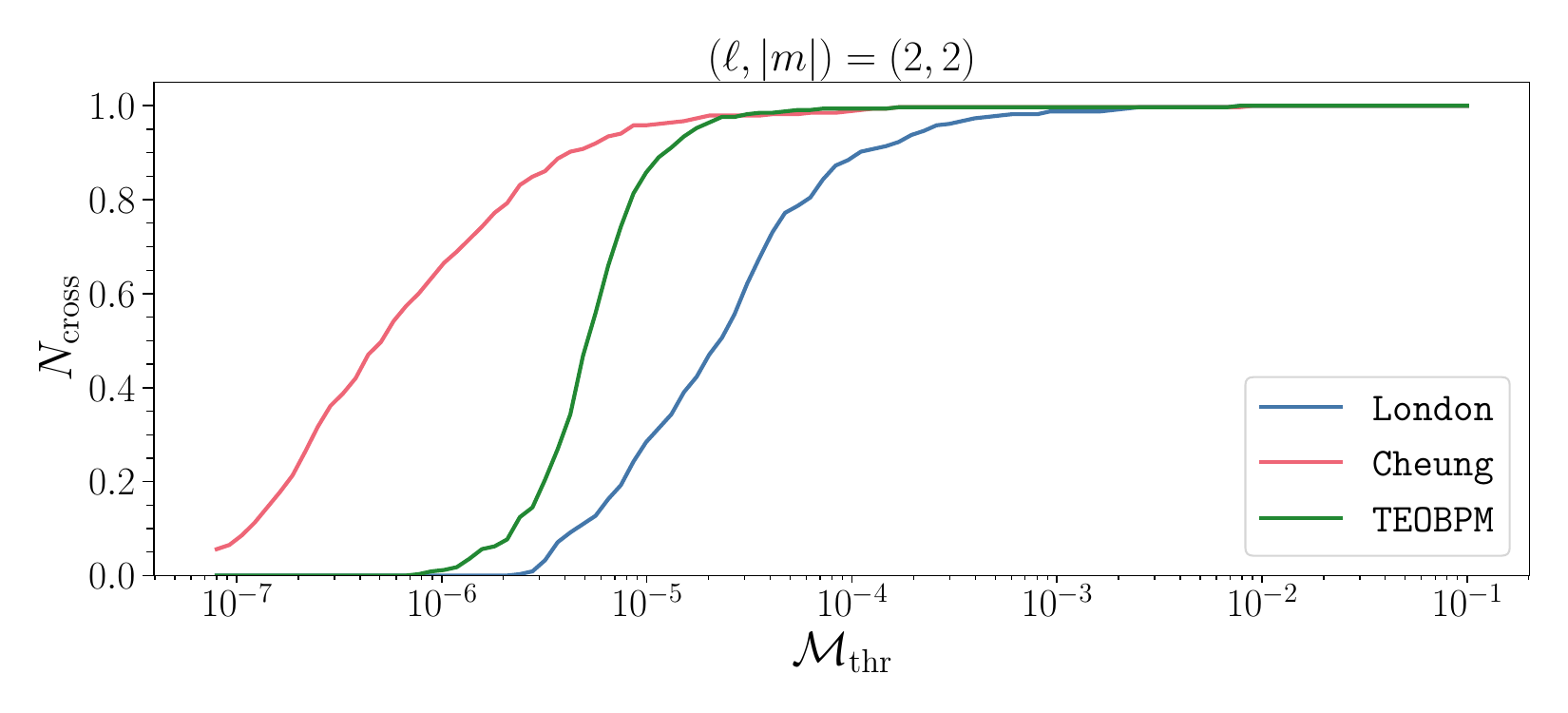}

    \includegraphics[width=0.9\linewidth]{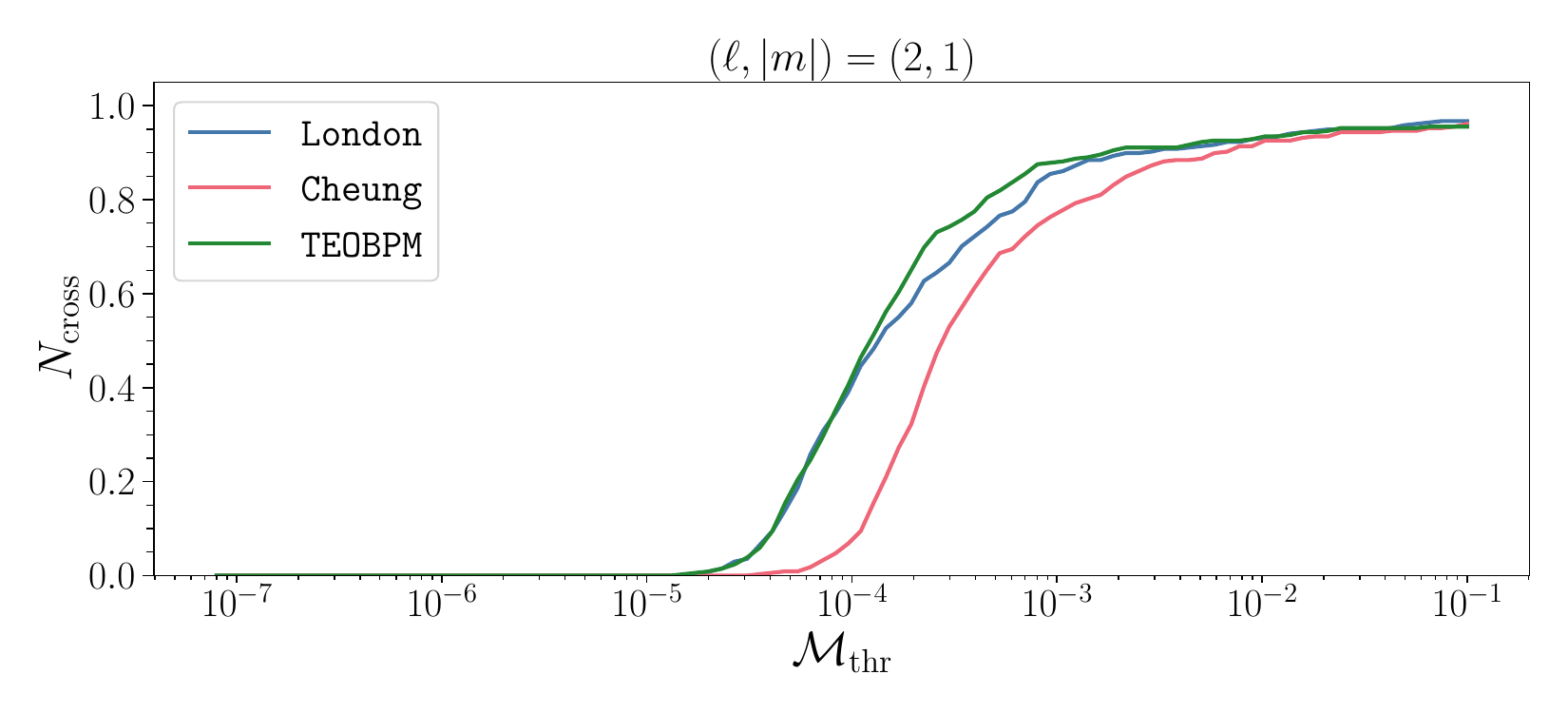}
    
    \includegraphics[width=0.9\linewidth]{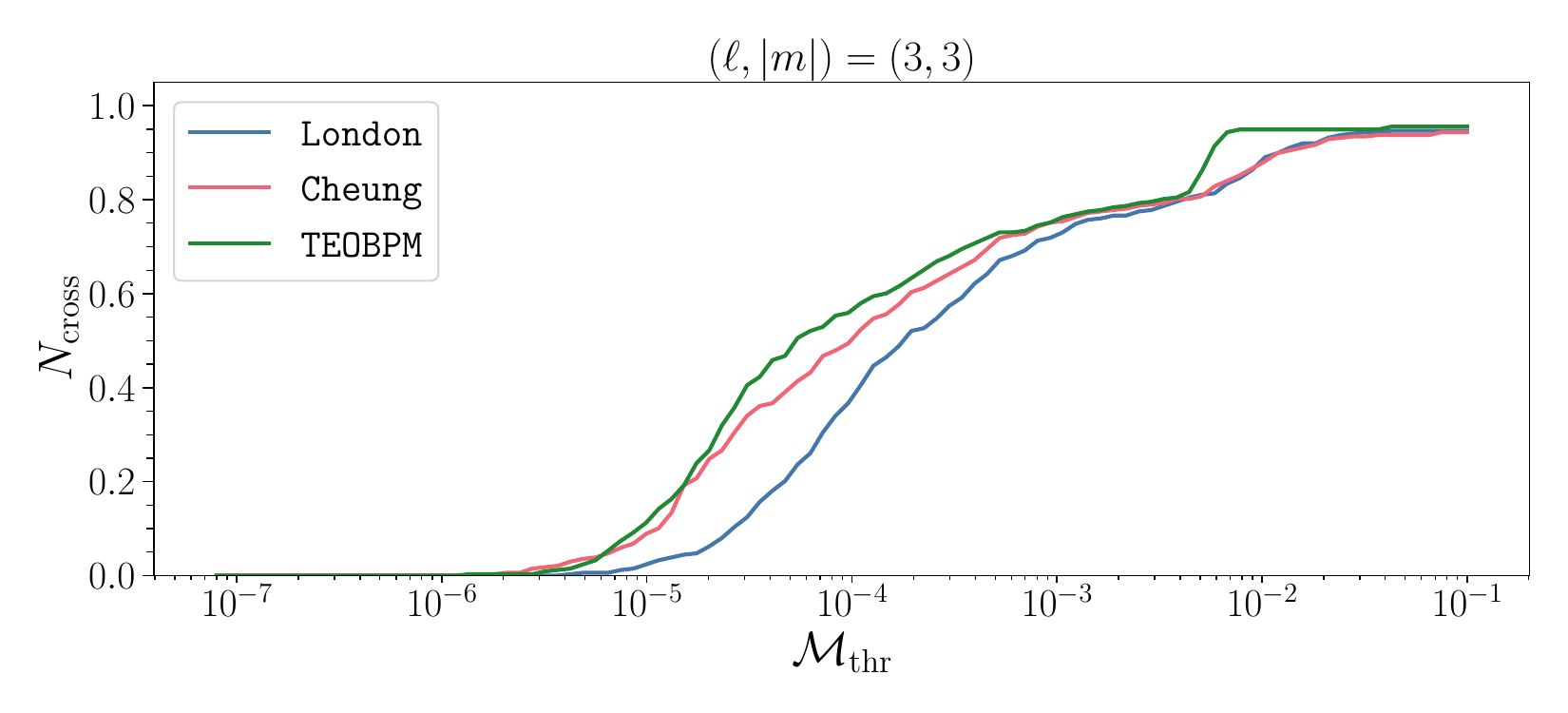}
    
    \includegraphics[width=0.9\linewidth]{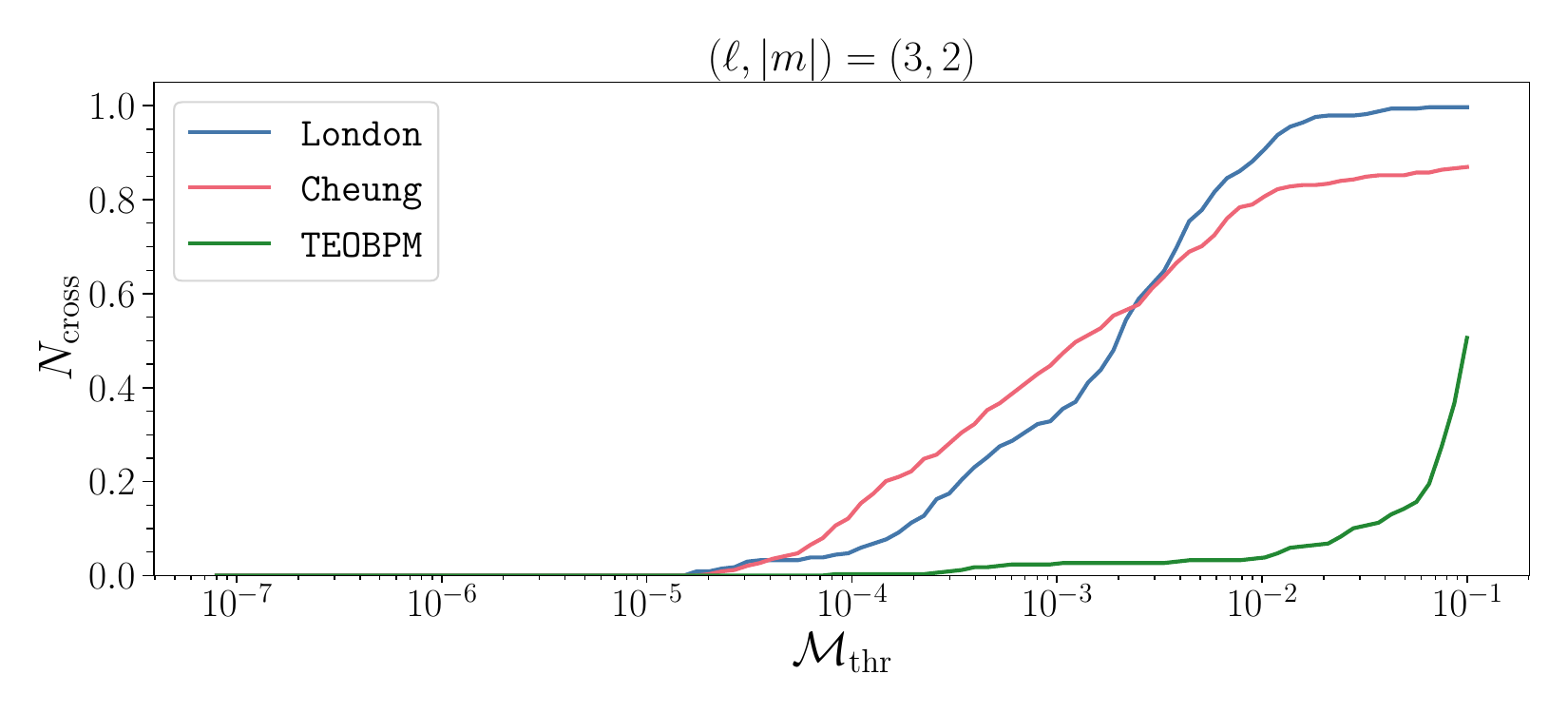}
    
    \includegraphics[width=0.9\linewidth]{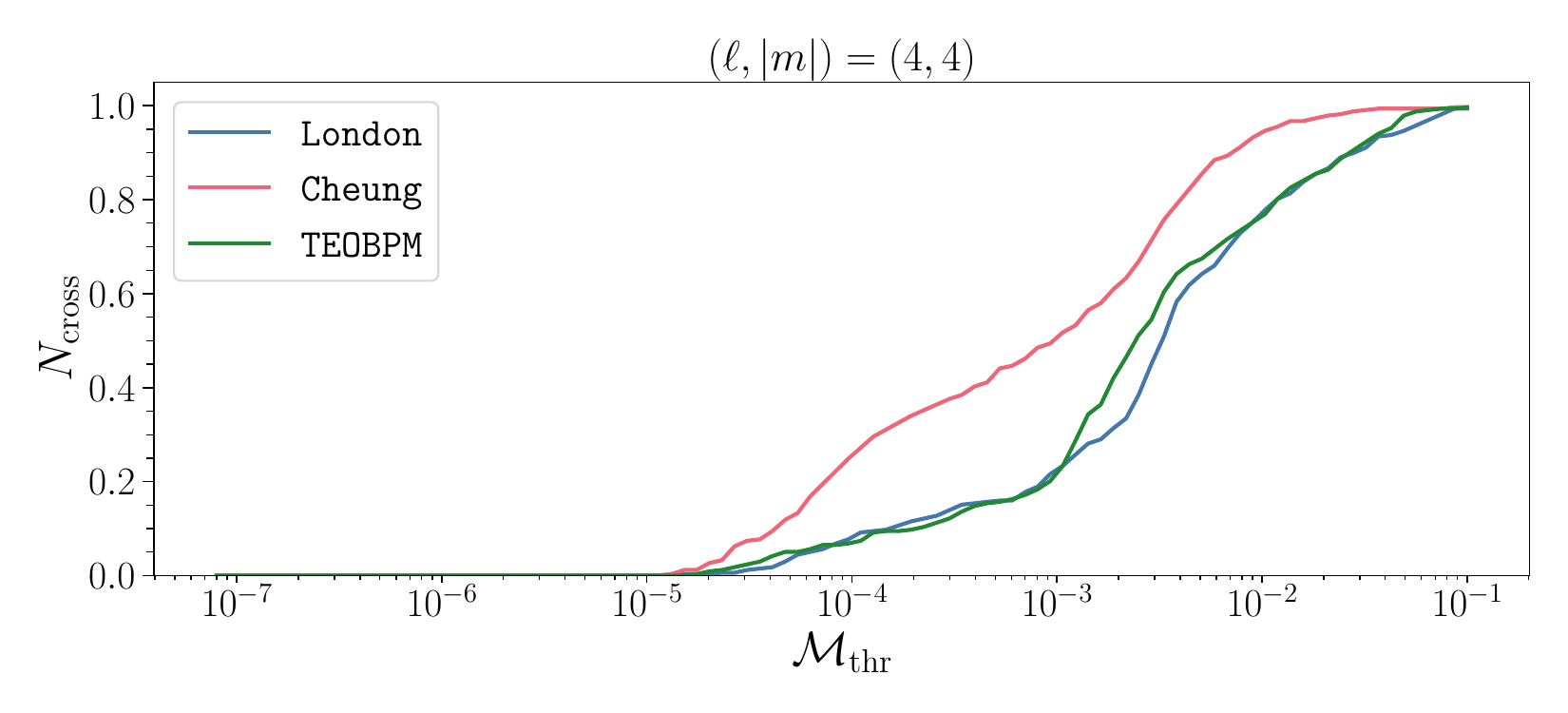}
    \caption{Number of simulations $N_{\rm cross}$ that can cross a given mismatch threshold for different $(\ell,|m|)$ harmonics, computed at the minimum starting time at which the mismatch threshold is achieved.
    }
    \label{fig:n_cross}
\end{figure}

It is also interesting to quantify the fraction of simulations that could reach a given mismatch threshold for a suitable choice of starting time. 
For each $(\ell,|m|)$ multipole, we compute the quantity
\begin{equation}
    N_{\rm cross}=\frac{\text{\{\#simulations | }\mathcal{M}\leq\mathcal{M}_{\rm thr}\}}{\text{\{\#total number of simulations\}}}
\end{equation}
as a function of $\mathcal{M}_{\rm thr}$, using the minimum starting time at which the threshold is met.
%
The result for the various models is shown in Fig.~\ref{fig:n_cross}.
The $(2,2)$ mode has $N_{\rm cross}$ very close to unity when $\mathcal{M}_{\rm thr}\gtrsim
10^{-4}$ for the \texttt{TEOBPM} and \texttt{Cheung} models, and at $\mathcal{M}_{\rm thr}\gtrsim 10^{-3}$ for the \texttt{London} model. The implication is that the models are accurate for all simulations, at least at that level. For lower mismatches, $N_{\rm cross}$ drops below unity: the required mismatch is never achieved for some of the simulations, regardless of the starting time. The mismatch drop of the \texttt{Cheung} model is slower than for the \texttt{TEOBPM} model (see the bottom left panel of Fig.~\ref{fig:mm_distrs_l_vs_c}). For $(\ell,|m|)\neq(2,2)$, $N_{\rm cross}$ drops below unity at higher mismatch thresholds, confirming that the models are less accurate for higher harmonics.

\subsubsection{Starting times such that \texttt{KerrBinary} models have the same mismatch as \texttt{TEOBPM}}

\begin{figure}[!t]
    \centering

    \includegraphics[width=0.4\textwidth]{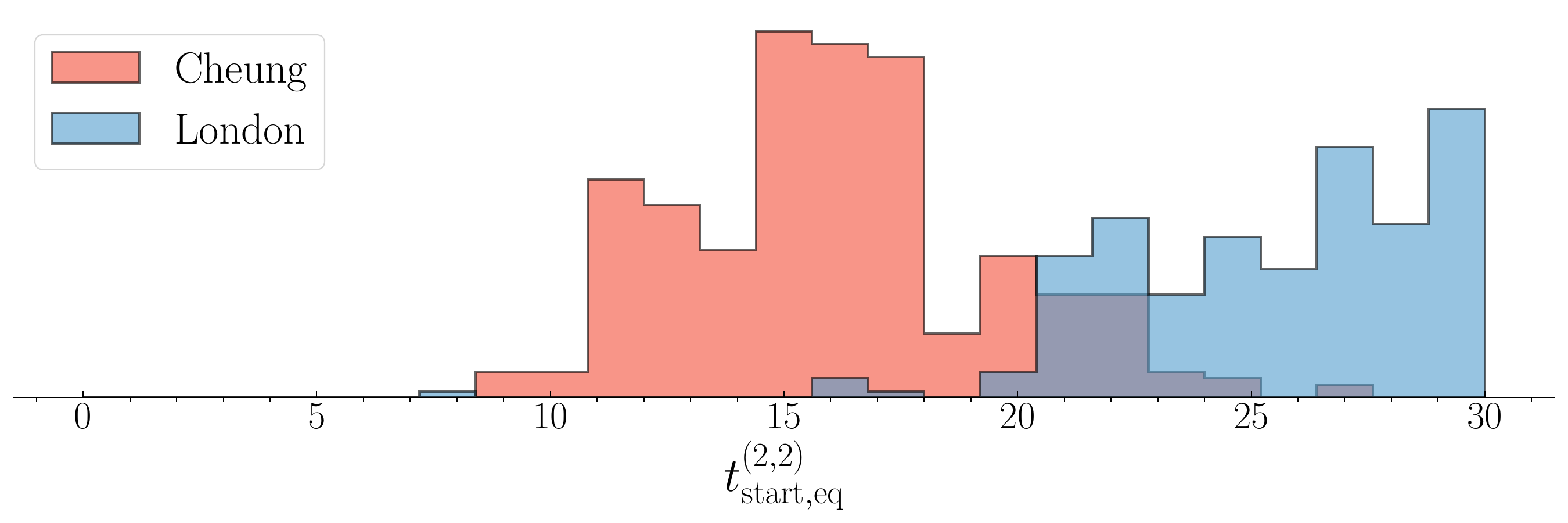}
    \hfill
    \includegraphics[width=0.4\textwidth]{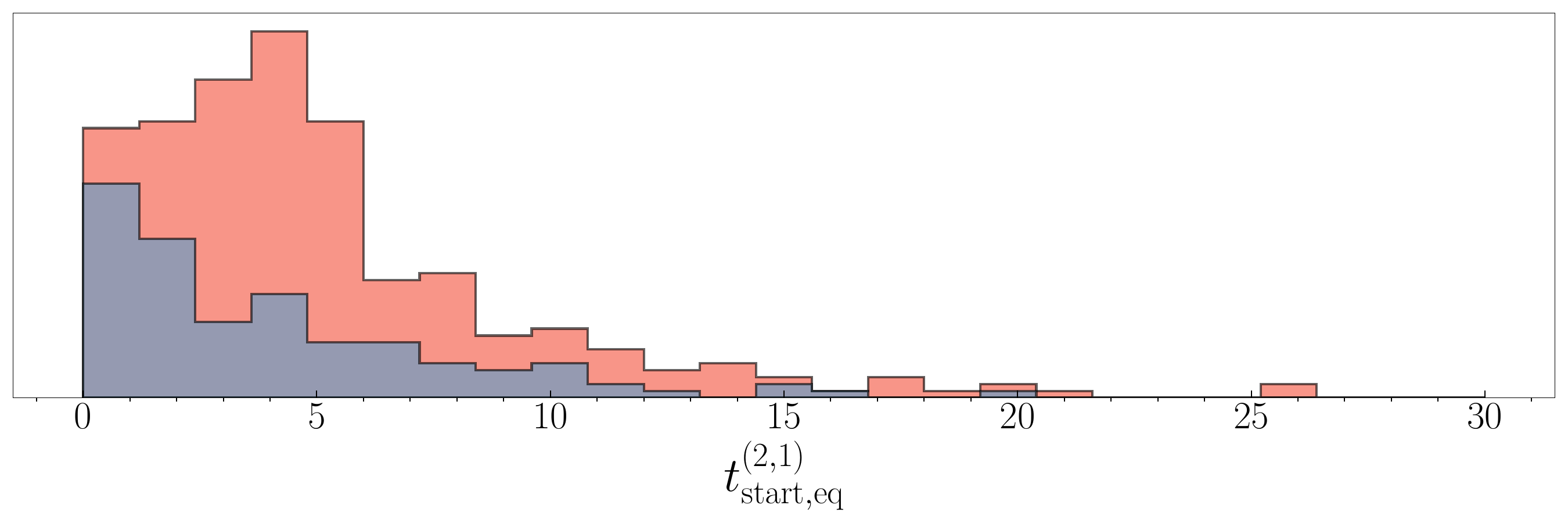}

    \vspace{1em}

    \includegraphics[width=0.4\textwidth]{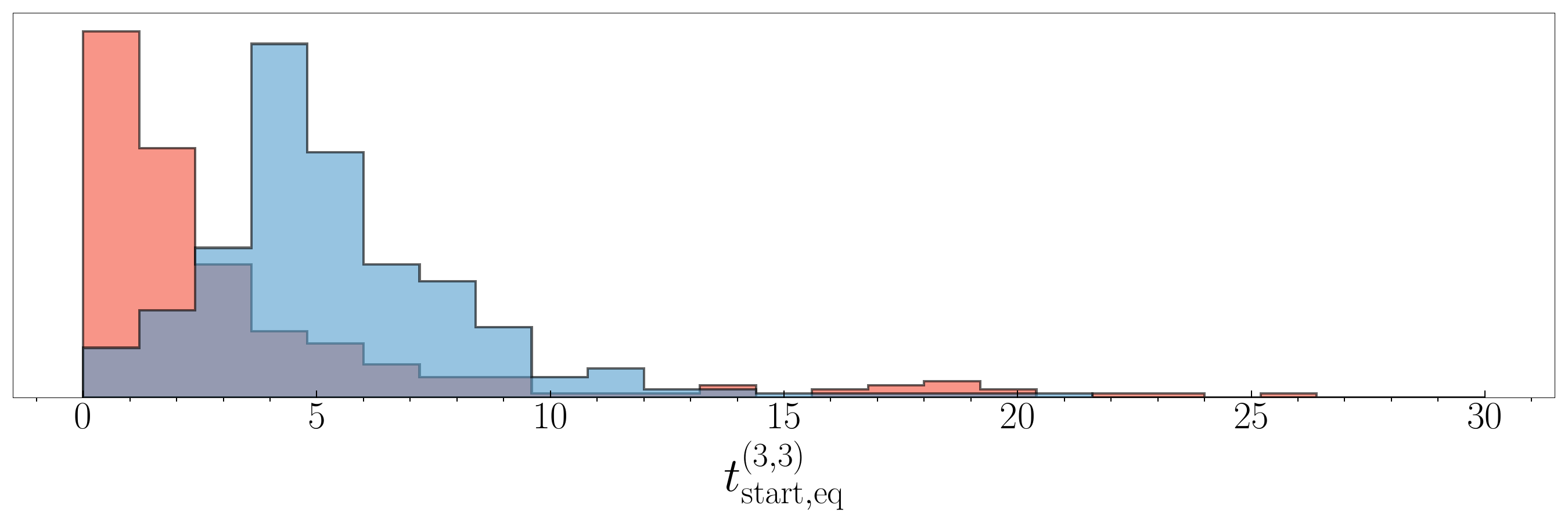}
    \hfill
    \includegraphics[width=0.4\textwidth]{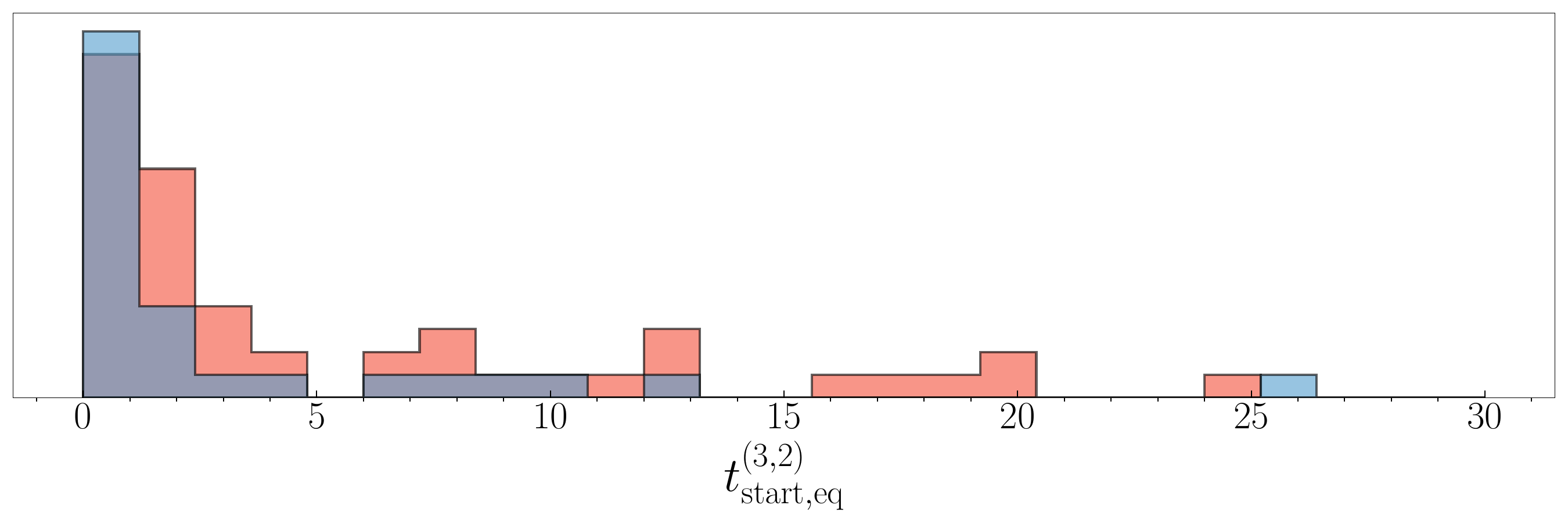}

    \vspace{1em}

    \includegraphics[width=0.4\textwidth]{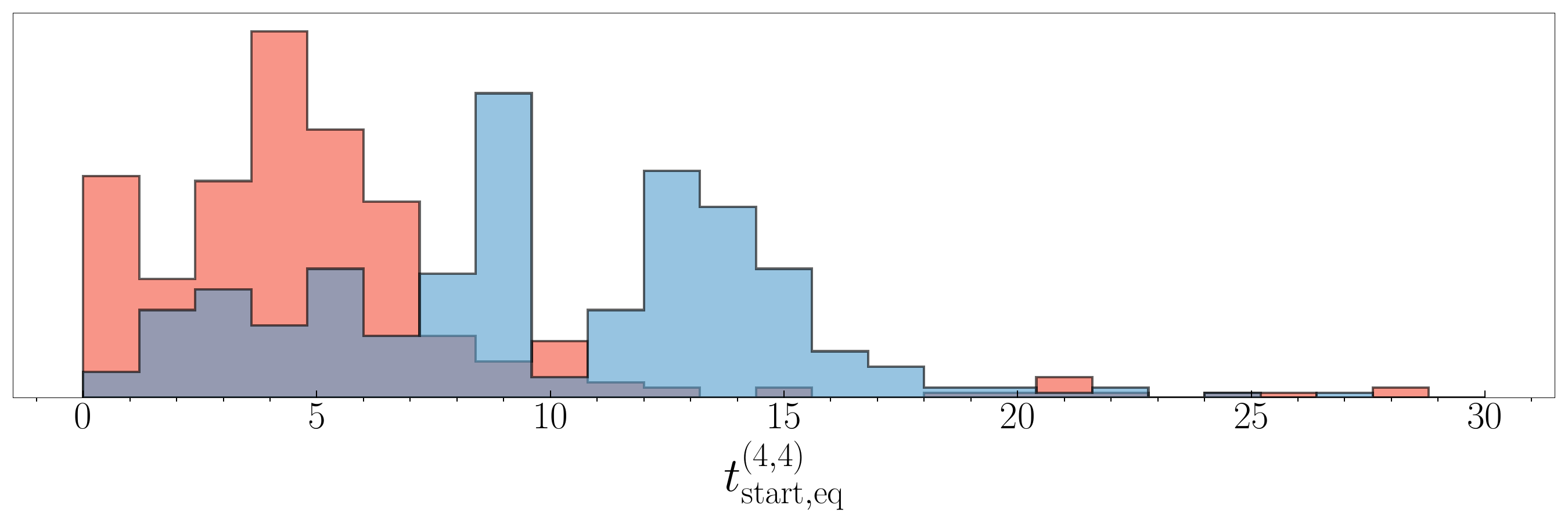}

    \caption{Histograms of the starting times $t_{\rm start, eq}^{(\ell,|m|)}$ such that the \texttt{Cheung} (red) and \texttt{London} (blue) model mismatches are equal to the \texttt{TEOBPM} model mismatches, as defined in Eq.~\eqref{cross_t_start}.}
    \label{fig:cheung_london_cross_t_start}
\end{figure}

As we pointed out when we discussed Fig.~\ref{fig:KerrBinary_mm_h+_hx_zero_spin}, there are two clear trends for any given SXS simulation: (1) the \texttt{TEOBPM} mismatches, expecially for the (2,2) mode, are nearly constant as a function of the starting time, since the effect of non-linearities and transient effects are included in the modeling from the peak; (2) the \texttt{London} and \texttt{Cheung} models mismatches decrease as functions of the starting time, because perturbation theory becomes a better approximation at late times. In general, there will be some critical starting time at which the mismatch of the QNM-based models becomes comparable with the mismatch of the \texttt{TEOBPM} model. Here we quantify the distribution of such starting times, defined as
\begin{equation}
  t^{(\ell,|m|)}_{\rm start,eq}=t^{(\ell,|m|)}_{\rm start}\Bigl(\mathcal{M}^{(\ell,|m|)}_{\rm \texttt{KerrBinary}}=\mathcal{M}^{(\ell,|m|)}_{\rm \texttt{TEOBPM}}\Bigr).
  \label{cross_t_start}
\end{equation}
At these times, roughly speaking, the QNM part of the signal should become dominant in the \texttt{TEOBPM} model.

In Fig.~\ref{fig:cheung_london_cross_t_start} we plot histograms of this quantity. In Table~\ref{fr_cross} we list the fraction of the total set of simulations $\mathcal{I}_{\rm tot}$ such that $\mathcal{M}^{(\ell,|m|)}_{\rm\texttt{KerrBinary}}=\mathcal{M}^{(\ell,|m|)}_{\rm \texttt{TEOBPM}}$. 
For the $(2,2)$ mode, the \texttt{Cheung} model distribution peaks around $(15 \pm 5)$M, with $\approx 99\%$ of the simulations such that $\mathcal{M}^{(\ell,|m|)}_{\rm \texttt{Cheung}}=\mathcal{M}^{(\ell,|m|)}_{\rm \texttt{TEOBPM}}$. By contrast, the \texttt{London} model achieves comparable accuracy to \texttt{TEOBPM} at later times, with  $\approx68\%$ of the simulations yielding $\mathcal{M}^{(\ell,|m|)}_{\rm \texttt{London}}=\mathcal{M}^{(\ell,|m|)}_{\rm \texttt{TEOBPM}}$.
The starting time distributions shift toward earlier times for the $(2,1)$, $(3,3)$, and $(4,4)$ multipoles. This reflects the lower accuracy of both the \texttt{TEOBPM} and the \texttt{KerrBinary} models at early times for these higher harmonics.
For the $(3,2)$ mode, the mismatches obtained with \texttt{TEOBPM} are systematically worse than those of the \texttt{London} and \texttt{Cheung} models across all times. This is why the starting time distributions found for this multipole in Fig.~\ref{fig:cheung_london_cross_t_start} are tightly concentrated at low values.

\begin{table}[t]
\centering
\begin{tabular}{ccc}
\hline
\hline
$(\ell,|m|)$ & $\%(\mathcal{M}^{(\ell,|m|)}_{\rm \texttt{Cheung}}=\mathcal{M}^{(\ell,|m|)}_{\rm \texttt{TEOBPM}})$ & $\%(\mathcal{M}^{(\ell,|m|)}_{\rm \texttt{London}}=\mathcal{M}^{(\ell,|m|)}_{\rm \texttt{TEOBPM}})$ \\
\hline
\hline
(2,2) & 99\% & 68\% \\
(2,1) & 97\% & 98\% \\
(3,3) & 98\% & 99\% \\
(3,2) & 35\% & 32\% \\
(4,4) & 98\% & 98\% \\
\hline
\end{tabular}
\caption{Fraction of the simulations such that the \texttt{Cheung} and \texttt{London} models have the same mismatch as \texttt{TEOBPM} at some starting time, for different multipoles.
}
\label{fr_cross}
\end{table}

\subsection{The role of mode exclusion for the \texttt{Cheung} model}

\begin{figure*}[!t]
    \centering
    \includegraphics[width=0.94\textwidth]{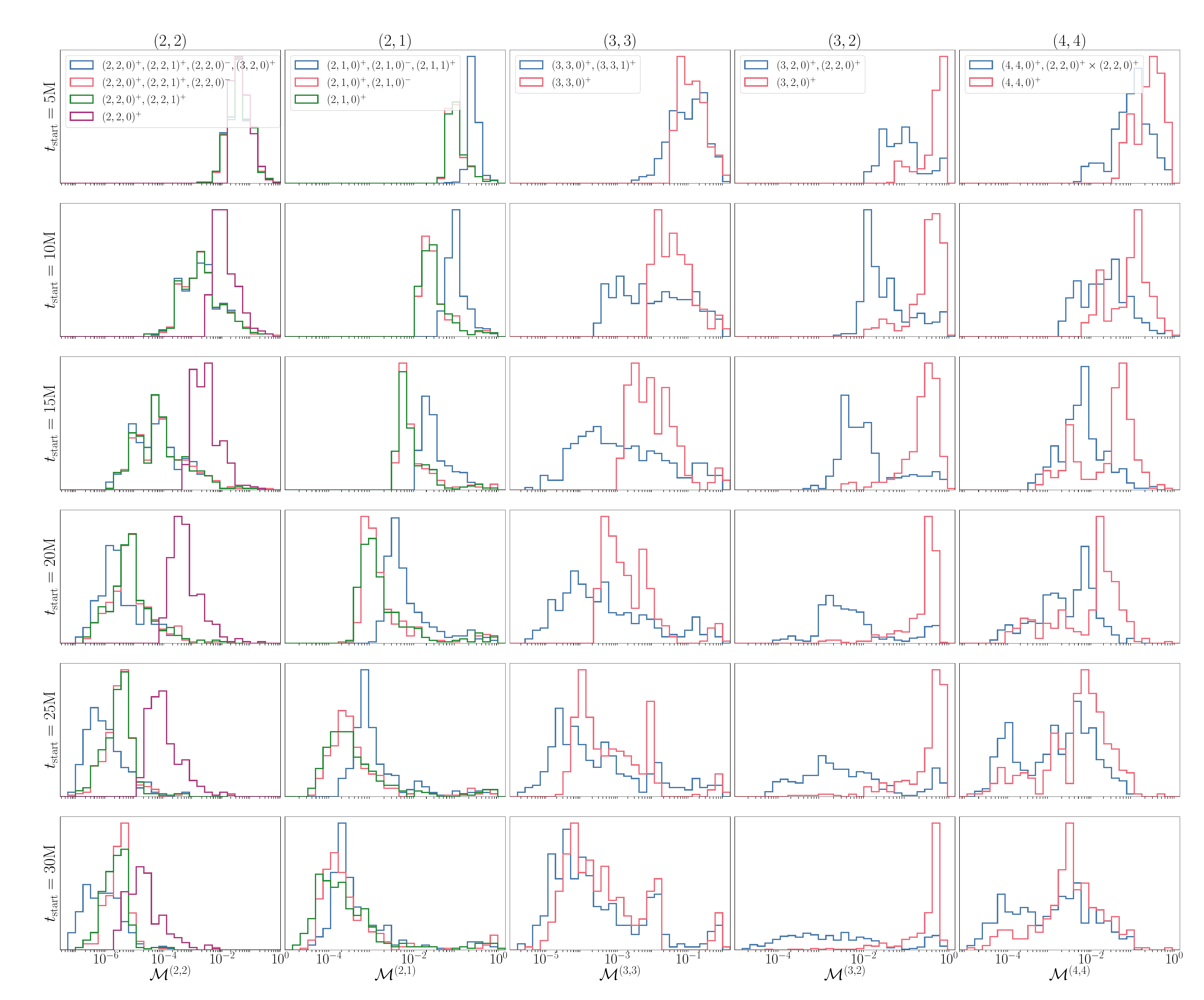}
    \caption{\texttt{Cheung} mismatch distributions for different $(\ell,|m|)$ multipoles (as indicated at the top of each column), computed at discrete values of $t_{\rm start}$ ranging from $0$M to $30$M (as indicated to the left of each row). Different colors denote different sets of QNMs (see Table~\ref{tab:modes_lm}).}
    \label{fig:Cheung_remove_modes}
\end{figure*}

\begin{table}[t]
\centering
\begin{tabular}{c|c}
\hline
\hline
$(\ell,|m|)$ & Contributing $(\ell, m, n, \pm)$  modes\\
\hline
\hline
(2,2) & (2,2,0)$^{\pm}$, (2,2,1)$^{+}$,(3,2,0)$^{+}$ \\
(2,1) & (2,1,0)$^{\pm}$, (2,1,1)$^{+}$ \\
(3,3) & (3,3,0)$^{+}$, (3,3,1)$^{+}$ \\
(3,2) & (3,2,0)$^{+}$, (2,2,0)$^{+}$ \\
(4,4) & (4, 4, 0)$^{+}$, ((2, 2, 0)$^+\times$(2, 2, 0)$^{+}$) \\
\hline
\end{tabular}
\caption{Dominant QNMs contributing to the various multipoles $(\ell, |m|)$ for the \texttt{Cheung} model.}
\label{tab:modes_lm}
\end{table}

We now address another important question:  
\textit{given a spherical multipole $(\ell,|m|)$, which QNMs contribute most significantly to the agreement between a given ringdown model and the NR signal?} 
From Eq.~\eqref{h_l|m|} we see that each $(\ell,|m|)$ multipole generally receives contributions from multiple QNMs $(\ell',m,n)$, coming from either higher overtones or mode mixing.

For a given $(\ell,|m|)$ mode, associated with a list of contributing QNMs, we systematically remove individual $(\ell,m,n)$ modes from the \texttt{Cheung} model waveform.
We list the QNMs implemented in \texttt{pyRing} for each $(\ell,|m|)$ multipole in Table~\ref{tab:modes_lm}. 
For each $(\ell,|m|)$, we remove the last $(\ell,m,n)^{\pm}$ QNM of the row: e.g., for the $(2,2)$ harmonic we first remove $(3,2,0)^+$, then $(2,2,0)^-$, and so on. 
We then quantify how much the mismatch distribution of the \texttt{Cheung} model with respect to the NR waveform is affected by each removal. If the mismatch distribution remains essentially unchanged upon removal, the corresponding QNM can be considered globally subdominant; conversely, if its removal leads to a significant degradation in the mismatch, the mode should be thought of as dominant. 

We perform this analysis for different starting times.
The results are presented in Fig.~\ref{fig:Cheung_remove_modes}. 
The findings are strongly dependent on the specific $(\ell,|m|)$ multipole considered, and we will discuss them on a case-by-case basis.

\begin{figure*}[!t]
    \centering
    \includegraphics[width=1.0\linewidth]{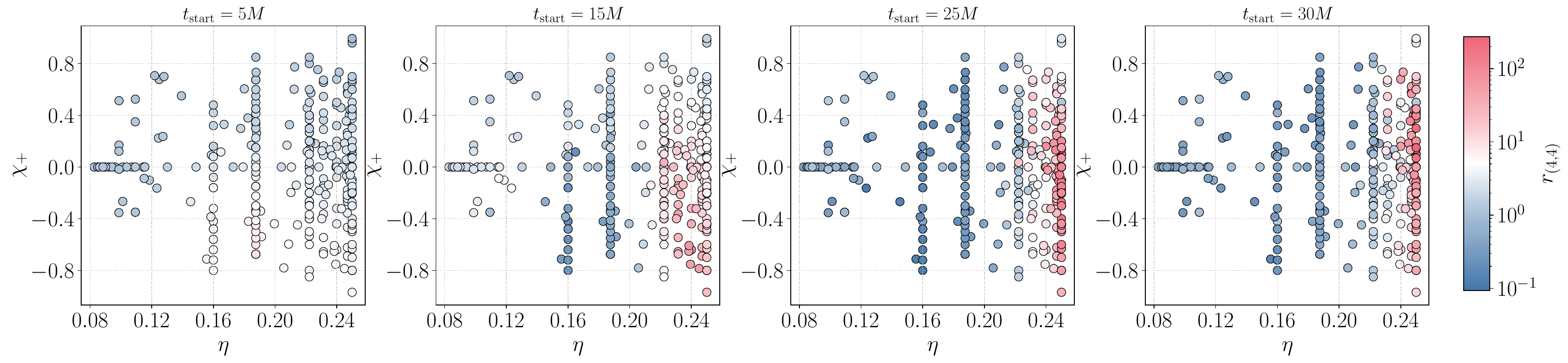}
    \caption{Ratio $r_{(4,4)}$ in the $(\eta,\chi_+)$ plane, as defined in Eq.~\eqref{ratio_44_eq}, for four selected starting times: $t_{\rm start}=[5, 15, 25, 30]$M.}
    \label{fig:ratio_44_t_start}
\end{figure*}

The $(2,2)$ multipole mismatch is only mildly affected by the removal of the $(3,2,0)^+$ QNM, particularly at late times.
Removing the counter-rotating mode $(2,2,0)^-$ does not produce any appreciable change. 
When we remove the $(2,2,1)^+$ overtone (purple histogram), the mismatch at early times increases; however, as $t_{\rm start}$ approaches $\sim 30$M the impact of removing the overtone diminishes, and the mismatch distributions of the $(2,2,0)^+$ mode alone and the combined $(2,2,0)^++(2,2,1)^+$ modes tend to overlap. 
This makes sense: the overtone decays faster than the fundamental mode, and thus its contribution becomes less dominant at late times.

On the contrary, for the $(2,1)$ mode, removing the first overtone $(2,1,1)^+$ worsens the mismatch at intermediate times. This is due to the poor accuracy of the $(2,1,1)^+$ amplitude fit discussed in Sec.~\ref {sec:res1}. At late times the overtone removal has a smaller impact, since the overtone has decayed.

The $(3,3)$ mode consists of a superposition of the $(3,3,0)^+$ and $(3,3,1)^+$ QNMs. The trend is similar to the previous cases: the removal of the overtone $(3,3,1)^+$ is significant at early times, but not so much at late times.

In the case of the $(3,2)$ multipole, removing the $(2,2,0)^+$ QNM results in a mismatch distribution approaching $\mathcal{O}(1)$. This dramatic loss of accuracy is expected: the \texttt{Cheung} model includes mode mixing between the $(3,2,0)^+$ and $(2,2,0)^+$ modes, which is the dominant contribution to this mode in non-precessing configurations.

Finally, the $(4,4)$ multipole has contributions from the linear mode $(4,4,0)^+$ and from the quadratic mode $(2,2,0)^{+}\times(2,2,0)^+$. When we remove the quadratic mode, the mismatch distribution gets somewhat worse at intermediate times (around $\approx 15$M). At later times the quadratic mode has decayed, and its removal has little to no impact on the mismatch.

To quantify the loss of mismatch due to removing the quadratic mode, in Fig.~\ref{fig:ratio_44_t_start} we plot the ratio
\begin{equation}
  r_{(4,4)}=\frac{\mathcal{M}_{(4,4,0)^+}}{\mathcal{M}_{((4,4,0)^+,(2,2,0)^+\times(2,2,0)^+)}}
  \label{ratio_44_eq}
\end{equation}
for different selected values of $t_{\rm start}$, in the subspace $(\eta,\chi_+)$. As the starting time increases, the ratio $r_{(4,4)}$ gets larger than unity for $\eta\sim 0.25$. 
This is consistent with Refs.~\cite{Cheung:2022rbm, Mitman:2022qdl}, where the inclusion of the quadratic mode was found to improve the agreement with the SXS:BBH:0305 simulation. 
The fact that the quadratic mode matters more at comparable masses is because its amplitude scales as $\sim A_{(2,2,0)^{+}}^2$, and that $(2,2,0)^+$ mode is more dominant for comparable-mass binaries. 

\begin{figure*}[!t]
    \centering
    \includegraphics[width=1.0\linewidth]{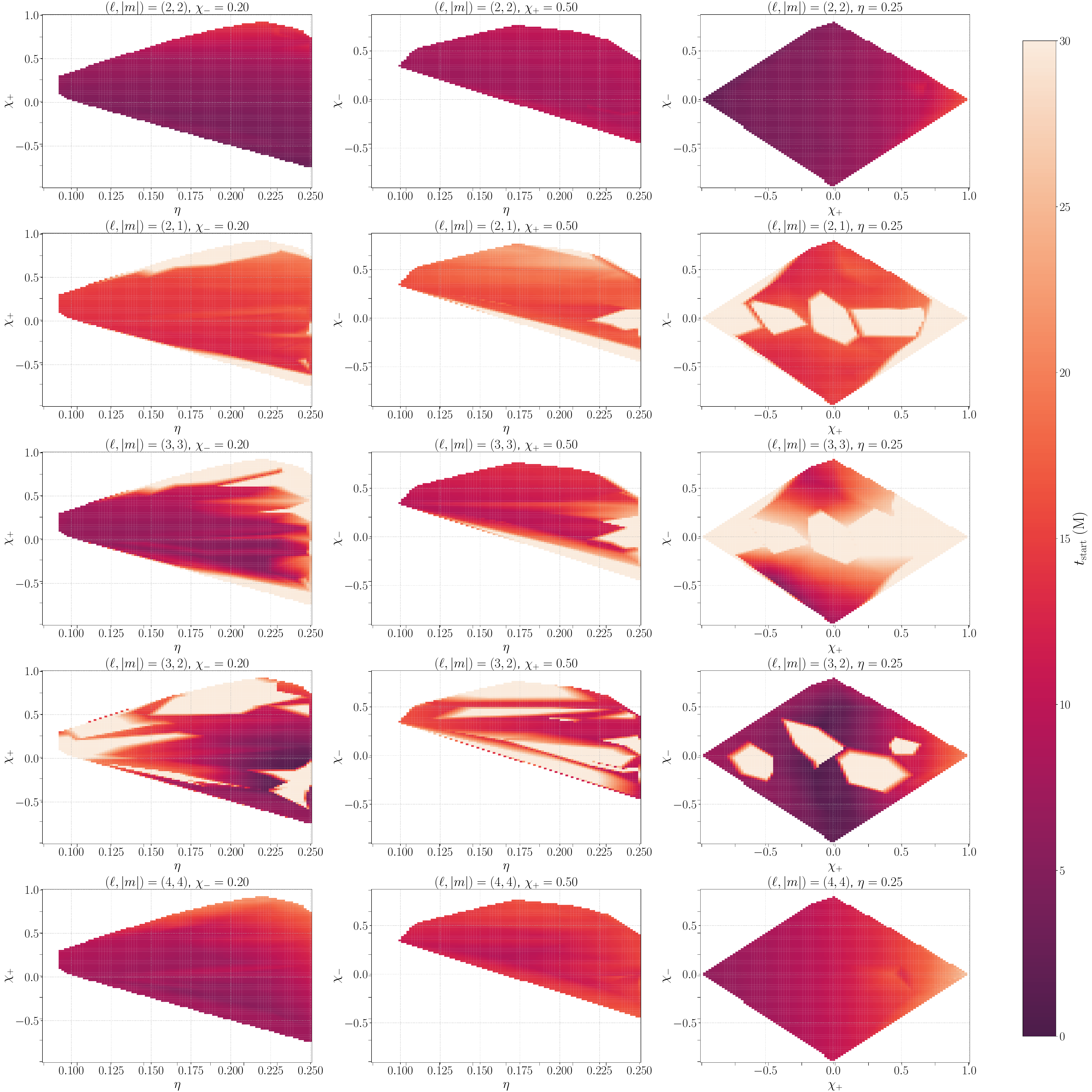}
    \caption{Colormaps of $t_{\rm start}(\mathcal{M}=3.5\%)$ for the \texttt{Cheung} model in different ``cuts'' of the parameter space: $(\eta,\chi_+)$ (left column), $(\eta,\chi_-)$ (middle column), and $(\chi_+,\chi_-)$ (right column). Different rows refer to different multipoles. In this case we use the whole set of SXS simulations, $\mathcal{I}_{\rm tot}$.}
    \label{interp_ts}
\end{figure*}

\subsection{Application to LIGO-Virgo-KAGRA analyses}

Our results can also find applications to observational analyses with LVK data.
In this context, one typically considers parameterized deviations in the QNM frequencies or amplitudes, as these parameters have a direct physical interpretation.
As starting at earlier times yields higher SNR, a key parameter is the minimum starting time at which the analysis can be applied, while still maintaining a sufficient level of accuracy against GR numerical solutions to avoid biases.

Here we provide a simple tool to answer this question.
We tabulate values of the starting time $t_{\rm start}(\eta, \chi_+, \chi_-|\mathcal{M}_{\rm th})$ at which a given mismatch threshold $\mathcal{M}_{\rm th}$ is reached for the entire SXS dataset under consideration.
We then interpolate across these discrete values, so we can perform this estimate over the entire parameter space.

To the simulations that do not achieve the given mismatch threshold, we assign a ``saturation'' starting time of $100 $M, indicating that in those regions we cannot reach the required accuracy.
To guarantee interpolation accuracy, we need as many points as possible, and therefore, we use the entire $\mathcal{I}_{\rm tot}$ dataset. 
These results are available online in the repository~\cite{crescimbeni2025interpolating} for models taken into exam.

An example of interpolation for the \texttt{Cheung} model is shown in Fig.~\ref{interp_ts}: the interpolant is well reconstructed for $(\ell,|m|)=(2,2)$, but fails for higher multipoles in certain regions of the parameter space and at sufficiently low mismatch, given the sparsity of available simulations at which the required accuracy can be achieved (see Fig.~\ref{fig:n_cross}). 

To test the accuracy of the interpolation across the parameter space, we first divided the SXS dataset $\mathcal{I}_{\rm tot}$ into a \textit{training} (\textit{validation}) set, containing $80\%$ ($20\%$) simulations randomly distributed in the parameter space. We performed the interpolation on the training set. Then we plotted the distributions of the residuals in $t_{\rm start}$ -- defined as the difference between the mismatch threshold's starting time and the starting time found by interpolation -- for the validation set. The results are shown in Fig.~\ref{fig:interpolation_check} of Appendix~\ref{sec:additional_results}.

\section{Discussion and conclusions}\label{sec:discussions_conclusions}

We have investigated the accuracy of several state-of-the-art ringdown models by comparing them to binary BH merger simulations from the SXS catalog~\cite{Stein:2019mop}. 
As a measure of accuracy, we have computed the mismatch between these models and the numerical waveforms, thus avoiding the conceptual and practical issues that arise when we analyze ringdown-only signals in the frequency domain.

We focused on the \texttt{London}~\cite{London:2018nxs}, \texttt{Cheung}~\cite{Cheung:2023vki}, and \texttt{TEOBPM}~\cite{Damour:2014yha,DelPozzo:2017rka,Nagar:2020eul,Nagar:2020xsk} models, all developed for non-precessing, quasi-circular progenitor systems. 
The first two are constructed as QNM superpositions with amplitudes and phases fitted to NR simulations, and are expected to be valid at sufficiently late times after merger.
In contrast, \texttt{TEOBPM} includes the merger and early post-merger regime, with a phenomenological time-domain parameterization calibrated to NR simulations.

We first analyzed the distribution of mismatches as a function of the starting time of the ringdown for different $(\ell,|m|)$ multipoles across an ensemble of aligned-spin, quasi-circular SXS simulations. We found that the \texttt{Cheung} model gives lower mismatches than the \texttt{London} model for most binaries. This is expected, as the \texttt{Cheung} model was built by fitting waveforms in the same SXS catalog used to perform the mismatch tests and because it includes overtones and mode-mixing effects, which improve the match for most waveform multipoles.
However, the \texttt{London} model outperforms \texttt{Cheung} for the $(2, 1)$ mode waveform.
This is because the \texttt{Cheung} model includes the $(2, 1, 1)$ overtone, which is robustly found only in a low number of simulations, resulting in a poor fitting of its overall amplitude.
Consistently with the findings of Ref.~\cite{Gennari:2023gmx}, the mismatch of the \texttt{TEOBPM} model is excellent for the dominant harmonic even at early times, but its accuracy is lower for some of the higher modes.
This is due to the lack of mode-mixing effects and additional late-time contributions (such as the $n=1$ overtone).
Excluding the $(3,2)$ multipole, in which the mode mixing is dominant, \texttt{TEOBPM} generally outperforms the other two models, except at late times ($t \gtrsim 15-20$M), where it does not fit the waveform as well because it only includes a single QNM.

Next, we studied the fraction of simulations $N_{\rm cross}$ for which the ringdown models cross a given mismatch threshold.
For the $(\ell,|m|) = (2,2)$ multipole and across all starting times considered,
the \texttt{London} model achieves a mismatch of $\gtrsim 10^{-3}$ for essentially all simulations;
the \texttt{TEOBPM} model has mismatches better then $\gtrsim 10^{-5}$, and
the \texttt{Cheung} model does slightly better than \texttt{TEOBPM} due to its extended QNM content (see Table~\ref{tab:modes_lm}).
For the $(3,2)$ multipole, the number of simulations that reach a given mismatch threshold for \texttt{TEOBPM} is always lower than for the two \texttt{KerrBinary} models.
We provide an interpolating function to estimate the earliest starting time for which a given QNM-based model meets a given mismatch threshold across the parameter space $(\eta, \chi_+, \chi_-)$ \cite{crescimbeni2025interpolating}. This should be valuable information for ringdown searches in LVK data.

Finally, we have investigated the relative importance of individual QNMs in the \texttt{Cheung} model by progressively removing overtones and evaluating their impact on the mismatch. We have found that excluding overtones degrades the model accuracy at early times, but not so much at late times. This is expected, as overtones decay faster than the fundamental mode. In the future, it will be important to repeat this sort of study by removing the modes in different combinations and assessing which modes are most relevant in other regions of the parameter space. 
An interesting case is the $(4,4)$ multipole, that contains a significant quadratic $(2,2,0)^+\times(2,2,0)^+$ QNM component in addition to the $(4,4,0)^+$ QNM.
We have found that removing the quadratic mode from the $(4,4)$ harmonic leads to higher inaccuracy at $\eta\simeq0.25$, but not for more asymmetric systems. This is mainly because $A_{(2,2,0)^{+}\times(2,2,0)^+}\propto A_{(2,2,0)^{+}}^2$, and $A_{2,2,0}$ is larger for comparable mass binaries, but it will be important to better understand possible instabilities in the quadratic mode fits.

We restricted attention to quasi-circular, non-precessing binary BHs, but our analysis should be extended to include precession~\cite{Finch:2021iip,Zhu:2023fnf,Nobili:2025ydt} and eccentricity~\cite{Carullo:2024smg}.
For instance, in precessing systems the $(2,2,0)^+$ mode may no longer dominate over some of the higher multipoles. 
Neglecting eccentricity or precession could also bias the determination of the optimal starting times obtained via the interpolating function. 
We leave these extensions to future work.

As a useful rule of thumb, for a signal with a given SNR, two waveforms are considered indistinguishable for parameter estimation purposes if their mismatch~\cite{Flanagan:1997kp,Lindblom:2008cm} 
\begin{equation}
  \mathcal{M} \lesssim \frac{1}{2\,{\rm SNR}^2}\,.
\end{equation}
Since this criterion is sufficient but not necessary, it is generally too conservative (see, e.g.~\cite{Chandramouli:2024vhw}).
Furthermore, its violation does not necessarily imply that differences are measurable, and detectable effects may be confined to a subset of parameters that could be of little interest~\cite{Thompson:2025hhc}.

Nonetheless, the above indistinguishability criterion suggests that events with ringdown ${\rm SNR}\approx20$, such as GW250114~\cite{LIGOScientific:2025obp, LIGOScientific:2025rid}, would require $\mathcal{M}\lesssim 10^{-3}$ to avoid waveform systematics.
Our results show that the \texttt{Cheung} and \texttt{TEOBPM} models typically have a smaller mismatch with respect to NR waveforms for $(\ell,|m|)=(2,2)$, when $t_{\rm start}\gtrsim 15$M and $t_{\rm start}\gtrsim 0$M, respectively, while this is not the case for the \texttt{London} model. We can conclude that using the \texttt{TEOBPM} model for large-SNR, GW250114-like postmerger detections with LIGO-Virgo-KAGRA should not introduce large modeling systematics, at least for the fundamental mode.
Instead, given the overall larger mismatches for higher harmonics, care should be applied when performing BH spectroscopy with those.

The outlook changes significantly when considering the landscape of future detectors.
Next-generation detectors such as ET and CE can reach ringdown ${\rm SNR}\approx100$ for several events per year~\cite{Bhagwat:2023jwv}, whereas space missions like LISA~\cite{LISA:2024hlh} can even achieve ringdown ${\rm SNR}\approx1000$ in certain optimistic scenarios~\cite{Berti:2016lat,Bhagwat:2021kwv}.
In these cases one would need $\mathcal{M}\lesssim 5\times 10^{-5}$ and $\mathcal{M}\lesssim 5\times 10^{-7}$, respectively, to make sure that the adopted waveform model is sufficiently accurate.
Our results show that this level of accuracy is hard to achieve even for $(\ell,|m|)=(2,2)$, and even more so for $(\ell,|m|)=(3,3)$. This calls for a drastic improvement of ringdown waveforms if one wishes to perform precision BH spectroscopy with future instruments.

\textbf{Note added.} While this work was being completed, the SXS collaboration updated the catalog by nearly doubling the total number of binary configurations from 2018 to 3756, covering more densely the region of parameter space involving asymmetric, precessing, and eccentric binaries~\cite{Scheel:2025jct}. Each waveform was also corrected to make sure that it refers to the binary's center-of-mass frame, and updated to include the memory effect (see~\cite{Mitman:2020bjf}). 
We have analyzed how these improvements affect the mismatch for the \texttt{Cheung}. The results of this comparison are reported in Appendix~\ref{sec:additional_results}, and they show that our main conclusions (obtained from the earlier version of the catalog) do not change appreciably.
This is not unexpected: at the level of accuracy we are interested in, the most significant catalog update concerns the inclusion of precessing simulations, which we do not consider in the present study. Finally, since we are using the last catalog version before the new release~\cite{Scheel:2025jct}, we also exclude simulation \texttt{SXS:BBH:1110} from our analysis due to significant numerical artifacts in the $(2,2)$ mode waveform (this simulation was removed in the latest release).


\let\oldaddcontentsline\addcontentsline
\renewcommand{\addcontentsline}[3]{}
\begin{acknowledgments}
We thank Fabrizio Corelli, Konstantinos Kritos, Francesco Iacovelli, Luca Reali, Sophia Yi, and Jay Wadekar for discussions. F.C.~acknowledges Johns Hopkins University and the University of Birmingham for the kind hospitality during the finalization of this project. F.C.~acknowledges the financial support provided under the “Progetti per Avvio alla Ricerca Tipo 1”, protocol number AR12419073C0A82B.
E.B.~is supported by NSF Grants No.~AST-2307146, No.~PHY-2513337, No.~PHY-090003, and No.~PHY-20043, by NASA Grant No.~21-ATP21-0010, by John Templeton Foundation Grant No.~62840, by the Simons Foundation [MPS-SIP-00001698, E.B.], by the Simons Foundation International [SFI-MPS-BH-00012593-02], and by Italian Ministry of Foreign Affairs and International Cooperation Grant No.~PGR01167.
M.H.Y.C.~is a Croucher Fellow supported by the Croucher Foundation.
M.H.Y.C.~acknowledges support by the Jonathan M. Nelson Center for Collaborative Research at the Institute for Advanced Study.
This work was carried out at the Advanced Research Computing at Hopkins (ARCH) core facility (\url{https://www.arch.jhu.edu/}), which is supported by the NSF Grant No.~OAC-1920103.
P.P.~is partially supported by the MUR FIS2 Advanced Grant ET-NOW (CUP:~B53C25001080001), the MUR PRIN Grant 2020KR4KN2 ``String
Theory as a bridge between Gauge Theories and Quantum Gravity'', by the MUR FARE programme (GW-NEXT, CUP:~B84I20000100001), and by the INFN TEONGRAV initiative.
Some numerical computations were performed at the Vera cluster supported by MUR and Sapienza University of Rome.
\FloatBarrier
\end{acknowledgments}


\appendix

\section{Equivalence between time domain and frequency domain and impact of the window on the PSD edges}
\label{sec:window_PSD}

In this Appendix, we investigate the impact of boundary conditions on the mismatch. 

We start by asking: at what precision does the time-domain formalism of Sec.~\ref{intro} give the same results as a frequency-domain approach, given standard choices for the treatment of boundary effects?

\begin{figure}[h!]
    \centering
    \includegraphics[width=1.0\linewidth]{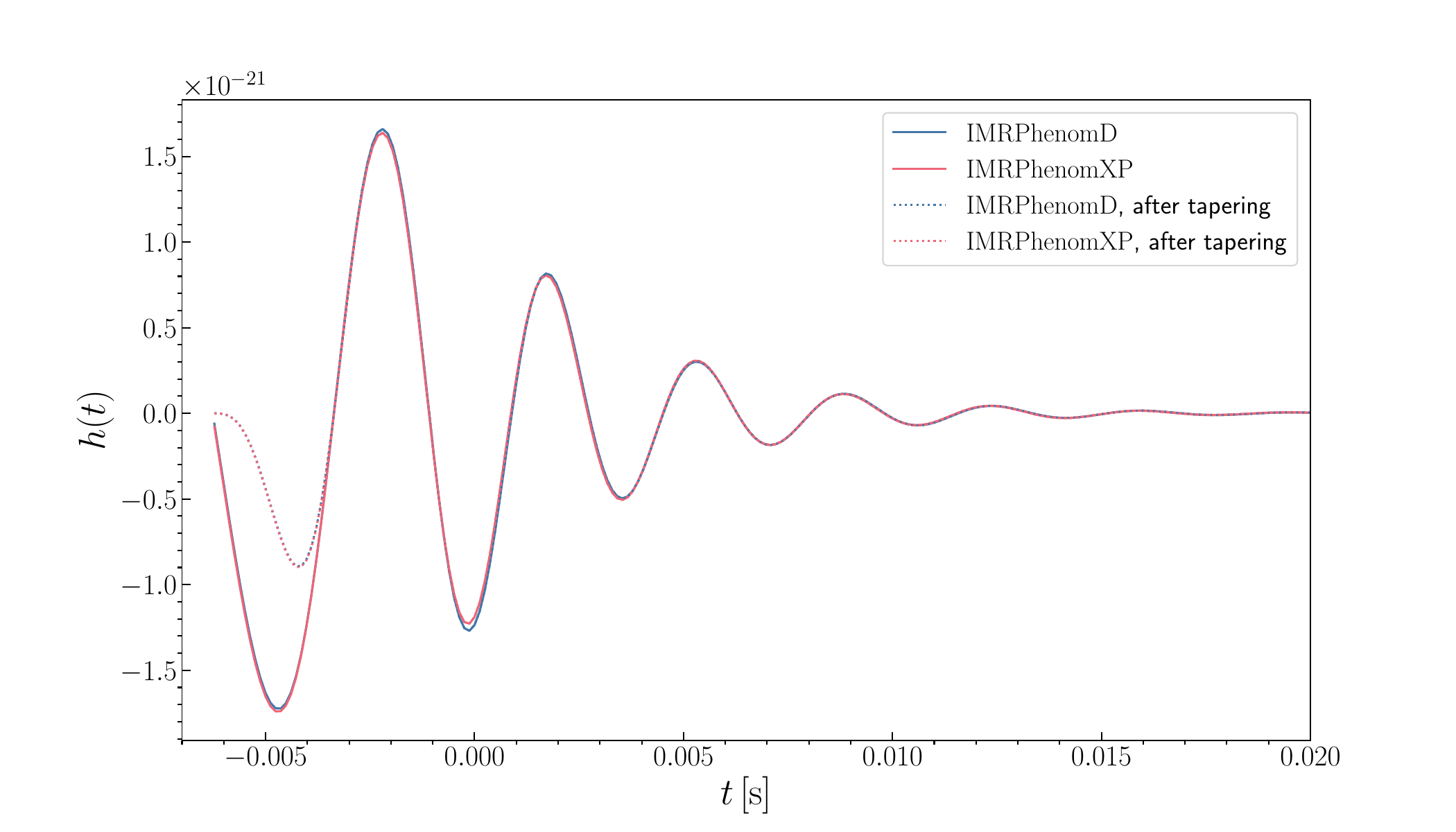}
    \caption{Waveforms used to compute the mismatches in the time and frequency domains, respectively, either without (continuous lines) or with (dashed lines) tapering at the lower edge.}
    \label{fig:aligned_wfs_RD}
\end{figure}

Consider two signals $h(t)$ and $s(t)$, with time-domain mismatch given by
\begin{equation}
        \mathcal{M}_{\rm TD}=1-\frac{\langle h|s\rangle}{\sqrt{\langle h|h\rangle\langle s|s\rangle}}\,,
    \label{TD_mismatch}
\end{equation}
where $\langle . \rangle$ denotes the scalar product in the time domain defined in Eq.~\eqref{sc_prod_full}.
The Fourier-domain mismatch is instead given by~\cite{Lindblom:2008cm, Blackman:2015pia, Boyle:2019kee}
\begin{equation}
        \mathcal{M}_{\rm FD}=1-\frac{(h|s)}{\sqrt{(h|h)(s|s)}}\,,
    \label{FD_mismatch}
\end{equation}
where the scalar product $(.|.)$ is defined as
\begin{equation}(h|s)=4\mathcal{R}\int_{f_{\rm min}}^{f_{\rm max}}\frac{\tilde{h}(f)\tilde{s}^*(f)}{S_n(f)}\,.
\end{equation}
Here $\mathcal{R}$ denotes the real part, and $\tilde{h}(f)$ is the Fourier transform of $h(t)$. We construct representative signals $h(t)$ and $s(t)$ as follows:
\begin{itemize}
    \item We generate waveforms using both \texttt{IMRPhenomD}~\cite{Khan:2015jqa} and \texttt{IMRPhenomXP}~\cite{Pratten:2020ceb}, choosing a detector-frame total mass $M=62M_{\odot}$, mass ratio $q=0.7$, $d_L=410$Mpc, and $\chi_1=\chi_2=0$. With this prescription, the frequency domain SNRs of the two signals are $\sim 33$.
    The two signals have been chosen to represent two waveforms with small differences, due in this case to the different baseline aligned-spin parameterization.
    
    \item We align the two waveforms at the peak, and isolate the ringdown parts by tapering the left edge of the signals, to zero-out the strain. This prevents artifacts when building the fast Fourier transform of the two signals.
    \item The right edge of the ringdown signal is padded with zeros to match the fast Fourier transform length. In other words, the signal will be zero for a stretch of length $N_{\rm pad} = N_{\rm ACF}-N_{\rm signal}$.
\end{itemize}
The two waveforms are plotted in Fig.~\ref{fig:aligned_wfs_RD} in the time domain.\\

We compute the mismatch on $h\equiv h_{+}$ both in the time and in the frequency domains. 
We use the PSD provided in Ref.~\cite{LIGOT1800044}, we set $f_{\rm min}=5$Hz and $f_{\rm max}=4096$Hz, and we compare the results of Eq.~\eqref{TD_mismatch} as implemented in \texttt{bayRing}~\cite{carullo_gregorio_2023_8284026} against Eq.~\eqref{FD_mismatch}, accessed through \texttt{pycbc}~\cite{Biwer:2018osg, pycbc_waveform}. 

\begin{figure}
    \centering
    \includegraphics[width=1.0\linewidth]{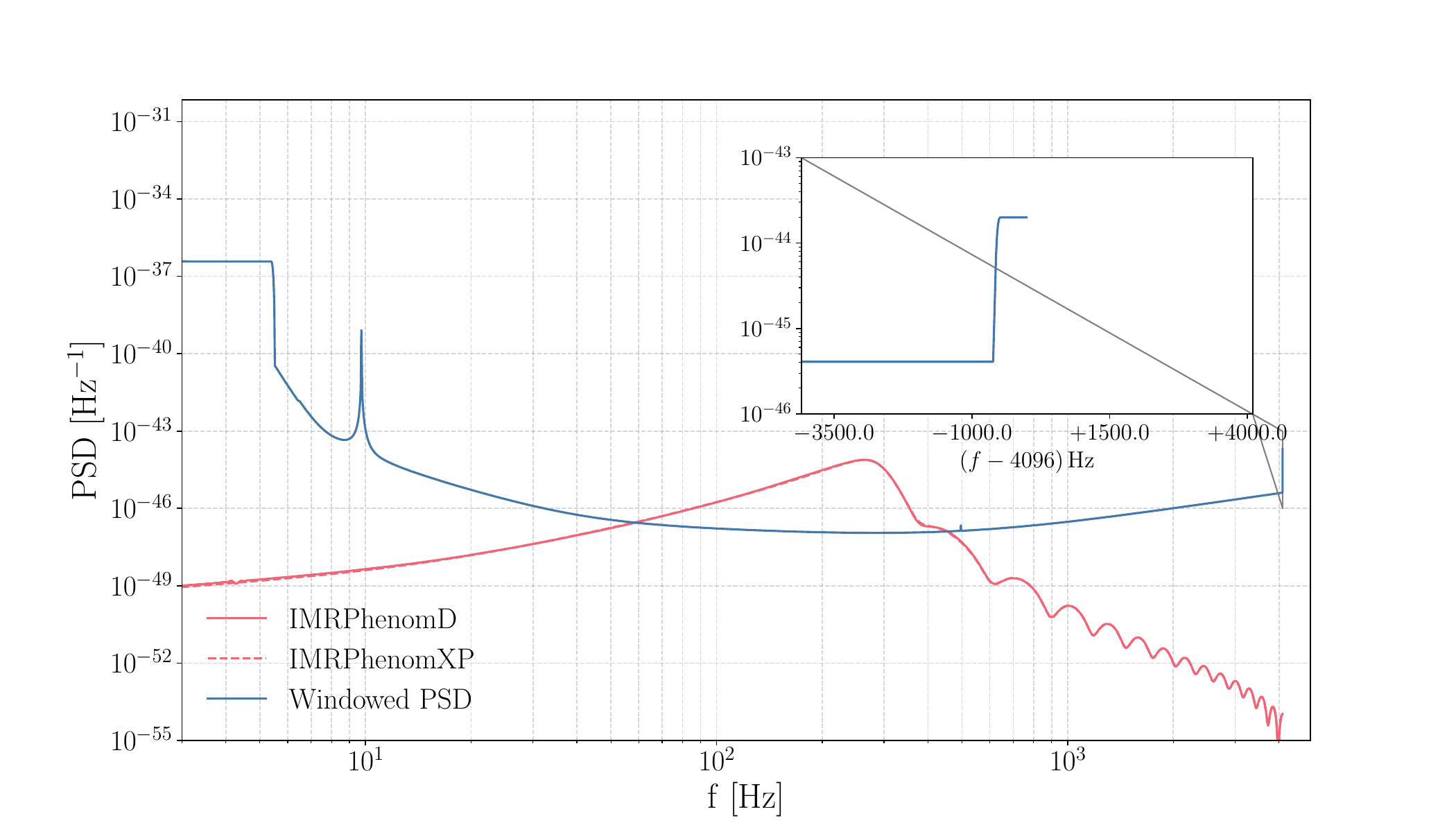}
    \caption{Illustrative windowed PSD as a function of the frequency (blue curve), generated starting from the PSD in Ref.~\cite{LIGOT1800044}. The window parameters are $w_l=w_h=0.5$Hz, $k=234$, $\mathcal{I}_l=100$, and $\mathcal{I}_h=49$; the inset is a zoom-in of the windowed PSD at high frequencies. The squared absolute value of two different waveform models in the frequency domain (\texttt{IMRPhenomD} and \texttt{IMRPhenomXP}) is shown in red.}
    \label{fig:FD_RD_and_PSD}
\end{figure}

We find:

\begin{equation}
\begin{cases}
\mathcal{M}_{FD}=0.00025582775\,,\\
\mathcal{M}_{TD}=0.00025582767\,,
\end{cases}
\label{opt_SNRs}
\end{equation}
with a relative difference
\begin{equation}
    p_{\rm TD-FD}=\frac{|\mathcal{M}_{TD}-\mathcal{M}_{FD}|}{\mathcal{M}_{FD}}=3.4\cdot10^{-7}\,.
    \label{prec_ref}
\end{equation}
Such a difference between the two mismatch values could be relevant at values of the SNR of order ${\cal O}(10^3)$~\cite{Purrer:2019jcp} typical of next-generation detectors, such as LISA.

As discussed in Sec.~\ref{sec:inference_framework}, it is important to apply a window function to ``smooth out'' the PSD edges to obtain the ACF through a fast Fourier transform. 
This is because simply setting the PSD to zero outside of the (finite) detector frequency band can have undesired consequences for signals whose frequency support extends beyond the sensitivity band, while setting it equal to a constant (see e.g. Fig.~1 of~\cite{Siegel:2024jqd}) can introduce Gibbs-like phenomena in the ACF.
Here we demonstrate that these effects are practically negligible in our case.
We replace the PSD $S(f)$ with:

\begin{equation}
\begin{cases}
    S(f)\cdot e^{(f-(f_{\min}+w_l))k}+T_l, & f \in [f_{\min}, f_{\min} + w_l] \\[1ex]
    S(f), & f \in [f_{\min} + w_l, f_{\max} - w_h] \\[1ex]
    S(f)\cdot e^{-(f-(f_{\max}-w_l))k}+T_h, & f \in [f_{\max} - w_h, f_{\max}]
\end{cases}
\label{window_shape}
\end{equation}
where 
\begin{equation}
\begin{cases}
T_l=S(f_{\min}) \cdot \mathcal{I}_l \cdot \dfrac{1 - e^{(f - (f_{\min} + w_l))k}}{1 - e^{-w_l \cdot k}}\,,\\
T_h=S(f_{\max}) \cdot \mathcal{I}_h \cdot \dfrac{1 - e^{-(f - (f_{\max} - w_l))k}}{1 - e^{w_l \cdot k}}\,,
\end{cases}    
\end{equation}
and $\mathcal{I}_l, \mathcal{I}_h\geq1$ are constants.
%
The parameters $(w_l,w_h)$ specify the window size, and $k$ determines its steepness.

In Fig.~\ref{fig:FD_RD_and_PSD} we plot an example of a windowed PSD, along with the frequency-domain signals introduced before. We can now try to determine a set of parameters $\underline W =(w_{l}, w_{h}, k, \mathcal{I}_l, \mathcal{I}_h)$ that maximizes the agreement between the frequency-domain and time-domain results by computing

\begin{equation}
    p_{\rm TD-FD}(\underline W)=\frac{|\mathcal{M}_{\rm TD}(\underline W)-\mathcal{M}_{\rm FD}|}{\mathcal{M}_{\rm FD}}\,,
    \label{prec_ref_w}
\end{equation}
where $\mathcal{M}_{\rm FD}$ was kept fixed, since the application of the window function only affects the time-domain calculation. 

To minimize changes in the PSD morphology we set $w_{l}=w_{h}=0.5$Hz (i.e., we make the window as small as possible, but sufficiently large to guarantee a smooth transition with a reasonable number of points). 
This can be problematic if, for instance, the ringdown signal in the frequency domain overlaps with one ``edge'' of the PSD (for instance, massive events have most power at low frequencies). 

We then scan the  $(k, \mathcal{I}_l, \mathcal{I}_h)$ parameter space.
We vary $k$ and $\mathcal{I}_{l,h}$ in the ranges $k\in [0.1,\,10^3]$, $\mathcal{I}_{l,h}\in[1,100]$ and we find that the quantity defined in Eq.~\eqref{prec_ref_w} has a minimum
\begin{equation}
    p_{\rm TD-FD}(\underline W_{\rm opt})=1.8\cdot10^{-8}
    \label{prec_ref_w_opt}
\end{equation}
for the window parameters $\underline W_{\rm opt}$ reported in Table~\ref{tab:opt_parameters} and used in the main text.
\begin{table}[t]
    \centering
    \begin{tabular}{|cc|ccc|}
        \hline
        \hline
        $w_{l}~[\mathrm{Hz}]$ & $w_{h}~[\mathrm{Hz}]$ & $k$ & $\mathcal{I}_l$ & $\mathcal{I}_h$ \\
        \hline
        \hline
        0.5 & 0.5 & 234.4 & 1.0 & 49.0 \\
        \hline
    \end{tabular}
    \caption{Optimal window parameters for the mismatch comparison between the ringdown-only signals of \texttt{IMRPhenomD} and \texttt{IMRPhenomXP}.}
    \label{tab:opt_parameters}
\end{table}
We find that varying $k$ has only a mild effect on $p_{\rm TD-FD}(\underline W)$.
However, the application of the window on a Welch PSD after bandpassing, where its edges are automatically set to lower values, can have a larger effect~\cite{Siegel:2024jqd}. 
We considered different binary parameters and found similar (small) improvements from the application of a window function, so our conclusions are expected to be general.

\section{Additional results}
\label{sec:additional_results}

In this Appendix, to improve readability, we collect additional results complementing the analysis presented in Sec.~\ref{sec:results}.

\begin{figure*}[!t]
\includegraphics[width=0.88\textwidth]{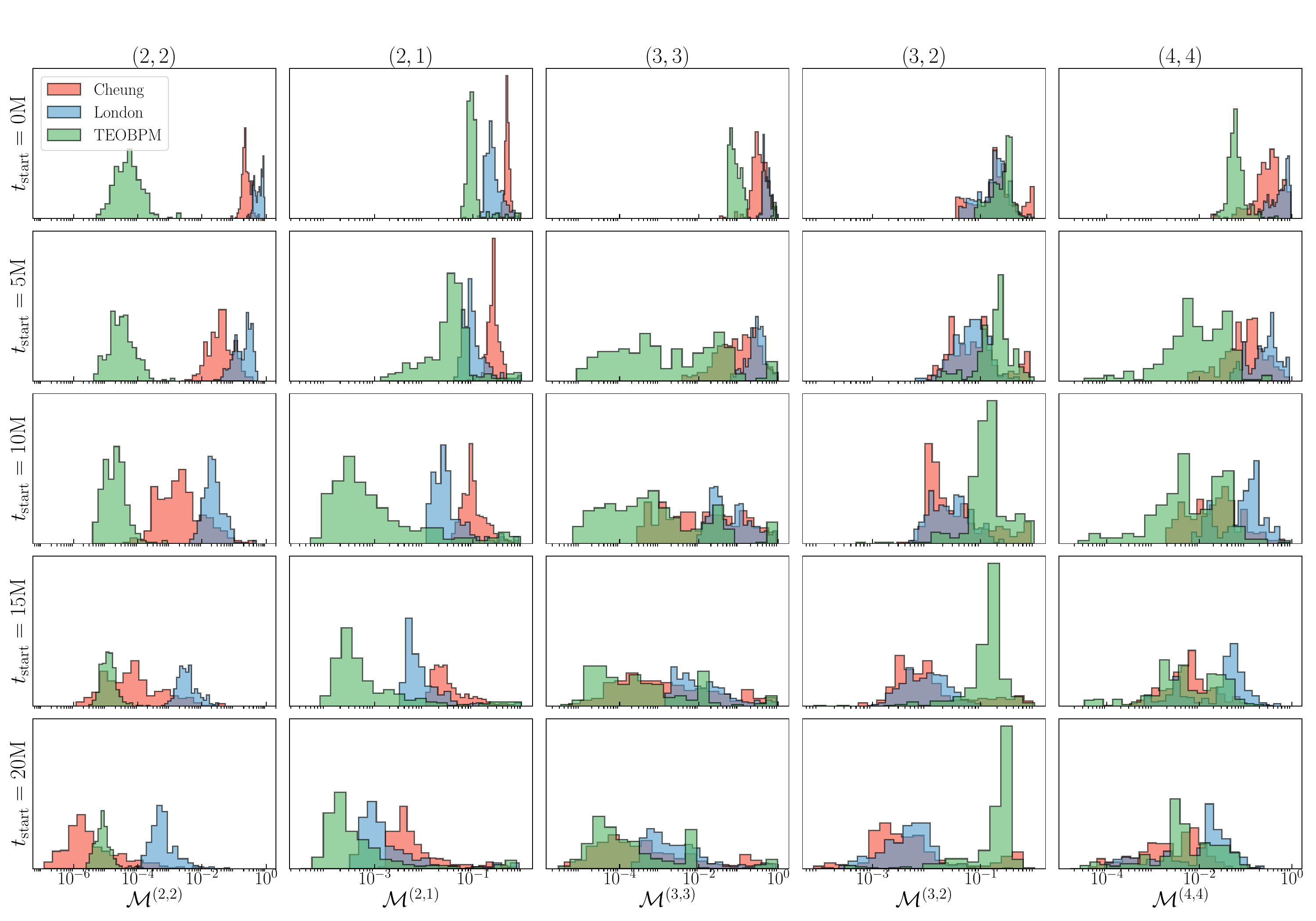}
\caption{Same as Fig.~\ref{fig:mm_distrs_l_vs_c}, but with $t_{\rm start}$ indicating the starting time defined with respect to the (2,2) peak.}
\label{fig:mm_distrs_l_vs_c_data_analysis}
\end{figure*}

In Fig.~\ref{fig:mm_distrs_l_vs_c_data_analysis} we show the mismatch distributions for different multipoles $(\ell,m)$ and different starting times for the \texttt{London}, \texttt{Cheung}, and \texttt{TEOBPM} models. This plot is similar to Fig.~\ref{fig:mm_distrs_l_vs_c}, but now the origin of the starting time $t_{\rm start}=0$ for each $(\ell,m)$ refers to the peak of the $(2,2)$ multipole. The plot shows that the \texttt{TEOBPM} mismatches are worse at early times for higher multipoles, because the specific mode is not excited yet and the pre-peak emission is not included.

\begin{figure*}
    \centering
    \includegraphics[width=\textwidth]{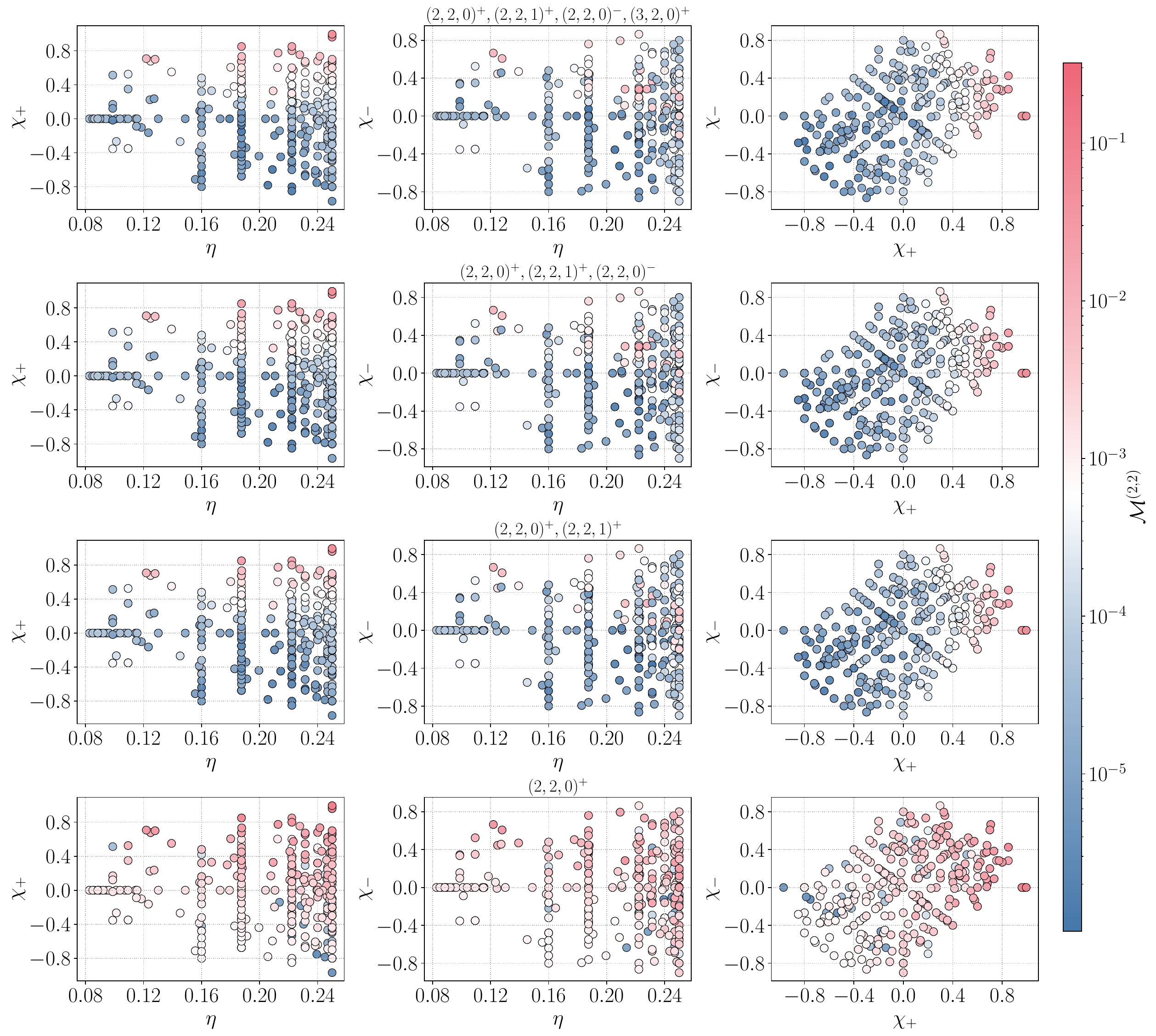}
    \caption{Scatter plots of the mismatch $\mathcal{M}^{(2,2)}$ of the \texttt{Cheung} model at $t_\mathrm{\rm start} = 15$M as a function of $(\eta, \chi_+)$, $(\eta, \chi_-)$, and $(\chi_+, \chi_-)$. Each row shows the effect of removing specific  QNMs combinations from the model.}
    \label{fig:projections_22}
\end{figure*}

\begin{figure*}
    \centering
    \includegraphics[width=\textwidth]{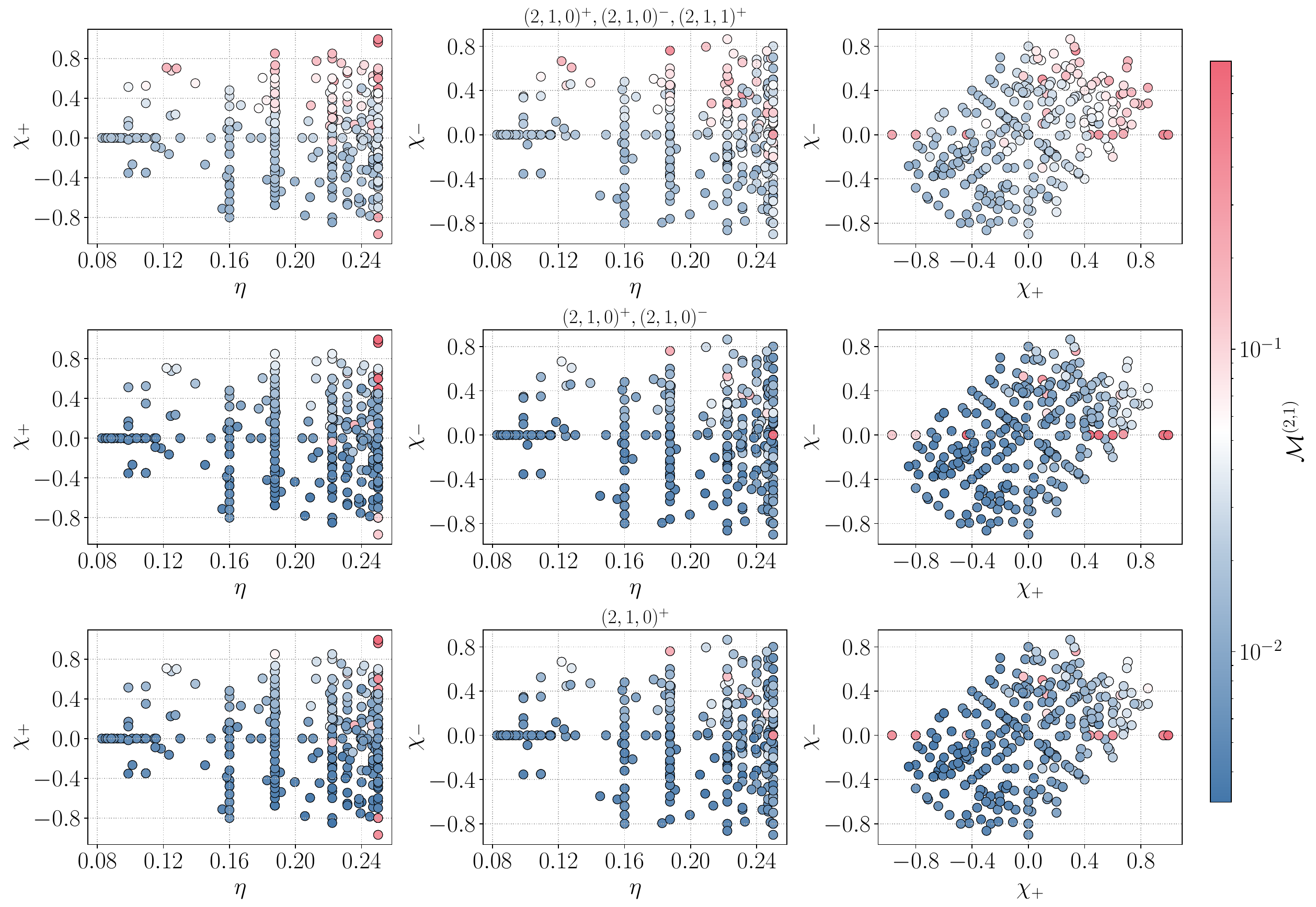}
    \caption{Same as Fig.~\ref{fig:projections_22}, but for $(\ell, m) = (2,1)$.}
    \label{fig:projections_21}
\end{figure*}

\begin{figure*}
    \centering
    \includegraphics[width=\textwidth]{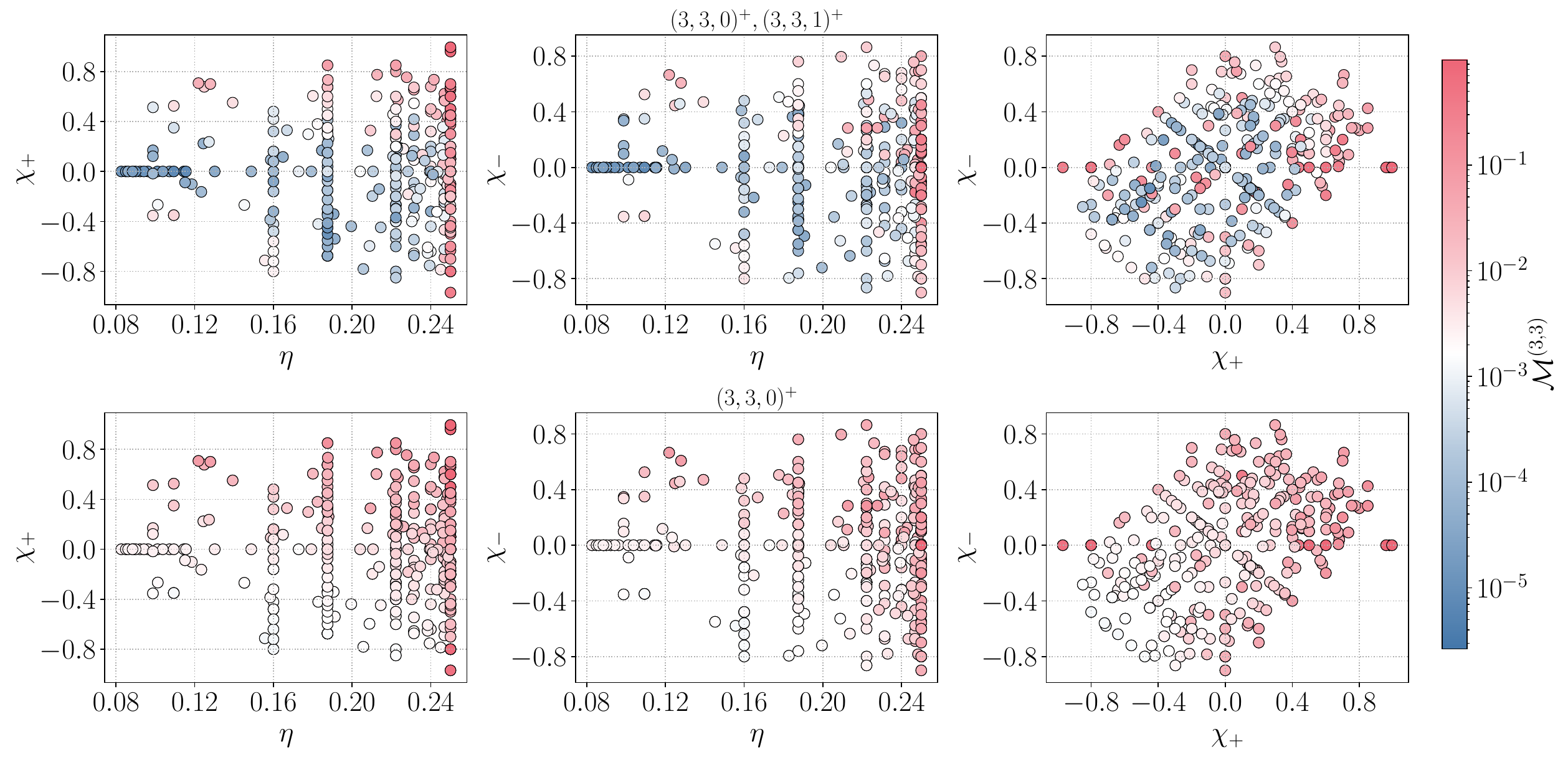}
    \caption{Same as Fig.~\ref{fig:projections_22}, but for $(\ell, m) = (3,3)$.}
    \label{fig:projections_33}
\end{figure*}

\begin{figure*}
    \centering
    \includegraphics[width=\textwidth]{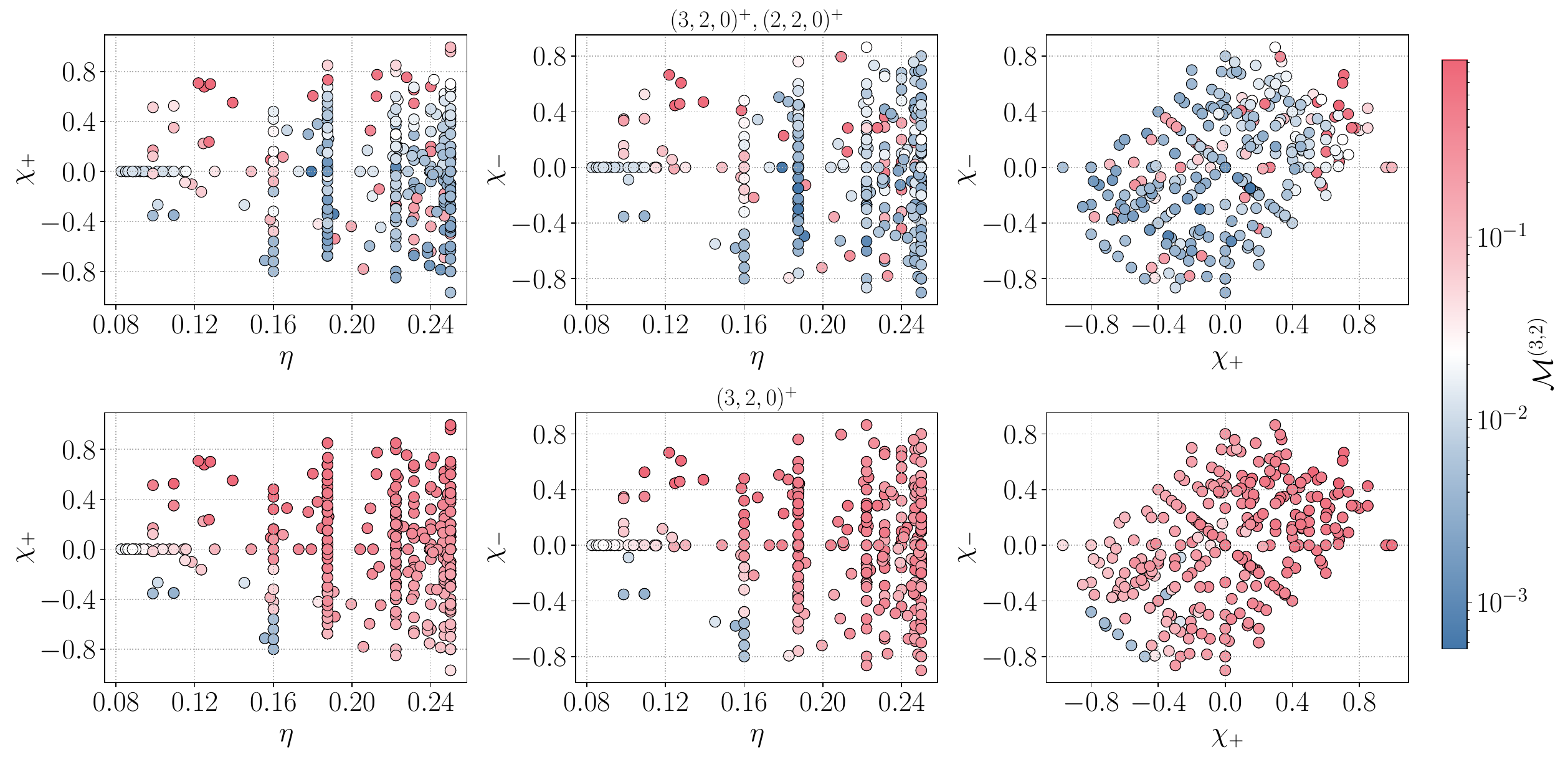}
    \caption{Same as Fig.~\ref{fig:projections_22}, but for $(\ell, m) = (3,2)$.}
    \label{fig:projections_32}
\end{figure*}

\begin{figure*}
    \centering
    \includegraphics[width=\textwidth]{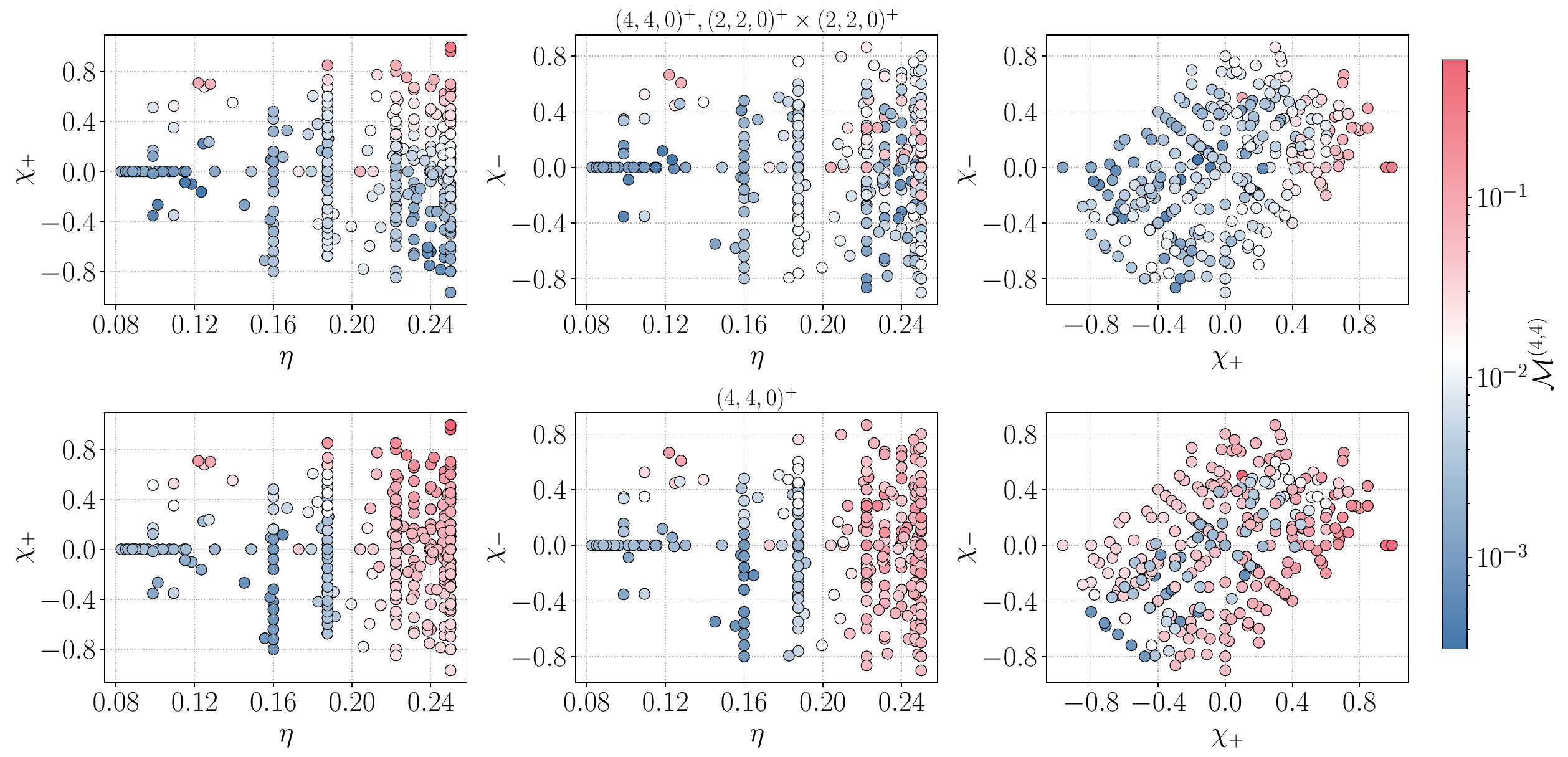}
    \caption{Same as Fig.~\ref{fig:projections_22}, but for $(\ell, m) = (4,4)$.}
    \label{fig:projections_44}
\end{figure*}

\begin{figure*}
    \centering
    \includegraphics[width=\textwidth]{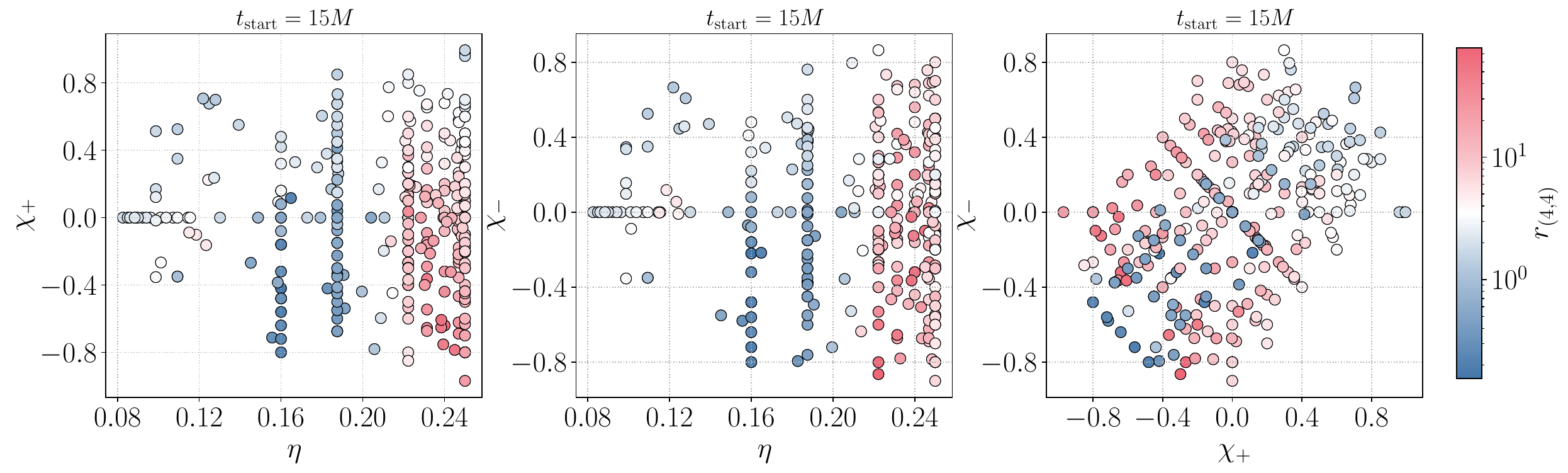}
    \caption{Same as Fig.~\ref{fig:ratio_44_t_start}, but for the ratio $r_{(4,4)}$ defined in Eq.~\eqref{ratio_44_eq}.}
    \label{fig:projections_44_ratio}
\end{figure*}

In Figs.~\ref{fig:projections_22}-\ref{fig:projections_44}, we plot the \texttt{Cheung} mismatch distributions for $t_{\rm start}=15$M in various two-dimensional projections of the $(\eta,\chi_+,\chi_-)$ parameter space. Different rows show the impact of removing specific QNMs. Note that: (1) removing the overtones has a mild impact on the $(2,2)$ and $(3,3)$ multipoles; (2) the overtone removal for the $(2,1)$ multipole actually improves the accuracy, as its fit is inaccurate; (3) removing the quadratic mode has a mild but noticeable effect. This is consistent with Fig.~\ref{fig:projections_44_ratio}, where we plot the ratio between the $(4,4,0)^{+}$ and the $(4,4,0)^{+}+(2,2,0)^{+}\times(2,2,0)^{+}$ mismatches at $t_{\rm start}=15$M.

\begin{figure*}
    \centering
    \includegraphics[width=0.45\linewidth]{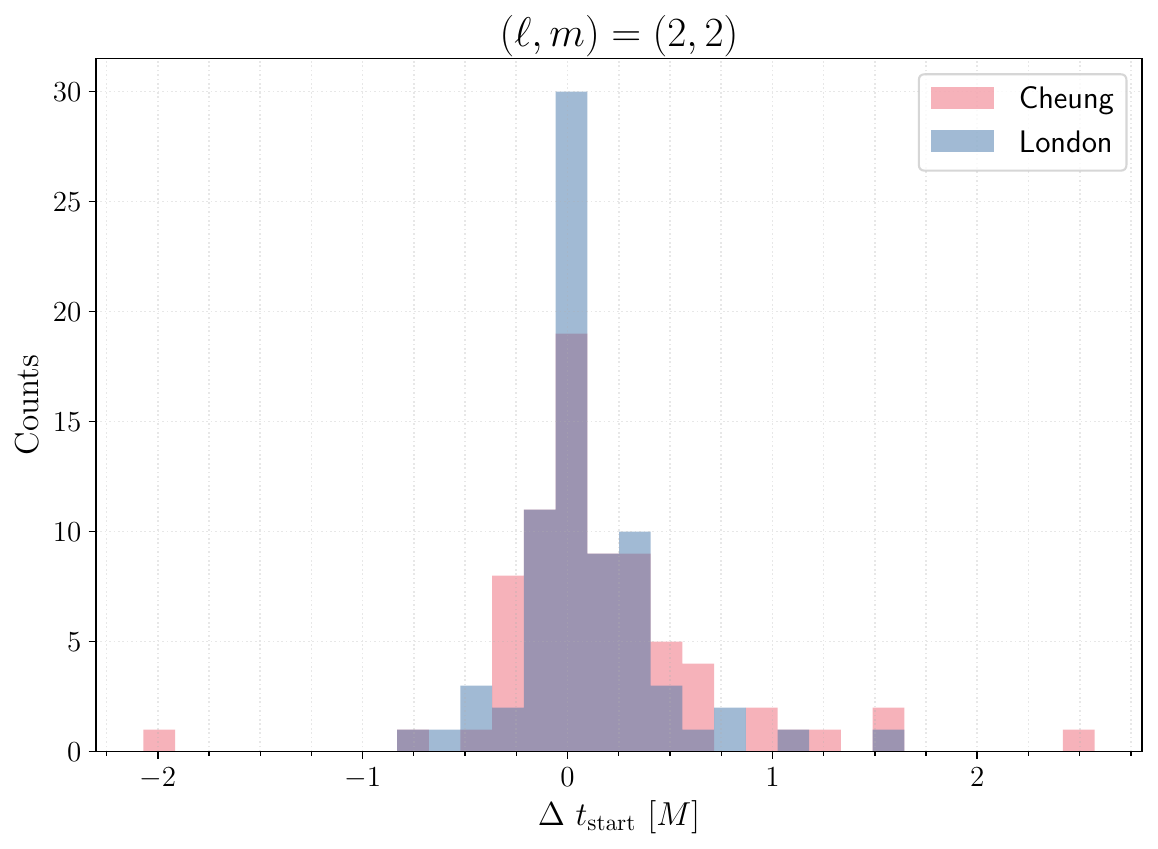}
    \includegraphics[width=0.45\linewidth]{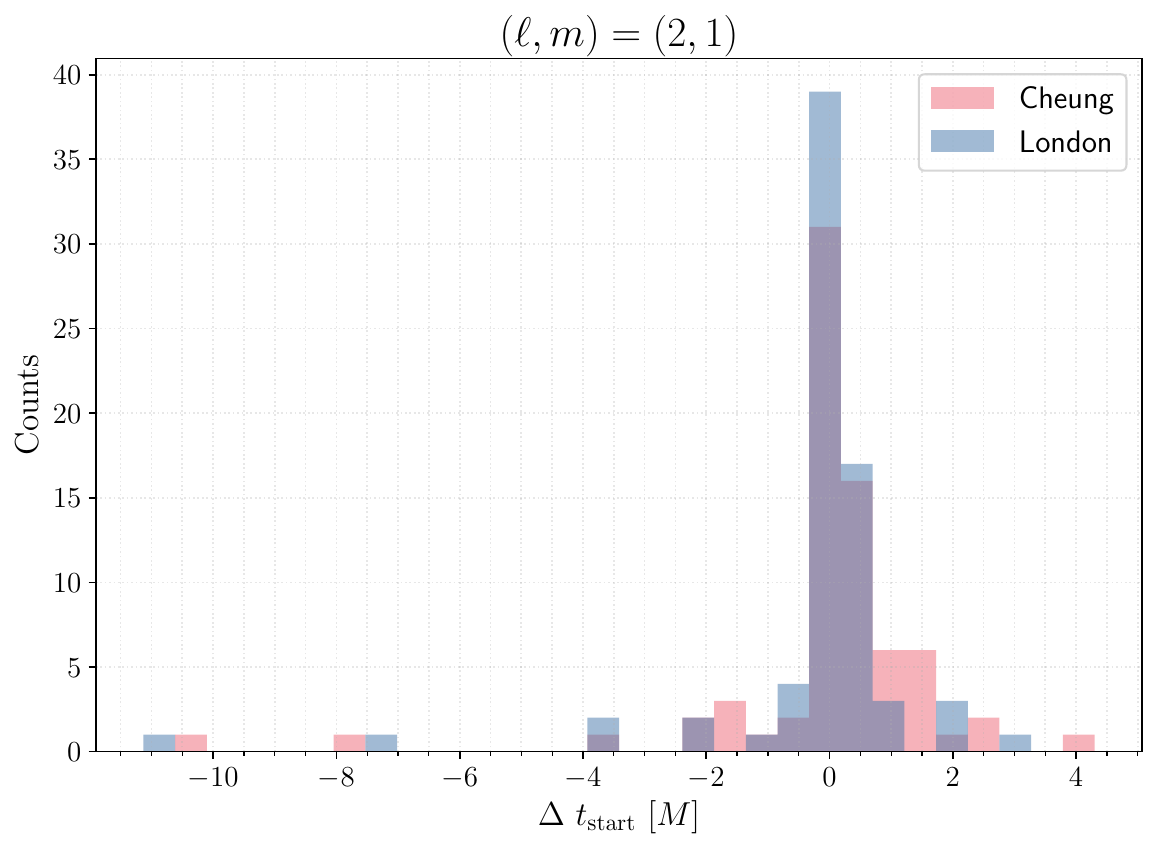}\\
    \includegraphics[width=0.45\linewidth]{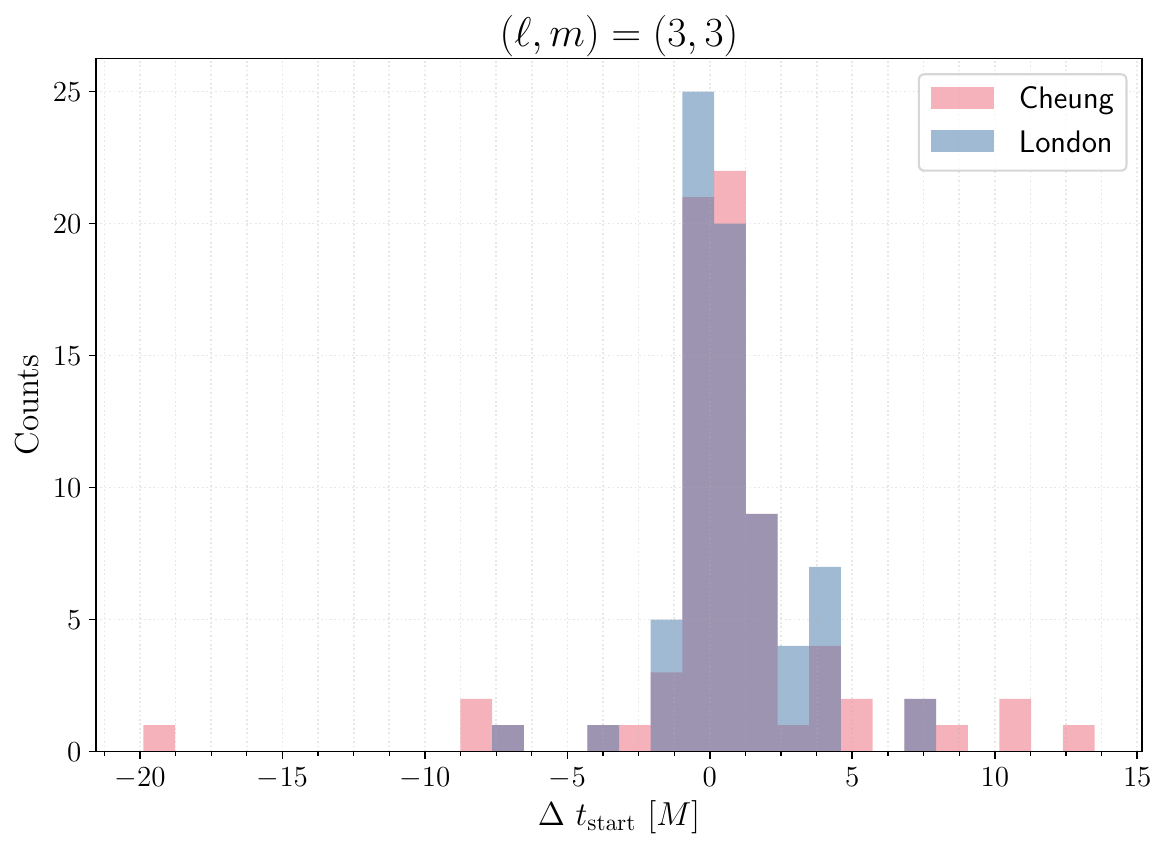}
    \includegraphics[width=0.45\linewidth]{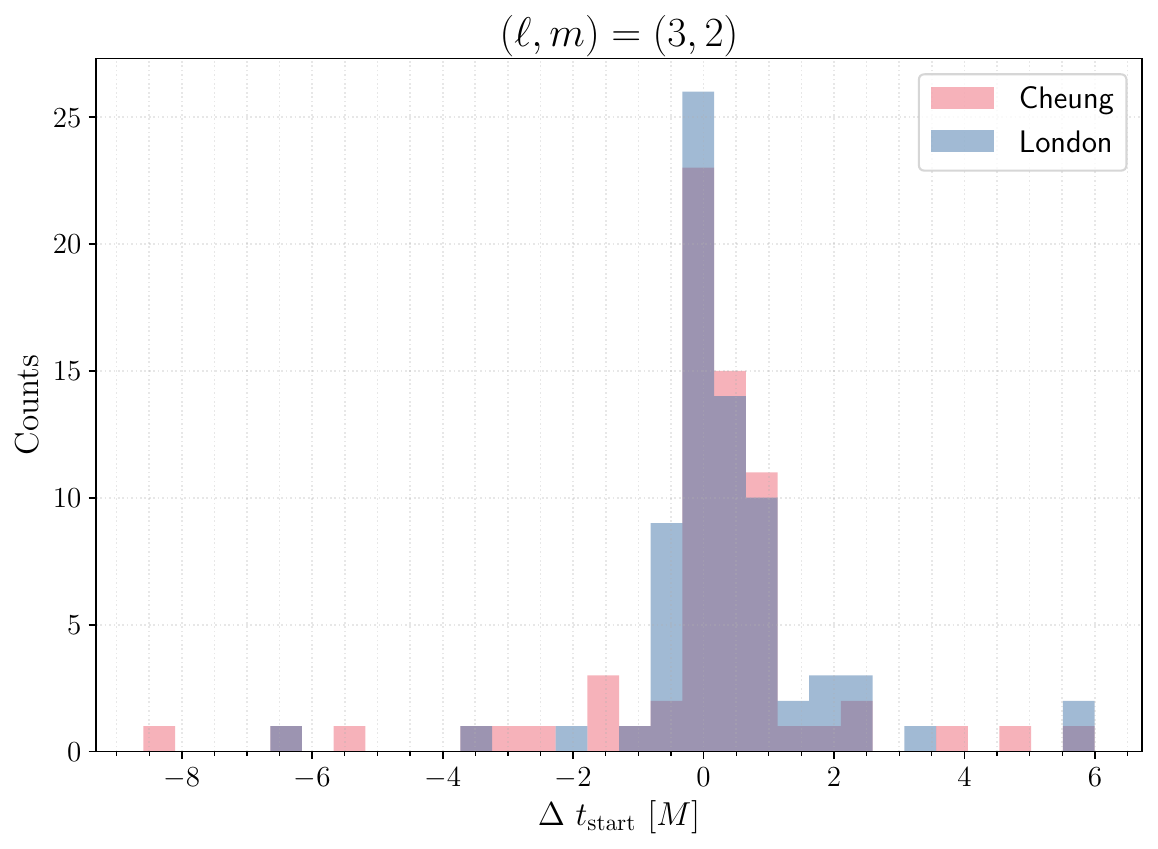}
    \includegraphics[width=0.45\linewidth]{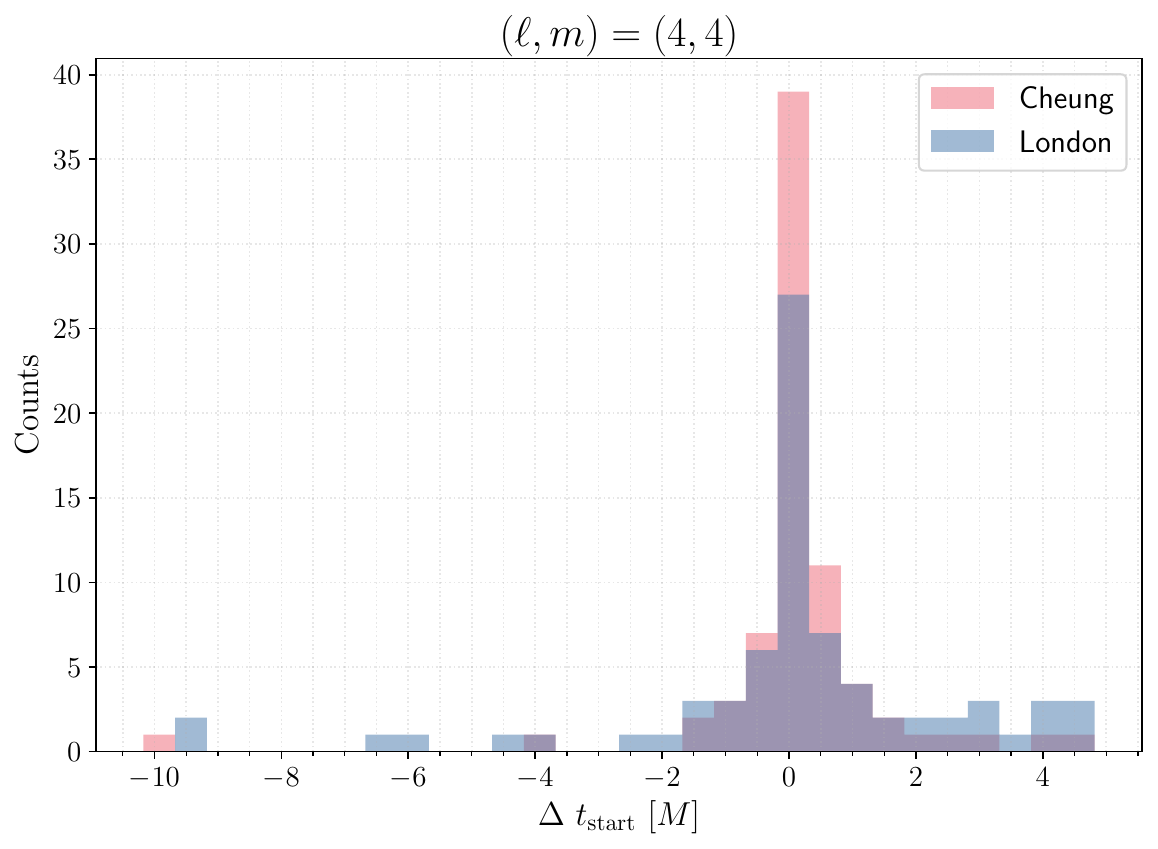}
    \caption{Distribution of the residual starting time $\Delta t_{\rm start}$, as defined in Eq.~\eqref{res_t_start_int}, for $20\%$ of the simulations of the $\mathcal{I}_{\rm tot}$ dataset, for both the \texttt{London} (blue) and \texttt{Cheung} (red) models and for different multipoles.}
    \label{fig:interpolation_check}
\end{figure*}

Then, in Fig.~\ref{fig:interpolation_check} we illustrate the accuracy of the interpolation across the parameter space by plotting the distributions of the residuals on $t_{\rm start}$, denoted as $\Delta t_{\rm start}$, for the validation set ($\sim 20\%$ of the SXS simulations). This quantity is defined as

\begin{equation}
  \Delta t_{\rm start} = t_{\rm start}^{\rm int}-t_{\rm start}^{\rm true}\,,
  \label{res_t_start_int}
\end{equation}
where $t_{\rm start}^{\rm int}$ and $t_{\rm start}^{\rm true}$ are the starting time provided by the interpolant and the true starting time, respectively. We plot the residual distributions for both the \texttt{London} and \texttt{Cheung} models. The histograms show that the residuals peak at 0. For the higher modes, however, some simulations present larger residuals.

\begin{figure}
    \centering
    \includegraphics[width=0.96\linewidth]{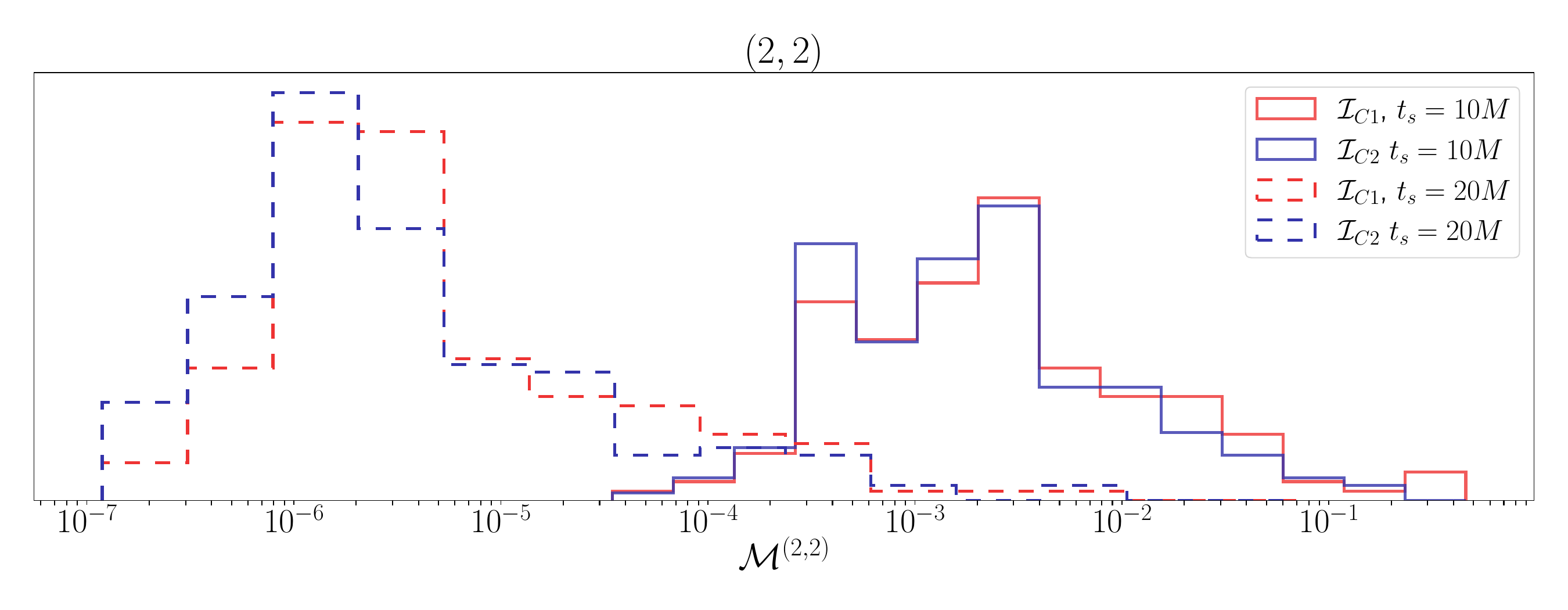}
    \includegraphics[width=0.96\linewidth]{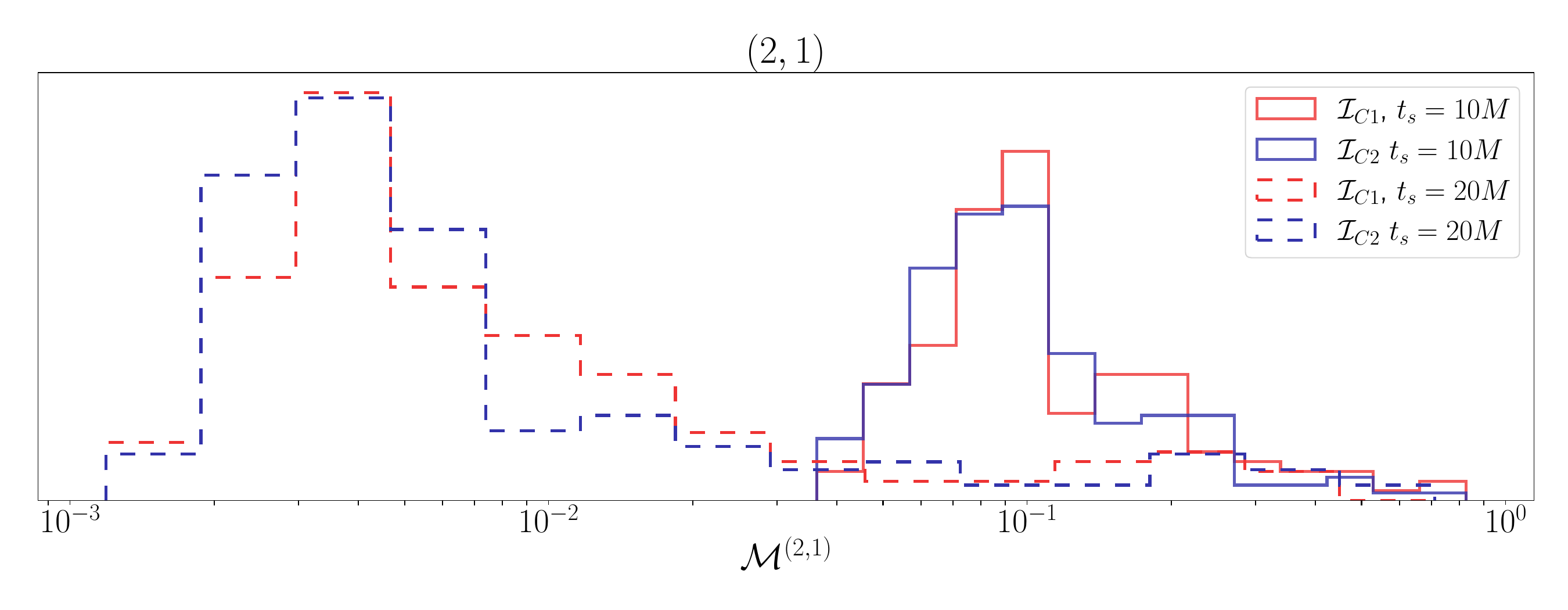}
    
    \includegraphics[width=0.96\linewidth]{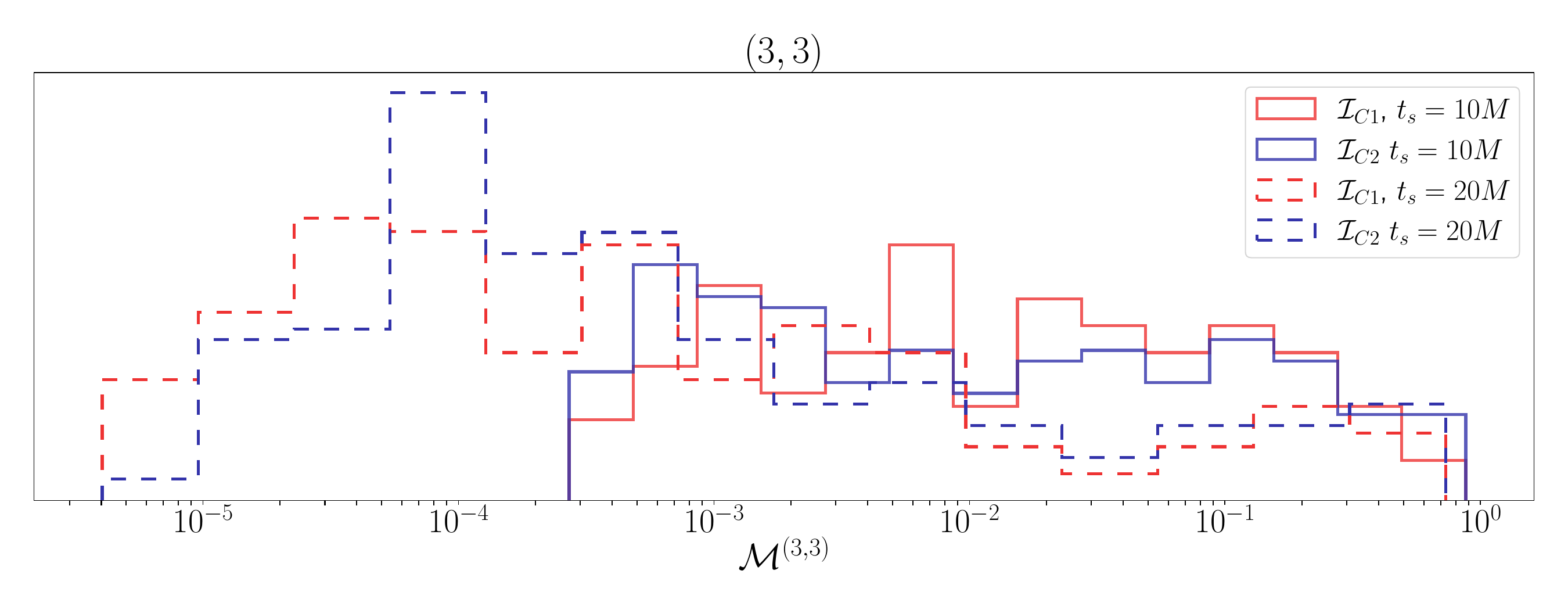}
    \includegraphics[width=0.96\linewidth]{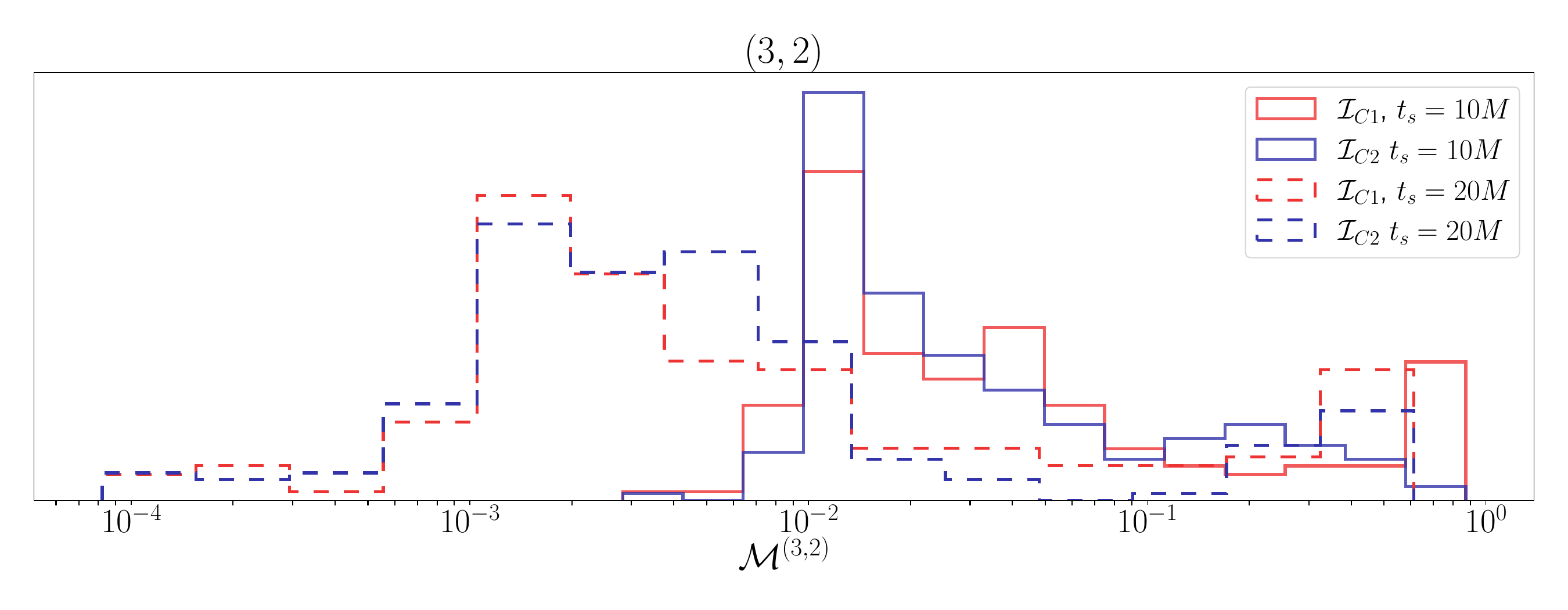}
    \includegraphics[width=0.96\linewidth]{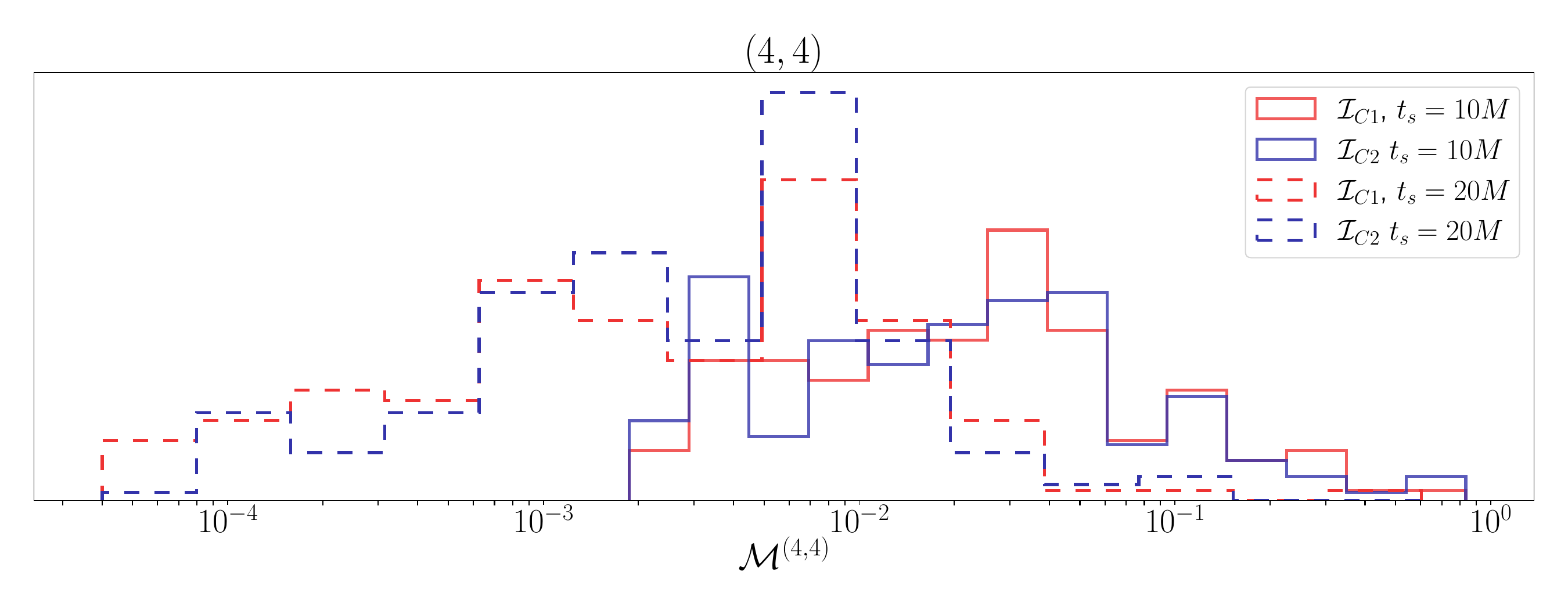}
    \caption{In this plot, we show the effect of using either the ensemble of simulations $\mathcal{I}_{\rm C}$ (red) or $\mathcal{I}_{\rm tot-C}$ (blue). Different rows correspond to different $(\ell, |m|)$ multipoles. For each panel, we report the results at $t_{\rm start}=10$M (continuous lines) and $20$M (dashed lines).}
    \label{mm_distrs_cheung_dataset}
\end{figure}

In Fig.~\ref{mm_distrs_cheung_dataset} we show the mismatch distributions of the \texttt{Cheung} model for $t_{\rm start}=10,20$M considering two different sets of simulation: (1) $\mathcal{I}_{\rm C}$, i.e., the non-precessing, quasi-circular SXS simulations used for calibrating the \texttt{Cheung} model; (2) $\mathcal{I}_{\rm tot-C}$, i.e., all the non-precessing, quasi-circular SXS simulations that were not used to calibrate the model. The two distributions have a similar behavior.

\begin{figure}
    \centering
    \includegraphics[width=0.86\linewidth]{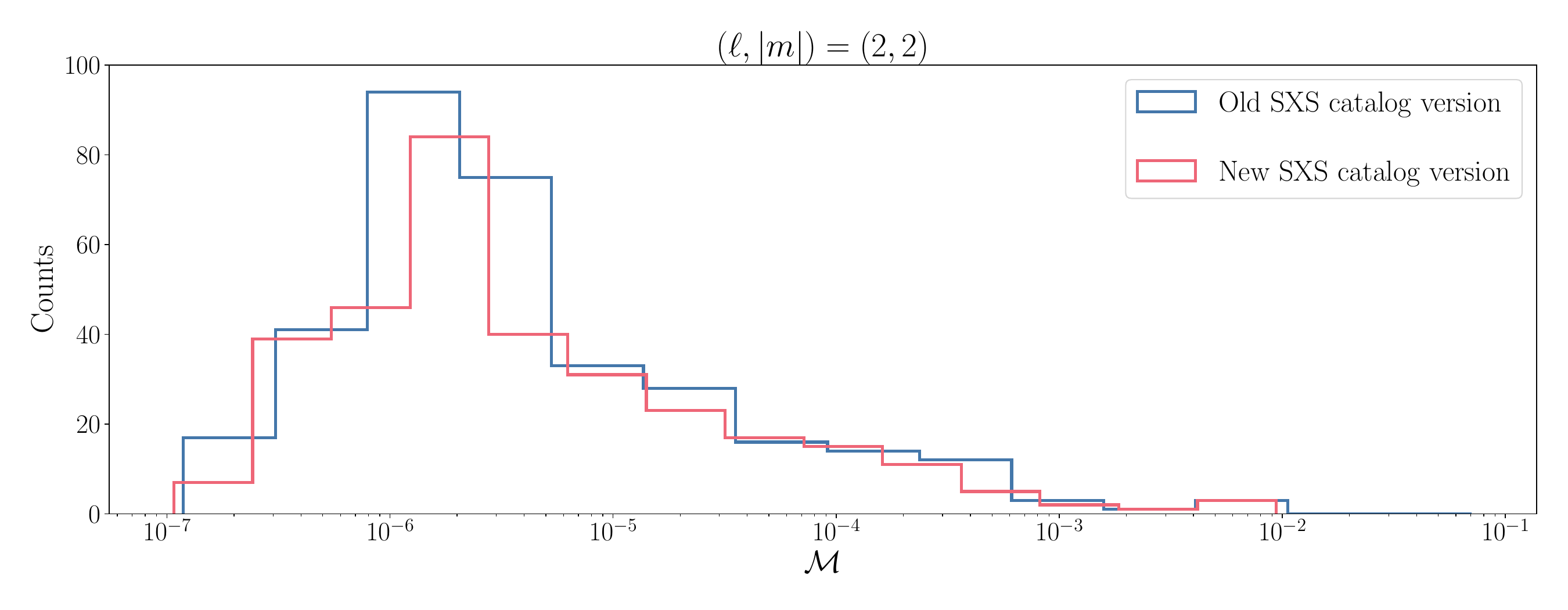}
    \includegraphics[width=0.86\linewidth]{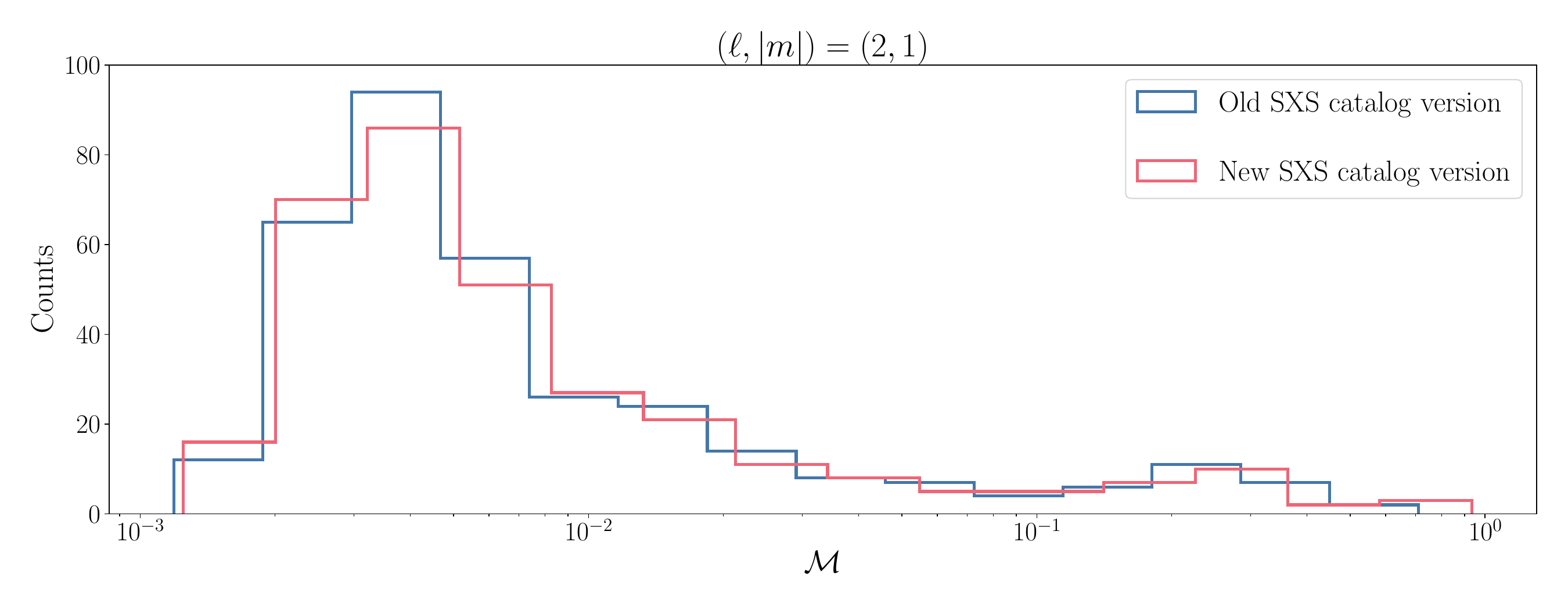}
    \includegraphics[width=0.86\linewidth]{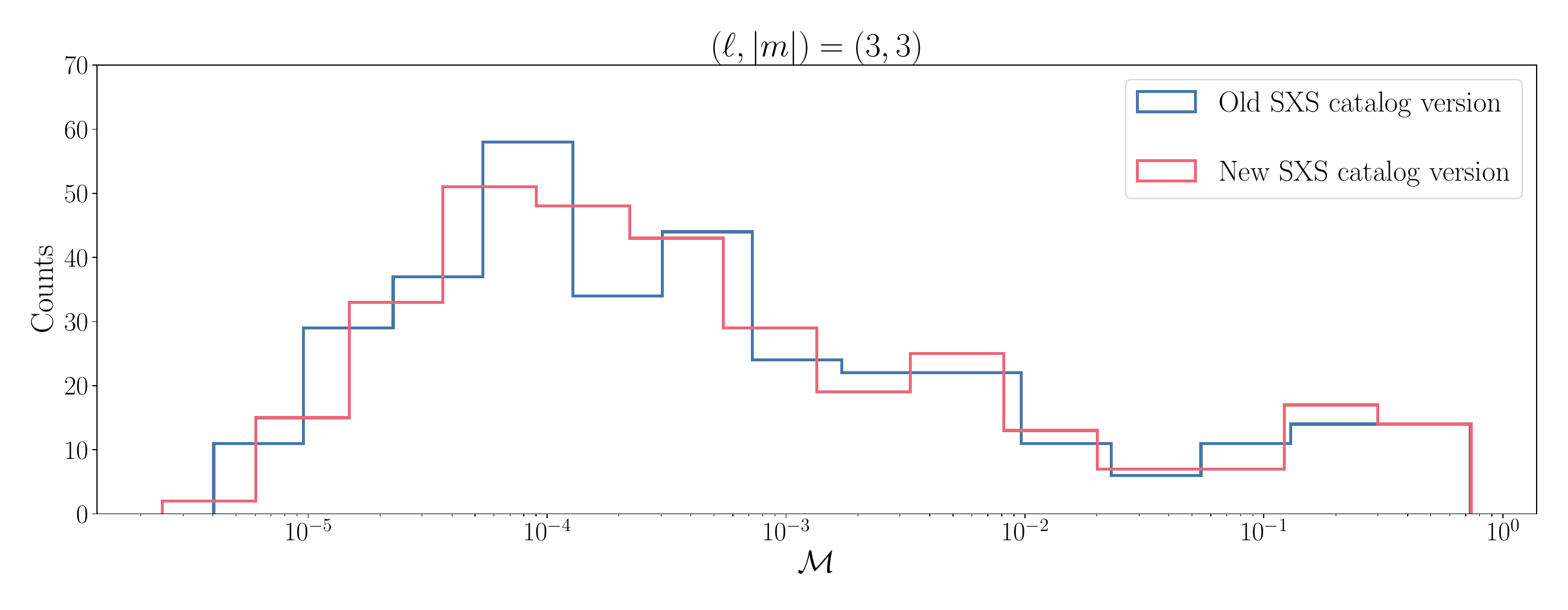}
    \includegraphics[width=0.86\linewidth]{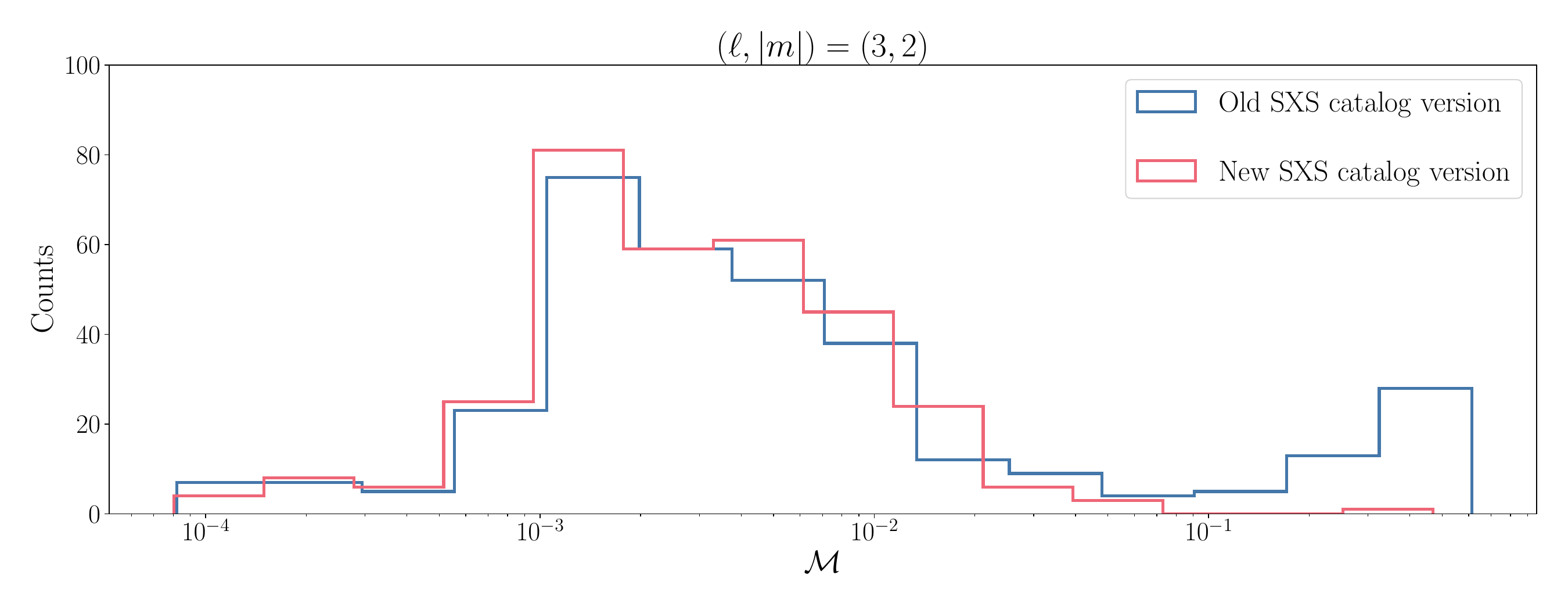}
    \includegraphics[width=0.86\linewidth]{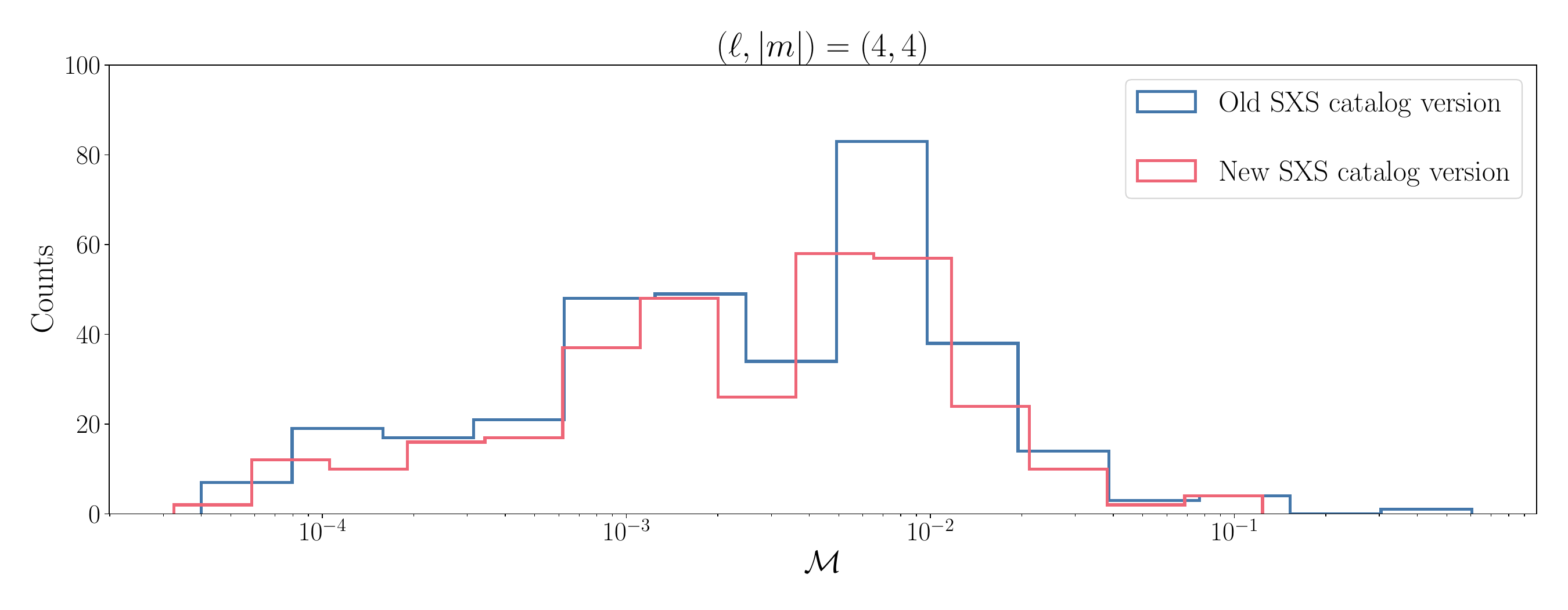}
    \caption{Mismatch distribution for the \texttt{Cheung} model by comparing the old (blue) and new (red) SXS catalog versions for the different $(\ell, |m|)$ modes. The simulations are plotted for $t_{\rm start}=20$M.}
    \label{mm_distrs_o_n}
\end{figure}

Finally, in Fig.~\ref{mm_distrs_o_n} we compare the \texttt{Cheung} model with the new, recently released version of the SXS catalog~\cite{Scheel:2025jct}. The distributions are roughly the same, because improvements in the new catalog are mostly related to the number of simulations used to span the parameter space. The main improvement in the new catalog concerns the $(3,2)$ mode, for which a subset of simulations in the old catalog have $\mathcal{M}={\cal O}(1)$.

\clearpage

\clearpage

\bibliography{main}

\begin{thebibliography}{123}%
\makeatletter
\providecommand \@ifxundefined [1]{%
 \@ifx{#1\undefined}
}%
\providecommand \@ifnum [1]{%
 \ifnum #1\expandafter \@firstoftwo
 \else \expandafter \@secondoftwo
 \fi
}%
\providecommand \@ifx [1]{%
 \ifx #1\expandafter \@firstoftwo
 \else \expandafter \@secondoftwo
 \fi
}%
\providecommand \natexlab [1]{#1}%
\providecommand \enquote  [1]{``#1''}%
\providecommand \bibnamefont  [1]{#1}%
\providecommand \bibfnamefont [1]{#1}%
\providecommand \citenamefont [1]{#1}%
\providecommand \href@noop [0]{\@secondoftwo}%
\providecommand \href [0]{\begingroup \@sanitize@url \@href}%
\providecommand \@href[1]{\@@startlink{#1}\@@href}%
\providecommand \@@href[1]{\endgroup#1\@@endlink}%
\providecommand \@sanitize@url [0]{\catcode `\\12\catcode `\$12\catcode `\&12\catcode `\#12\catcode `\^12\catcode `\_12\catcode `\%12\relax}%
\providecommand \@@startlink[1]{}%
\providecommand \@@endlink[0]{}%
\providecommand \url  [0]{\begingroup\@sanitize@url \@url }%
\providecommand \@url [1]{\endgroup\@href {#1}{\urlprefix }}%
\providecommand \urlprefix  [0]{URL }%
\providecommand \Eprint [0]{\href }%
\providecommand \doibase [0]{http://dx.doi.org/}%
\providecommand \selectlanguage [0]{\@gobble}%
\providecommand \bibinfo  [0]{\@secondoftwo}%
\providecommand \bibfield  [0]{\@secondoftwo}%
\providecommand \translation [1]{[#1]}%
\providecommand \BibitemOpen [0]{}%
\providecommand \bibitemStop [0]{}%
\providecommand \bibitemNoStop [0]{.\EOS\space}%
\providecommand \EOS [0]{\spacefactor3000\relax}%
\providecommand \BibitemShut  [1]{\csname bibitem#1\endcsname}%
\let\auto@bib@innerbib\@empty
\bibitem [{\citenamefont {Abbott}\ \emph {et~al.}(2016)\citenamefont {Abbott} \emph {et~al.}}]{Abbott:2016blz}%
  \BibitemOpen
  \bibfield  {author} {\bibinfo {author} {\bibfnamefont {B.~P.}\ \bibnamefont {Abbott}} \emph {et~al.} (\bibinfo {collaboration} {LIGO Scientific, Virgo}),\ }\href {\doibase 10.1103/PhysRevLett.116.221101} {\bibfield  {journal} {\bibinfo  {journal} {Phys. Rev. Lett.}\ }\textbf {\bibinfo {volume} {116}},\ \bibinfo {pages} {221101} (\bibinfo {year} {2016})},\ \Eprint {http://arxiv.org/abs/1602.03841} {arXiv:1602.03841 [gr-qc]} \BibitemShut {NoStop}%
\bibitem [{\citenamefont {Abbott}\ \emph {et~al.}(2023{\natexlab{a}})\citenamefont {Abbott} \emph {et~al.}}]{Abbott:2023aa}%
  \BibitemOpen
  \bibfield  {author} {\bibinfo {author} {\bibfnamefont {R.}~\bibnamefont {Abbott}} \emph {et~al.} (\bibinfo {collaboration} {KAGRA, Virgo, LIGO Scientific}),\ }\href@noop {} {\bibfield  {journal} {\bibinfo  {journal} {Phys. Rev. X}\ }\textbf {\bibinfo {volume} {13}},\ \bibinfo {pages} {041039} (\bibinfo {year} {2023}{\natexlab{a}})},\ \Eprint {http://arxiv.org/abs/2111.03606} {arXiv:2111.03606 [gr-qc]} \BibitemShut {NoStop}%
\bibitem [{\citenamefont {Abbott}\ \emph {et~al.}(2021{\natexlab{a}})\citenamefont {Abbott} \emph {et~al.}}]{Abbott:2021xx}%
  \BibitemOpen
  \bibfield  {author} {\bibinfo {author} {\bibfnamefont {R.}~\bibnamefont {Abbott}} \emph {et~al.} (\bibinfo {collaboration} {LIGO Scientific, Virgo, KAGRA}),\ }\href@noop {} {\  (\bibinfo {year} {2021}{\natexlab{a}})},\ \Eprint {http://arxiv.org/abs/2112.06861} {arXiv:2112.06861 [gr-qc]} \BibitemShut {NoStop}%
\bibitem [{\citenamefont {Abbott}\ \emph {et~al.}(2023{\natexlab{b}})\citenamefont {Abbott} \emph {et~al.}}]{Abbott:2023yy}%
  \BibitemOpen
  \bibfield  {author} {\bibinfo {author} {\bibfnamefont {R.}~\bibnamefont {Abbott}} \emph {et~al.} (\bibinfo {collaboration} {KAGRA, Virgo, LIGO Scientific}),\ }\href@noop {} {\bibfield  {journal} {\bibinfo  {journal} {Phys. Rev. X}\ }\textbf {\bibinfo {volume} {13}},\ \bibinfo {pages} {011048} (\bibinfo {year} {2023}{\natexlab{b}})},\ \Eprint {http://arxiv.org/abs/2111.03634} {arXiv:2111.03634 [astro-ph.HE]} \BibitemShut {NoStop}%
\bibitem [{\citenamefont {Abbott}\ \emph {et~al.}(2023{\natexlab{c}})\citenamefont {Abbott} \emph {et~al.}}]{Abbott:2023zz}%
  \BibitemOpen
  \bibfield  {author} {\bibinfo {author} {\bibfnamefont {R.}~\bibnamefont {Abbott}} \emph {et~al.} (\bibinfo {collaboration} {LIGO Scientific, Virgo, KAGRA}),\ }\href@noop {} {\bibfield  {journal} {\bibinfo  {journal} {Astrophys. J.}\ }\textbf {\bibinfo {volume} {949}},\ \bibinfo {pages} {76} (\bibinfo {year} {2023}{\natexlab{c}})},\ \Eprint {http://arxiv.org/abs/2111.03604} {arXiv:2111.03604 [astro-ph.CO]} \BibitemShut {NoStop}%
\bibitem [{\citenamefont {Abbott}\ \emph {et~al.}(2019)\citenamefont {Abbott} \emph {et~al.}}]{Abbott:2019prv}%
  \BibitemOpen
  \bibfield  {author} {\bibinfo {author} {\bibfnamefont {B.~P.}\ \bibnamefont {Abbott}} \emph {et~al.} (\bibinfo {collaboration} {LIGO Scientific, Virgo}),\ }\href {\doibase 10.1103/PhysRevD.100.104036} {\bibfield  {journal} {\bibinfo  {journal} {Phys. Rev. D}\ }\textbf {\bibinfo {volume} {100}},\ \bibinfo {pages} {104036} (\bibinfo {year} {2019})},\ \Eprint {http://arxiv.org/abs/1903.04467} {arXiv:1903.04467 [gr-qc]} \BibitemShut {NoStop}%
\bibitem [{\citenamefont {Abbott}\ \emph {et~al.}(2021{\natexlab{b}})\citenamefont {Abbott} \emph {et~al.}}]{Abbott:2020jks}%
  \BibitemOpen
  \bibfield  {author} {\bibinfo {author} {\bibfnamefont {R.}~\bibnamefont {Abbott}} \emph {et~al.} (\bibinfo {collaboration} {LIGO Scientific, Virgo}),\ }\href {\doibase 10.1103/PhysRevD.103.122002} {\bibfield  {journal} {\bibinfo  {journal} {Phys. Rev. D}\ }\textbf {\bibinfo {volume} {103}},\ \bibinfo {pages} {122002} (\bibinfo {year} {2021}{\natexlab{b}})},\ \Eprint {http://arxiv.org/abs/2010.14529} {arXiv:2010.14529 [gr-qc]} \BibitemShut {NoStop}%
\bibitem [{\citenamefont {Berti}\ \emph {et~al.}(2015)\citenamefont {Berti} \emph {et~al.}}]{Berti:2015itd}%
  \BibitemOpen
  \bibfield  {author} {\bibinfo {author} {\bibfnamefont {E.}~\bibnamefont {Berti}} \emph {et~al.},\ }\href {\doibase 10.1088/0264-9381/32/24/243001} {\bibfield  {journal} {\bibinfo  {journal} {Class. Quant. Grav.}\ }\textbf {\bibinfo {volume} {32}},\ \bibinfo {pages} {243001} (\bibinfo {year} {2015})},\ \Eprint {http://arxiv.org/abs/1501.07274} {arXiv:1501.07274 [gr-qc]} \BibitemShut {NoStop}%
\bibitem [{\citenamefont {Berti}\ \emph {et~al.}(2018)\citenamefont {Berti}, \citenamefont {Yagi}, \citenamefont {Yang},\ and\ \citenamefont {Yunes}}]{Berti:2018vdi}%
  \BibitemOpen
  \bibfield  {author} {\bibinfo {author} {\bibfnamefont {E.}~\bibnamefont {Berti}}, \bibinfo {author} {\bibfnamefont {K.}~\bibnamefont {Yagi}}, \bibinfo {author} {\bibfnamefont {H.}~\bibnamefont {Yang}}, \ and\ \bibinfo {author} {\bibfnamefont {N.}~\bibnamefont {Yunes}},\ }\href {\doibase 10.1007/s10714-018-2372-6} {\bibfield  {journal} {\bibinfo  {journal} {Gen. Rel. Grav.}\ }\textbf {\bibinfo {volume} {50}},\ \bibinfo {pages} {49} (\bibinfo {year} {2018})},\ \Eprint {http://arxiv.org/abs/1801.03587} {arXiv:1801.03587 [gr-qc]} \BibitemShut {NoStop}%
\bibitem [{\citenamefont {Cardoso}\ and\ \citenamefont {Pani}(2019)}]{Cardoso:2019rvt}%
  \BibitemOpen
  \bibfield  {author} {\bibinfo {author} {\bibfnamefont {V.}~\bibnamefont {Cardoso}}\ and\ \bibinfo {author} {\bibfnamefont {P.}~\bibnamefont {Pani}},\ }\href {\doibase 10.1007/s41114-019-0020-4} {\bibfield  {journal} {\bibinfo  {journal} {Living Rev. Rel.}\ }\textbf {\bibinfo {volume} {22}},\ \bibinfo {pages} {4} (\bibinfo {year} {2019})},\ \Eprint {http://arxiv.org/abs/1904.05363} {arXiv:1904.05363 [gr-qc]} \BibitemShut {NoStop}%
\bibitem [{\citenamefont {Regge}\ and\ \citenamefont {Wheeler}(1957)}]{Regge:1957td}%
  \BibitemOpen
  \bibfield  {author} {\bibinfo {author} {\bibfnamefont {T.}~\bibnamefont {Regge}}\ and\ \bibinfo {author} {\bibfnamefont {J.~A.}\ \bibnamefont {Wheeler}},\ }\href {\doibase 10.1103/PhysRev.108.1063} {\bibfield  {journal} {\bibinfo  {journal} {Phys. Rev.}\ }\textbf {\bibinfo {volume} {108}},\ \bibinfo {pages} {1063} (\bibinfo {year} {1957})}\BibitemShut {NoStop}%
\bibitem [{\citenamefont {Zerilli}(1970)}]{Zerilli:1970se}%
  \BibitemOpen
  \bibfield  {author} {\bibinfo {author} {\bibfnamefont {F.~J.}\ \bibnamefont {Zerilli}},\ }\href {\doibase 10.1103/PhysRevD.2.2141} {\bibfield  {journal} {\bibinfo  {journal} {Phys. Rev. D}\ }\textbf {\bibinfo {volume} {2}},\ \bibinfo {pages} {2141} (\bibinfo {year} {1970})}\BibitemShut {NoStop}%
\bibitem [{\citenamefont {Teukolsky}(1972)}]{Teukolsky:1972my}%
  \BibitemOpen
  \bibfield  {author} {\bibinfo {author} {\bibfnamefont {S.~A.}\ \bibnamefont {Teukolsky}},\ }\href {\doibase 10.1103/PhysRevLett.29.1114} {\bibfield  {journal} {\bibinfo  {journal} {Phys. Rev. Lett.}\ }\textbf {\bibinfo {volume} {29}},\ \bibinfo {pages} {1114} (\bibinfo {year} {1972})}\BibitemShut {NoStop}%
\bibitem [{\citenamefont {Teukolsky}(1973)}]{Teukolsky:1973ha}%
  \BibitemOpen
  \bibfield  {author} {\bibinfo {author} {\bibfnamefont {S.~A.}\ \bibnamefont {Teukolsky}},\ }\href {\doibase 10.1086/152444} {\bibfield  {journal} {\bibinfo  {journal} {Astrophys. J.}\ }\textbf {\bibinfo {volume} {185}},\ \bibinfo {pages} {635} (\bibinfo {year} {1973})}\BibitemShut {NoStop}%
\bibitem [{\citenamefont {Vishveshwara}(1970)}]{Vishveshwara:1970zz}%
  \BibitemOpen
  \bibfield  {author} {\bibinfo {author} {\bibfnamefont {C.~V.}\ \bibnamefont {Vishveshwara}},\ }\href {\doibase 10.1038/227936a0} {\bibfield  {journal} {\bibinfo  {journal} {Nature}\ }\textbf {\bibinfo {volume} {227}},\ \bibinfo {pages} {936} (\bibinfo {year} {1970})}\BibitemShut {NoStop}%
\bibitem [{\citenamefont {Press}(1971)}]{Press:1971wr}%
  \BibitemOpen
  \bibfield  {author} {\bibinfo {author} {\bibfnamefont {W.~H.}\ \bibnamefont {Press}},\ }\href {\doibase 10.1086/180849} {\bibfield  {journal} {\bibinfo  {journal} {Astrophys. J. Lett.}\ }\textbf {\bibinfo {volume} {170}},\ \bibinfo {pages} {L105} (\bibinfo {year} {1971})}\BibitemShut {NoStop}%
\bibitem [{\citenamefont {Leaver}(1986)}]{Leaver:1986gd}%
  \BibitemOpen
  \bibfield  {author} {\bibinfo {author} {\bibfnamefont {E.~W.}\ \bibnamefont {Leaver}},\ }\href {\doibase 10.1103/PhysRevD.34.384} {\bibfield  {journal} {\bibinfo  {journal} {Phys. Rev. D}\ }\textbf {\bibinfo {volume} {34}},\ \bibinfo {pages} {384} (\bibinfo {year} {1986})}\BibitemShut {NoStop}%
\bibitem [{\citenamefont {Israel}(1967)}]{Israel:1967wq}%
  \BibitemOpen
  \bibfield  {author} {\bibinfo {author} {\bibfnamefont {W.}~\bibnamefont {Israel}},\ }\href {\doibase 10.1103/PhysRev.164.1776} {\bibfield  {journal} {\bibinfo  {journal} {Phys. Rev.}\ }\textbf {\bibinfo {volume} {164}},\ \bibinfo {pages} {1776} (\bibinfo {year} {1967})}\BibitemShut {NoStop}%
\bibitem [{\citenamefont {Hawking}(1972)}]{Hawking:1972qk}%
  \BibitemOpen
  \bibfield  {author} {\bibinfo {author} {\bibfnamefont {S.~W.}\ \bibnamefont {Hawking}},\ }\href {\doibase 10.1007/BF01877517} {\bibfield  {journal} {\bibinfo  {journal} {Commun. Math. Phys.}\ }\textbf {\bibinfo {volume} {25}},\ \bibinfo {pages} {152} (\bibinfo {year} {1972})}\BibitemShut {NoStop}%
\bibitem [{\citenamefont {Carter}(1971)}]{Carter:1971zc}%
  \BibitemOpen
  \bibfield  {author} {\bibinfo {author} {\bibfnamefont {B.}~\bibnamefont {Carter}},\ }\href {\doibase 10.1103/PhysRevLett.26.331} {\bibfield  {journal} {\bibinfo  {journal} {Phys. Rev. Lett.}\ }\textbf {\bibinfo {volume} {26}},\ \bibinfo {pages} {331} (\bibinfo {year} {1971})}\BibitemShut {NoStop}%
\bibitem [{\citenamefont {Robinson}(1975)}]{Robinson:1975bv}%
  \BibitemOpen
  \bibfield  {author} {\bibinfo {author} {\bibfnamefont {D.~C.}\ \bibnamefont {Robinson}},\ }\href {\doibase 10.1103/PhysRevLett.34.905} {\bibfield  {journal} {\bibinfo  {journal} {Phys. Rev. Lett.}\ }\textbf {\bibinfo {volume} {34}},\ \bibinfo {pages} {905} (\bibinfo {year} {1975})}\BibitemShut {NoStop}%
\bibitem [{\citenamefont {Detweiler}(1980)}]{Detweiler:1980gk}%
  \BibitemOpen
  \bibfield  {author} {\bibinfo {author} {\bibfnamefont {S.~L.}\ \bibnamefont {Detweiler}},\ }\href {\doibase 10.1086/158109} {\bibfield  {journal} {\bibinfo  {journal} {Astrophys. J.}\ }\textbf {\bibinfo {volume} {239}},\ \bibinfo {pages} {292} (\bibinfo {year} {1980})}\BibitemShut {NoStop}%
\bibitem [{\citenamefont {Dreyer}\ \emph {et~al.}(2004)\citenamefont {Dreyer}, \citenamefont {Kelly}, \citenamefont {Krishnan}, \citenamefont {Finn}, \citenamefont {Garrison},\ and\ \citenamefont {Lopez-Aleman}}]{Dreyer:2003bv}%
  \BibitemOpen
  \bibfield  {author} {\bibinfo {author} {\bibfnamefont {O.}~\bibnamefont {Dreyer}}, \bibinfo {author} {\bibfnamefont {B.~J.}\ \bibnamefont {Kelly}}, \bibinfo {author} {\bibfnamefont {B.}~\bibnamefont {Krishnan}}, \bibinfo {author} {\bibfnamefont {L.~S.}\ \bibnamefont {Finn}}, \bibinfo {author} {\bibfnamefont {D.}~\bibnamefont {Garrison}}, \ and\ \bibinfo {author} {\bibfnamefont {R.}~\bibnamefont {Lopez-Aleman}},\ }\href {\doibase 10.1088/0264-9381/21/4/003} {\bibfield  {journal} {\bibinfo  {journal} {Class. Quant. Grav.}\ }\textbf {\bibinfo {volume} {21}},\ \bibinfo {pages} {787} (\bibinfo {year} {2004})},\ \Eprint {http://arxiv.org/abs/gr-qc/0309007} {arXiv:gr-qc/0309007 [gr-qc]} \BibitemShut {NoStop}%
\bibitem [{\citenamefont {Berti}\ \emph {et~al.}(2006{\natexlab{a}})\citenamefont {Berti}, \citenamefont {Cardoso},\ and\ \citenamefont {Will}}]{Berti:2005ys}%
  \BibitemOpen
  \bibfield  {author} {\bibinfo {author} {\bibfnamefont {E.}~\bibnamefont {Berti}}, \bibinfo {author} {\bibfnamefont {V.}~\bibnamefont {Cardoso}}, \ and\ \bibinfo {author} {\bibfnamefont {C.~M.}\ \bibnamefont {Will}},\ }\href {\doibase 10.1103/PhysRevD.73.064030} {\bibfield  {journal} {\bibinfo  {journal} {Phys. Rev. D}\ }\textbf {\bibinfo {volume} {73}},\ \bibinfo {pages} {064030} (\bibinfo {year} {2006}{\natexlab{a}})},\ \Eprint {http://arxiv.org/abs/gr-qc/0512160} {arXiv:gr-qc/0512160 [gr-qc]} \BibitemShut {NoStop}%
\bibitem [{\citenamefont {Berti}\ \emph {et~al.}(2007{\natexlab{a}})\citenamefont {Berti}, \citenamefont {Cardoso}, \citenamefont {Cardoso},\ and\ \citenamefont {Cavaglia}}]{Berti:2007zu}%
  \BibitemOpen
  \bibfield  {author} {\bibinfo {author} {\bibfnamefont {E.}~\bibnamefont {Berti}}, \bibinfo {author} {\bibfnamefont {J.}~\bibnamefont {Cardoso}}, \bibinfo {author} {\bibfnamefont {V.}~\bibnamefont {Cardoso}}, \ and\ \bibinfo {author} {\bibfnamefont {M.}~\bibnamefont {Cavaglia}},\ }\href {\doibase 10.1103/PhysRevD.76.104044} {\bibfield  {journal} {\bibinfo  {journal} {Phys. Rev. D}\ }\textbf {\bibinfo {volume} {76}},\ \bibinfo {pages} {104044} (\bibinfo {year} {2007}{\natexlab{a}})},\ \Eprint {http://arxiv.org/abs/0707.1202} {arXiv:0707.1202 [gr-qc]} \BibitemShut {NoStop}%
\bibitem [{\citenamefont {Meidam}\ \emph {et~al.}(2014)\citenamefont {Meidam}, \citenamefont {Agathos}, \citenamefont {Van Den~Broeck}, \citenamefont {Veitch},\ and\ \citenamefont {Sathyaprakash}}]{Meidam:2014jpa}%
  \BibitemOpen
  \bibfield  {author} {\bibinfo {author} {\bibfnamefont {J.}~\bibnamefont {Meidam}}, \bibinfo {author} {\bibfnamefont {M.}~\bibnamefont {Agathos}}, \bibinfo {author} {\bibfnamefont {C.}~\bibnamefont {Van Den~Broeck}}, \bibinfo {author} {\bibfnamefont {J.}~\bibnamefont {Veitch}}, \ and\ \bibinfo {author} {\bibfnamefont {B.~S.}\ \bibnamefont {Sathyaprakash}},\ }\href {\doibase 10.1103/PhysRevD.90.064009} {\bibfield  {journal} {\bibinfo  {journal} {Phys. Rev. D}\ }\textbf {\bibinfo {volume} {90}},\ \bibinfo {pages} {064009} (\bibinfo {year} {2014})},\ \Eprint {http://arxiv.org/abs/1406.3201} {arXiv:1406.3201 [gr-qc]} \BibitemShut {NoStop}%
\bibitem [{\citenamefont {Berti}\ \emph {et~al.}(2025)\citenamefont {Berti}, \citenamefont {Cardoso}, \citenamefont {Carullo} \emph {et~al.}}]{Berti:2025hly}%
  \BibitemOpen
  \bibfield  {author} {\bibinfo {author} {\bibfnamefont {E.}~\bibnamefont {Berti}}, \bibinfo {author} {\bibfnamefont {V.}~\bibnamefont {Cardoso}}, \bibinfo {author} {\bibfnamefont {G.}~\bibnamefont {Carullo}},  \emph {et~al.},\ }\href@noop {} {\  (\bibinfo {year} {2025})},\ \Eprint {http://arxiv.org/abs/2505.23895} {arXiv:2505.23895 [gr-qc]} \BibitemShut {NoStop}%
\bibitem [{\citenamefont {Abac}\ \emph {et~al.}(2025{\natexlab{a}})\citenamefont {Abac} \emph {et~al.}}]{LIGOScientific:2025obp}%
  \BibitemOpen
  \bibfield  {author} {\bibinfo {author} {\bibfnamefont {A.~G.}\ \bibnamefont {Abac}} \emph {et~al.} (\bibinfo {collaboration} {LIGO Scientific, Virgo, KAGRA}),\ }\href@noop {} {\  (\bibinfo {year} {2025}{\natexlab{a}})},\ \Eprint {http://arxiv.org/abs/2509.08099} {arXiv:2509.08099 [gr-qc]} \BibitemShut {NoStop}%
\bibitem [{\citenamefont {Abac}\ \emph {et~al.}(2025{\natexlab{b}})\citenamefont {Abac} \emph {et~al.}}]{LIGOScientific:2025rid}%
  \BibitemOpen
  \bibfield  {author} {\bibinfo {author} {\bibfnamefont {A.~G.}\ \bibnamefont {Abac}} \emph {et~al.} (\bibinfo {collaboration} {LIGO Scientific, Virgo, KAGRA}),\ }\href {\doibase 10.1103/kw5g-d732} {\bibfield  {journal} {\bibinfo  {journal} {Phys. Rev. Lett.}\ }\textbf {\bibinfo {volume} {135}},\ \bibinfo {pages} {111403} (\bibinfo {year} {2025}{\natexlab{b}})},\ \Eprint {http://arxiv.org/abs/2509.08054} {arXiv:2509.08054 [gr-qc]} \BibitemShut {NoStop}%
\bibitem [{\citenamefont {De~Amicis}\ \emph {et~al.}(2025)\citenamefont {De~Amicis}, \citenamefont {Cannizzaro}, \citenamefont {Carullo},\ and\ \citenamefont {Sberna}}]{DeAmicis:2025xuh}%
  \BibitemOpen
  \bibfield  {author} {\bibinfo {author} {\bibfnamefont {M.}~\bibnamefont {De~Amicis}}, \bibinfo {author} {\bibfnamefont {E.}~\bibnamefont {Cannizzaro}}, \bibinfo {author} {\bibfnamefont {G.}~\bibnamefont {Carullo}}, \ and\ \bibinfo {author} {\bibfnamefont {L.}~\bibnamefont {Sberna}},\ }\href@noop {} {\  (\bibinfo {year} {2025})},\ \Eprint {http://arxiv.org/abs/2506.21668} {arXiv:2506.21668 [gr-qc]} \BibitemShut {NoStop}%
\bibitem [{\citenamefont {Kamaretsos}\ \emph {et~al.}(2012{\natexlab{a}})\citenamefont {Kamaretsos}, \citenamefont {Hannam}, \citenamefont {Husa},\ and\ \citenamefont {Sathyaprakash}}]{Kamaretsos:2011um}%
  \BibitemOpen
  \bibfield  {author} {\bibinfo {author} {\bibfnamefont {I.}~\bibnamefont {Kamaretsos}}, \bibinfo {author} {\bibfnamefont {M.}~\bibnamefont {Hannam}}, \bibinfo {author} {\bibfnamefont {S.}~\bibnamefont {Husa}}, \ and\ \bibinfo {author} {\bibfnamefont {B.~S.}\ \bibnamefont {Sathyaprakash}},\ }\href {\doibase 10.1103/PhysRevD.85.024018} {\bibfield  {journal} {\bibinfo  {journal} {Phys. Rev. D}\ }\textbf {\bibinfo {volume} {85}},\ \bibinfo {pages} {024018} (\bibinfo {year} {2012}{\natexlab{a}})},\ \Eprint {http://arxiv.org/abs/1107.0854} {arXiv:1107.0854 [gr-qc]} \BibitemShut {NoStop}%
\bibitem [{\citenamefont {Kamaretsos}\ \emph {et~al.}(2012{\natexlab{b}})\citenamefont {Kamaretsos}, \citenamefont {Hannam},\ and\ \citenamefont {Sathyaprakash}}]{Kamaretsos:2012bs}%
  \BibitemOpen
  \bibfield  {author} {\bibinfo {author} {\bibfnamefont {I.}~\bibnamefont {Kamaretsos}}, \bibinfo {author} {\bibfnamefont {M.}~\bibnamefont {Hannam}}, \ and\ \bibinfo {author} {\bibfnamefont {B.}~\bibnamefont {Sathyaprakash}},\ }\href {\doibase 10.1103/PhysRevLett.109.141102} {\bibfield  {journal} {\bibinfo  {journal} {Phys. Rev. Lett.}\ }\textbf {\bibinfo {volume} {109}},\ \bibinfo {pages} {141102} (\bibinfo {year} {2012}{\natexlab{b}})},\ \Eprint {http://arxiv.org/abs/1207.0399} {arXiv:1207.0399 [gr-qc]} \BibitemShut {NoStop}%
\bibitem [{\citenamefont {Barausse}\ \emph {et~al.}(2015)\citenamefont {Barausse}, \citenamefont {Bellovary}, \citenamefont {Berti}, \citenamefont {Holley-Bockelmann}, \citenamefont {Farris}, \citenamefont {Sathyaprakash},\ and\ \citenamefont {Sesana}}]{Barausse:2014oca}%
  \BibitemOpen
  \bibfield  {author} {\bibinfo {author} {\bibfnamefont {E.}~\bibnamefont {Barausse}}, \bibinfo {author} {\bibfnamefont {J.}~\bibnamefont {Bellovary}}, \bibinfo {author} {\bibfnamefont {E.}~\bibnamefont {Berti}}, \bibinfo {author} {\bibfnamefont {K.}~\bibnamefont {Holley-Bockelmann}}, \bibinfo {author} {\bibfnamefont {B.}~\bibnamefont {Farris}}, \bibinfo {author} {\bibfnamefont {B.}~\bibnamefont {Sathyaprakash}}, \ and\ \bibinfo {author} {\bibfnamefont {A.}~\bibnamefont {Sesana}},\ }\href {\doibase 10.1088/1742-6596/610/1/012001} {\bibfield  {journal} {\bibinfo  {journal} {J. Phys. Conf. Ser.}\ }\textbf {\bibinfo {volume} {610}},\ \bibinfo {pages} {012001} (\bibinfo {year} {2015})},\ \Eprint {http://arxiv.org/abs/1410.2907} {arXiv:1410.2907 [astro-ph.HE]} \BibitemShut {NoStop}%
\bibitem [{\citenamefont {Miller}\ \emph {et~al.}(2025)\citenamefont {Miller}, \citenamefont {Isi}, \citenamefont {Chatziioannou}, \citenamefont {Varma},\ and\ \citenamefont {Hourihane}}]{Miller:2025eak}%
  \BibitemOpen
  \bibfield  {author} {\bibinfo {author} {\bibfnamefont {S.~J.}\ \bibnamefont {Miller}}, \bibinfo {author} {\bibfnamefont {M.}~\bibnamefont {Isi}}, \bibinfo {author} {\bibfnamefont {K.}~\bibnamefont {Chatziioannou}}, \bibinfo {author} {\bibfnamefont {V.}~\bibnamefont {Varma}}, \ and\ \bibinfo {author} {\bibfnamefont {S.}~\bibnamefont {Hourihane}},\ }\href@noop {} {\  (\bibinfo {year} {2025})},\ \Eprint {http://arxiv.org/abs/2505.14573} {arXiv:2505.14573 [gr-qc]} \BibitemShut {NoStop}%
\bibitem [{\citenamefont {Hadar}\ and\ \citenamefont {Kol}(2011)}]{Hadar:2009ip}%
  \BibitemOpen
  \bibfield  {author} {\bibinfo {author} {\bibfnamefont {S.}~\bibnamefont {Hadar}}\ and\ \bibinfo {author} {\bibfnamefont {B.}~\bibnamefont {Kol}},\ }\href {\doibase 10.1103/PhysRevD.84.044019} {\bibfield  {journal} {\bibinfo  {journal} {Phys. Rev. D}\ }\textbf {\bibinfo {volume} {84}},\ \bibinfo {pages} {044019} (\bibinfo {year} {2011})},\ \Eprint {http://arxiv.org/abs/0911.3899} {arXiv:0911.3899 [gr-qc]} \BibitemShut {NoStop}%
\bibitem [{\citenamefont {Zhang}\ \emph {et~al.}(2013)\citenamefont {Zhang}, \citenamefont {Berti},\ and\ \citenamefont {Cardoso}}]{Zhang:2013ksa}%
  \BibitemOpen
  \bibfield  {author} {\bibinfo {author} {\bibfnamefont {Z.}~\bibnamefont {Zhang}}, \bibinfo {author} {\bibfnamefont {E.}~\bibnamefont {Berti}}, \ and\ \bibinfo {author} {\bibfnamefont {V.}~\bibnamefont {Cardoso}},\ }\href {\doibase 10.1103/PhysRevD.88.044018} {\bibfield  {journal} {\bibinfo  {journal} {Phys. Rev. D}\ }\textbf {\bibinfo {volume} {88}},\ \bibinfo {pages} {044018} (\bibinfo {year} {2013})},\ \Eprint {http://arxiv.org/abs/1305.4306} {arXiv:1305.4306 [gr-qc]} \BibitemShut {NoStop}%
\bibitem [{\citenamefont {K\"uchler}\ \emph {et~al.}(2025)\citenamefont {K\"uchler}, \citenamefont {Comp\`ere},\ and\ \citenamefont {Pound}}]{Kuchler:2025hwx}%
  \BibitemOpen
  \bibfield  {author} {\bibinfo {author} {\bibfnamefont {L.}~\bibnamefont {K\"uchler}}, \bibinfo {author} {\bibfnamefont {G.}~\bibnamefont {Comp\`ere}}, \ and\ \bibinfo {author} {\bibfnamefont {A.}~\bibnamefont {Pound}},\ }\href@noop {} {\  (\bibinfo {year} {2025})},\ \Eprint {http://arxiv.org/abs/2506.02189} {arXiv:2506.02189 [gr-qc]} \BibitemShut {NoStop}%
\bibitem [{\citenamefont {Maggio}\ \emph {et~al.}(2023)\citenamefont {Maggio}, \citenamefont {Silva}, \citenamefont {Buonanno},\ and\ \citenamefont {Ghosh}}]{Maggio:2022hre}%
  \BibitemOpen
  \bibfield  {author} {\bibinfo {author} {\bibfnamefont {E.}~\bibnamefont {Maggio}}, \bibinfo {author} {\bibfnamefont {H.~O.}\ \bibnamefont {Silva}}, \bibinfo {author} {\bibfnamefont {A.}~\bibnamefont {Buonanno}}, \ and\ \bibinfo {author} {\bibfnamefont {A.}~\bibnamefont {Ghosh}},\ }\href {\doibase 10.1103/PhysRevD.108.024043} {\bibfield  {journal} {\bibinfo  {journal} {Phys. Rev. D}\ }\textbf {\bibinfo {volume} {108}},\ \bibinfo {pages} {024043} (\bibinfo {year} {2023})},\ \Eprint {http://arxiv.org/abs/2212.09655} {arXiv:2212.09655 [gr-qc]} \BibitemShut {NoStop}%
\bibitem [{\citenamefont {Forteza}\ \emph {et~al.}(2023)\citenamefont {Forteza}, \citenamefont {Bhagwat}, \citenamefont {Kumar},\ and\ \citenamefont {Pani}}]{Forteza:2023abc}%
  \BibitemOpen
  \bibfield  {author} {\bibinfo {author} {\bibfnamefont {X.~J.}\ \bibnamefont {Forteza}}, \bibinfo {author} {\bibfnamefont {S.}~\bibnamefont {Bhagwat}}, \bibinfo {author} {\bibfnamefont {S.}~\bibnamefont {Kumar}}, \ and\ \bibinfo {author} {\bibfnamefont {P.}~\bibnamefont {Pani}},\ }\href {\doibase 10.1103/PhysRevLett.130.021001} {\bibfield  {journal} {\bibinfo  {journal} {Phys. Rev. Lett.}\ }\textbf {\bibinfo {volume} {130}},\ \bibinfo {pages} {021001} (\bibinfo {year} {2023})},\ \Eprint {http://arxiv.org/abs/2205.14910} {arXiv:2205.14910 [gr-qc]} \BibitemShut {NoStop}%
\bibitem [{\citenamefont {Toubiana}\ \emph {et~al.}(2024)\citenamefont {Toubiana}, \citenamefont {Pompili}, \citenamefont {Buonanno}, \citenamefont {Gair},\ and\ \citenamefont {Katz}}]{Toubiana:2023cwr}%
  \BibitemOpen
  \bibfield  {author} {\bibinfo {author} {\bibfnamefont {A.}~\bibnamefont {Toubiana}}, \bibinfo {author} {\bibfnamefont {L.}~\bibnamefont {Pompili}}, \bibinfo {author} {\bibfnamefont {A.}~\bibnamefont {Buonanno}}, \bibinfo {author} {\bibfnamefont {J.~R.}\ \bibnamefont {Gair}}, \ and\ \bibinfo {author} {\bibfnamefont {M.~L.}\ \bibnamefont {Katz}},\ }\href {\doibase 10.1103/PhysRevD.109.104019} {\bibfield  {journal} {\bibinfo  {journal} {Phys. Rev. D}\ }\textbf {\bibinfo {volume} {109}},\ \bibinfo {pages} {104019} (\bibinfo {year} {2024})},\ \Eprint {http://arxiv.org/abs/2307.15086} {arXiv:2307.15086 [gr-qc]} \BibitemShut {NoStop}%
\bibitem [{\citenamefont {Toubiana}\ and\ \citenamefont {Gair}(2024)}]{Toubiana:2024abc}%
  \BibitemOpen
  \bibfield  {author} {\bibinfo {author} {\bibfnamefont {A.}~\bibnamefont {Toubiana}}\ and\ \bibinfo {author} {\bibfnamefont {J.~R.}\ \bibnamefont {Gair}},\ }\href@noop {} {\  (\bibinfo {year} {2024})},\ \Eprint {http://arxiv.org/abs/2401.06845} {arXiv:2401.06845 [gr-qc]} \BibitemShut {NoStop}%
\bibitem [{\citenamefont {Gupta}\ \emph {et~al.}(2024)\citenamefont {Gupta} \emph {et~al.}}]{Gupta:2024def}%
  \BibitemOpen
  \bibfield  {author} {\bibinfo {author} {\bibfnamefont {A.}~\bibnamefont {Gupta}} \emph {et~al.},\ }\href@noop {} {\  (\bibinfo {year} {2024})},\ \Eprint {http://arxiv.org/abs/2405.02197} {arXiv:2405.02197 [gr-qc]} \BibitemShut {NoStop}%
\bibitem [{\citenamefont {Yi}\ \emph {et~al.}(2025)\citenamefont {Yi}, \citenamefont {Iacovelli}, \citenamefont {Marsat}, \citenamefont {Wadekar},\ and\ \citenamefont {Berti}}]{Yi:2025pxe}%
  \BibitemOpen
  \bibfield  {author} {\bibinfo {author} {\bibfnamefont {S.}~\bibnamefont {Yi}}, \bibinfo {author} {\bibfnamefont {F.}~\bibnamefont {Iacovelli}}, \bibinfo {author} {\bibfnamefont {S.}~\bibnamefont {Marsat}}, \bibinfo {author} {\bibfnamefont {D.}~\bibnamefont {Wadekar}}, \ and\ \bibinfo {author} {\bibfnamefont {E.}~\bibnamefont {Berti}},\ }\href@noop {} {\  (\bibinfo {year} {2025})},\ \Eprint {http://arxiv.org/abs/2502.12237} {arXiv:2502.12237 [gr-qc]} \BibitemShut {NoStop}%
\bibitem [{\citenamefont {Pompili}\ \emph {et~al.}(2025)\citenamefont {Pompili}, \citenamefont {Maggio}, \citenamefont {Silva},\ and\ \citenamefont {Buonanno}}]{Pompili:2025cdc}%
  \BibitemOpen
  \bibfield  {author} {\bibinfo {author} {\bibfnamefont {L.}~\bibnamefont {Pompili}}, \bibinfo {author} {\bibfnamefont {E.}~\bibnamefont {Maggio}}, \bibinfo {author} {\bibfnamefont {H.~O.}\ \bibnamefont {Silva}}, \ and\ \bibinfo {author} {\bibfnamefont {A.}~\bibnamefont {Buonanno}},\ }\href@noop {} {\  (\bibinfo {year} {2025})},\ \Eprint {http://arxiv.org/abs/2504.10130} {arXiv:2504.10130 [gr-qc]} \BibitemShut {NoStop}%
\bibitem [{\citenamefont {Hu}\ \emph {et~al.}(2025)\citenamefont {Hu}, \citenamefont {Doneva}, \citenamefont {Wang}, \citenamefont {Paschalidis}, \citenamefont {Bozzola}, \citenamefont {Yazadjiev},\ and\ \citenamefont {Shao}}]{Hu:2025sea}%
  \BibitemOpen
  \bibfield  {author} {\bibinfo {author} {\bibfnamefont {Z.}~\bibnamefont {Hu}}, \bibinfo {author} {\bibfnamefont {D.~D.}\ \bibnamefont {Doneva}}, \bibinfo {author} {\bibfnamefont {Z.}~\bibnamefont {Wang}}, \bibinfo {author} {\bibfnamefont {V.}~\bibnamefont {Paschalidis}}, \bibinfo {author} {\bibfnamefont {G.}~\bibnamefont {Bozzola}}, \bibinfo {author} {\bibfnamefont {S.~S.}\ \bibnamefont {Yazadjiev}}, \ and\ \bibinfo {author} {\bibfnamefont {L.}~\bibnamefont {Shao}},\ }\href@noop {} {\  (\bibinfo {year} {2025})},\ \Eprint {http://arxiv.org/abs/2509.07111} {arXiv:2509.07111 [gr-qc]} \BibitemShut {NoStop}%
\bibitem [{\citenamefont {Colpi}\ \emph {et~al.}(2024)\citenamefont {Colpi} \emph {et~al.}}]{LISA:2024hlh}%
  \BibitemOpen
  \bibfield  {author} {\bibinfo {author} {\bibfnamefont {M.}~\bibnamefont {Colpi}} \emph {et~al.} (\bibinfo {collaboration} {LISA}),\ }\href@noop {} {\  (\bibinfo {year} {2024})},\ \Eprint {http://arxiv.org/abs/2402.07571} {arXiv:2402.07571 [astro-ph.CO]} \BibitemShut {NoStop}%
\bibitem [{\citenamefont {Abac}\ \emph {et~al.}(2025{\natexlab{c}})\citenamefont {Abac} \emph {et~al.}}]{Abac:2025saz}%
  \BibitemOpen
  \bibfield  {author} {\bibinfo {author} {\bibfnamefont {A.}~\bibnamefont {Abac}} \emph {et~al.},\ }\href@noop {} {\  (\bibinfo {year} {2025}{\natexlab{c}})},\ \Eprint {http://arxiv.org/abs/2503.12263} {arXiv:2503.12263 [gr-qc]} \BibitemShut {NoStop}%
\bibitem [{\citenamefont {Evans}\ \emph {et~al.}(2023)\citenamefont {Evans} \emph {et~al.}}]{Evans:2023euw}%
  \BibitemOpen
  \bibfield  {author} {\bibinfo {author} {\bibfnamefont {M.}~\bibnamefont {Evans}} \emph {et~al.},\ }\href@noop {} {\  (\bibinfo {year} {2023})},\ \Eprint {http://arxiv.org/abs/2306.13745} {arXiv:2306.13745 [astro-ph.IM]} \BibitemShut {NoStop}%
\bibitem [{\citenamefont {London}\ and\ \citenamefont {Fauchon-Jones}(2019)}]{London:2018nxs}%
  \BibitemOpen
  \bibfield  {author} {\bibinfo {author} {\bibfnamefont {L.}~\bibnamefont {London}}\ and\ \bibinfo {author} {\bibfnamefont {E.}~\bibnamefont {Fauchon-Jones}},\ }\href {\doibase 10.1088/1361-6382/ab2f11} {\bibfield  {journal} {\bibinfo  {journal} {Class. Quant. Grav.}\ }\textbf {\bibinfo {volume} {36}},\ \bibinfo {pages} {235015} (\bibinfo {year} {2019})},\ \Eprint {http://arxiv.org/abs/1810.03550} {arXiv:1810.03550 [gr-qc]} \BibitemShut {NoStop}%
\bibitem [{\citenamefont {Cheung}\ \emph {et~al.}(2024)\citenamefont {Cheung}, \citenamefont {Berti}, \citenamefont {Baibhav},\ and\ \citenamefont {Cotesta}}]{Cheung:2023vki}%
  \BibitemOpen
  \bibfield  {author} {\bibinfo {author} {\bibfnamefont {M.~H.-Y.}\ \bibnamefont {Cheung}}, \bibinfo {author} {\bibfnamefont {E.}~\bibnamefont {Berti}}, \bibinfo {author} {\bibfnamefont {V.}~\bibnamefont {Baibhav}}, \ and\ \bibinfo {author} {\bibfnamefont {R.}~\bibnamefont {Cotesta}},\ }\href {\doibase 10.1103/PhysRevD.109.044069} {\bibfield  {journal} {\bibinfo  {journal} {Phys. Rev. D}\ }\textbf {\bibinfo {volume} {109}},\ \bibinfo {pages} {044069} (\bibinfo {year} {2024})},\ \bibinfo {note} {[Erratum: Phys.Rev.D 110, 049902 (2024)]},\ \Eprint {http://arxiv.org/abs/2310.04489} {arXiv:2310.04489 [gr-qc]} \BibitemShut {NoStop}%
\bibitem [{\citenamefont {Gleiser}\ \emph {et~al.}(1996)\citenamefont {Gleiser}, \citenamefont {Nicasio}, \citenamefont {Price},\ and\ \citenamefont {Pullin}}]{Gleiser:1996yc}%
  \BibitemOpen
  \bibfield  {author} {\bibinfo {author} {\bibfnamefont {R.~J.}\ \bibnamefont {Gleiser}}, \bibinfo {author} {\bibfnamefont {C.~O.}\ \bibnamefont {Nicasio}}, \bibinfo {author} {\bibfnamefont {R.~H.}\ \bibnamefont {Price}}, \ and\ \bibinfo {author} {\bibfnamefont {J.}~\bibnamefont {Pullin}},\ }\href {\doibase 10.1103/PhysRevLett.77.4483} {\bibfield  {journal} {\bibinfo  {journal} {Phys. Rev. Lett.}\ }\textbf {\bibinfo {volume} {77}},\ \bibinfo {pages} {4483} (\bibinfo {year} {1996})},\ \Eprint {http://arxiv.org/abs/gr-qc/9609022} {arXiv:gr-qc/9609022} \BibitemShut {NoStop}%
\bibitem [{\citenamefont {London}\ \emph {et~al.}(2014)\citenamefont {London}, \citenamefont {Shoemaker},\ and\ \citenamefont {Healy}}]{London:2014cma}%
  \BibitemOpen
  \bibfield  {author} {\bibinfo {author} {\bibfnamefont {L.}~\bibnamefont {London}}, \bibinfo {author} {\bibfnamefont {D.}~\bibnamefont {Shoemaker}}, \ and\ \bibinfo {author} {\bibfnamefont {J.}~\bibnamefont {Healy}},\ }\href {\doibase 10.1103/PhysRevD.90.124032} {\bibfield  {journal} {\bibinfo  {journal} {Phys. Rev. D}\ }\textbf {\bibinfo {volume} {90}},\ \bibinfo {pages} {124032} (\bibinfo {year} {2014})},\ \bibinfo {note} {[Erratum: Phys.Rev.D 94, 069902 (2016)]},\ \Eprint {http://arxiv.org/abs/1404.3197} {arXiv:1404.3197 [gr-qc]} \BibitemShut {NoStop}%
\bibitem [{\citenamefont {Cheung}\ \emph {et~al.}(2023)\citenamefont {Cheung} \emph {et~al.}}]{Cheung:2022rbm}%
  \BibitemOpen
  \bibfield  {author} {\bibinfo {author} {\bibfnamefont {M.~H.-Y.}\ \bibnamefont {Cheung}} \emph {et~al.},\ }\href {\doibase 10.1103/PhysRevLett.130.081401} {\bibfield  {journal} {\bibinfo  {journal} {Phys. Rev. Lett.}\ }\textbf {\bibinfo {volume} {130}},\ \bibinfo {pages} {081401} (\bibinfo {year} {2023})},\ \Eprint {http://arxiv.org/abs/2208.07374} {arXiv:2208.07374 [gr-qc]} \BibitemShut {NoStop}%
\bibitem [{\citenamefont {Mitman}\ \emph {et~al.}(2023)\citenamefont {Mitman} \emph {et~al.}}]{Mitman:2022qdl}%
  \BibitemOpen
  \bibfield  {author} {\bibinfo {author} {\bibfnamefont {K.}~\bibnamefont {Mitman}} \emph {et~al.},\ }\href {\doibase 10.1103/PhysRevLett.130.081402} {\bibfield  {journal} {\bibinfo  {journal} {Phys. Rev. Lett.}\ }\textbf {\bibinfo {volume} {130}},\ \bibinfo {pages} {081402} (\bibinfo {year} {2023})},\ \Eprint {http://arxiv.org/abs/2208.07380} {arXiv:2208.07380 [gr-qc]} \BibitemShut {NoStop}%
\bibitem [{\citenamefont {Baibhav}\ \emph {et~al.}(2023)\citenamefont {Baibhav}, \citenamefont {Cheung}, \citenamefont {Berti}, \citenamefont {Cardoso}, \citenamefont {Carullo}, \citenamefont {Cotesta}, \citenamefont {Del~Pozzo},\ and\ \citenamefont {Duque}}]{Baibhav:2023clw}%
  \BibitemOpen
  \bibfield  {author} {\bibinfo {author} {\bibfnamefont {V.}~\bibnamefont {Baibhav}}, \bibinfo {author} {\bibfnamefont {M.~H.-Y.}\ \bibnamefont {Cheung}}, \bibinfo {author} {\bibfnamefont {E.}~\bibnamefont {Berti}}, \bibinfo {author} {\bibfnamefont {V.}~\bibnamefont {Cardoso}}, \bibinfo {author} {\bibfnamefont {G.}~\bibnamefont {Carullo}}, \bibinfo {author} {\bibfnamefont {R.}~\bibnamefont {Cotesta}}, \bibinfo {author} {\bibfnamefont {W.}~\bibnamefont {Del~Pozzo}}, \ and\ \bibinfo {author} {\bibfnamefont {F.}~\bibnamefont {Duque}},\ }\href {\doibase 10.1103/PhysRevD.108.104020} {\bibfield  {journal} {\bibinfo  {journal} {Phys. Rev. D}\ }\textbf {\bibinfo {volume} {108}},\ \bibinfo {pages} {104020} (\bibinfo {year} {2023})},\ \Eprint {http://arxiv.org/abs/2302.03050} {arXiv:2302.03050 [gr-qc]} \BibitemShut {NoStop}%
\bibitem [{\citenamefont {Bucciotti}\ \emph {et~al.}(2023)\citenamefont {Bucciotti}, \citenamefont {Kuntz}, \citenamefont {Serra},\ and\ \citenamefont {Trincherini}}]{Bucciotti:2023ets}%
  \BibitemOpen
  \bibfield  {author} {\bibinfo {author} {\bibfnamefont {B.}~\bibnamefont {Bucciotti}}, \bibinfo {author} {\bibfnamefont {A.}~\bibnamefont {Kuntz}}, \bibinfo {author} {\bibfnamefont {F.}~\bibnamefont {Serra}}, \ and\ \bibinfo {author} {\bibfnamefont {E.}~\bibnamefont {Trincherini}},\ }\href {\doibase 10.1007/JHEP12(2023)048} {\bibfield  {journal} {\bibinfo  {journal} {JHEP}\ }\textbf {\bibinfo {volume} {12}},\ \bibinfo {pages} {048} (\bibinfo {year} {2023})},\ \Eprint {http://arxiv.org/abs/2309.08501} {arXiv:2309.08501 [hep-th]} \BibitemShut {NoStop}%
\bibitem [{\citenamefont {Perrone}\ \emph {et~al.}(2024)\citenamefont {Perrone}, \citenamefont {Barreira}, \citenamefont {Kehagias},\ and\ \citenamefont {Riotto}}]{Perrone:2023jzq}%
  \BibitemOpen
  \bibfield  {author} {\bibinfo {author} {\bibfnamefont {D.}~\bibnamefont {Perrone}}, \bibinfo {author} {\bibfnamefont {T.}~\bibnamefont {Barreira}}, \bibinfo {author} {\bibfnamefont {A.}~\bibnamefont {Kehagias}}, \ and\ \bibinfo {author} {\bibfnamefont {A.}~\bibnamefont {Riotto}},\ }\href {\doibase 10.1016/j.nuclphysb.2023.116432} {\bibfield  {journal} {\bibinfo  {journal} {Nucl. Phys. B}\ }\textbf {\bibinfo {volume} {999}},\ \bibinfo {pages} {116432} (\bibinfo {year} {2024})},\ \Eprint {http://arxiv.org/abs/2308.15886} {arXiv:2308.15886 [gr-qc]} \BibitemShut {NoStop}%
\bibitem [{\citenamefont {Redondo-Yuste}\ \emph {et~al.}(2024{\natexlab{a}})\citenamefont {Redondo-Yuste}, \citenamefont {Carullo}, \citenamefont {Ripley}, \citenamefont {Berti},\ and\ \citenamefont {Cardoso}}]{Redondo-Yuste:2023seq}%
  \BibitemOpen
  \bibfield  {author} {\bibinfo {author} {\bibfnamefont {J.}~\bibnamefont {Redondo-Yuste}}, \bibinfo {author} {\bibfnamefont {G.}~\bibnamefont {Carullo}}, \bibinfo {author} {\bibfnamefont {J.~L.}\ \bibnamefont {Ripley}}, \bibinfo {author} {\bibfnamefont {E.}~\bibnamefont {Berti}}, \ and\ \bibinfo {author} {\bibfnamefont {V.}~\bibnamefont {Cardoso}},\ }\href {\doibase 10.1103/PhysRevD.109.L101503} {\bibfield  {journal} {\bibinfo  {journal} {Phys. Rev. D}\ }\textbf {\bibinfo {volume} {109}},\ \bibinfo {pages} {L101503} (\bibinfo {year} {2024}{\natexlab{a}})},\ \Eprint {http://arxiv.org/abs/2308.14796} {arXiv:2308.14796 [gr-qc]} \BibitemShut {NoStop}%
\bibitem [{\citenamefont {Ma}\ and\ \citenamefont {Yang}(2024)}]{Ma:2024qcv}%
  \BibitemOpen
  \bibfield  {author} {\bibinfo {author} {\bibfnamefont {S.}~\bibnamefont {Ma}}\ and\ \bibinfo {author} {\bibfnamefont {H.}~\bibnamefont {Yang}},\ }\href {\doibase 10.1103/PhysRevD.109.104070} {\bibfield  {journal} {\bibinfo  {journal} {Phys. Rev. D}\ }\textbf {\bibinfo {volume} {109}},\ \bibinfo {pages} {104070} (\bibinfo {year} {2024})},\ \Eprint {http://arxiv.org/abs/2401.15516} {arXiv:2401.15516 [gr-qc]} \BibitemShut {NoStop}%
\bibitem [{\citenamefont {Bucciotti}\ \emph {et~al.}(2024)\citenamefont {Bucciotti}, \citenamefont {Juliano}, \citenamefont {Kuntz},\ and\ \citenamefont {Trincherini}}]{Bucciotti:2024zyp}%
  \BibitemOpen
  \bibfield  {author} {\bibinfo {author} {\bibfnamefont {B.}~\bibnamefont {Bucciotti}}, \bibinfo {author} {\bibfnamefont {L.}~\bibnamefont {Juliano}}, \bibinfo {author} {\bibfnamefont {A.}~\bibnamefont {Kuntz}}, \ and\ \bibinfo {author} {\bibfnamefont {E.}~\bibnamefont {Trincherini}},\ }\href {\doibase 10.1103/PhysRevD.110.104048} {\bibfield  {journal} {\bibinfo  {journal} {Phys. Rev. D}\ }\textbf {\bibinfo {volume} {110}},\ \bibinfo {pages} {104048} (\bibinfo {year} {2024})},\ \Eprint {http://arxiv.org/abs/2405.06012} {arXiv:2405.06012 [gr-qc]} \BibitemShut {NoStop}%
\bibitem [{\citenamefont {Bourg}\ \emph {et~al.}(2025)\citenamefont {Bourg}, \citenamefont {Panosso~Macedo}, \citenamefont {Spiers}, \citenamefont {Leather}, \citenamefont {Bonga},\ and\ \citenamefont {Pound}}]{Bourg:2024jme}%
  \BibitemOpen
  \bibfield  {author} {\bibinfo {author} {\bibfnamefont {P.}~\bibnamefont {Bourg}}, \bibinfo {author} {\bibfnamefont {R.}~\bibnamefont {Panosso~Macedo}}, \bibinfo {author} {\bibfnamefont {A.}~\bibnamefont {Spiers}}, \bibinfo {author} {\bibfnamefont {B.}~\bibnamefont {Leather}}, \bibinfo {author} {\bibfnamefont {B.}~\bibnamefont {Bonga}}, \ and\ \bibinfo {author} {\bibfnamefont {A.}~\bibnamefont {Pound}},\ }\href {\doibase 10.1103/PhysRevLett.134.061401} {\bibfield  {journal} {\bibinfo  {journal} {Phys. Rev. Lett.}\ }\textbf {\bibinfo {volume} {134}},\ \bibinfo {pages} {061401} (\bibinfo {year} {2025})},\ \Eprint {http://arxiv.org/abs/2405.10270} {arXiv:2405.10270 [gr-qc]} \BibitemShut {NoStop}%
\bibitem [{\citenamefont {Zhu}\ \emph {et~al.}(2024)\citenamefont {Zhu} \emph {et~al.}}]{Zhu:2024rej}%
  \BibitemOpen
  \bibfield  {author} {\bibinfo {author} {\bibfnamefont {H.}~\bibnamefont {Zhu}} \emph {et~al.},\ }\href {\doibase 10.1103/PhysRevD.109.104050} {\bibfield  {journal} {\bibinfo  {journal} {Phys. Rev. D}\ }\textbf {\bibinfo {volume} {109}},\ \bibinfo {pages} {104050} (\bibinfo {year} {2024})},\ \Eprint {http://arxiv.org/abs/2401.00805} {arXiv:2401.00805 [gr-qc]} \BibitemShut {NoStop}%
\bibitem [{\citenamefont {Nobili}\ \emph {et~al.}(2025)\citenamefont {Nobili}, \citenamefont {Bhagwat}, \citenamefont {Pacilio},\ and\ \citenamefont {Gerosa}}]{Nobili:2025ydt}%
  \BibitemOpen
  \bibfield  {author} {\bibinfo {author} {\bibfnamefont {F.}~\bibnamefont {Nobili}}, \bibinfo {author} {\bibfnamefont {S.}~\bibnamefont {Bhagwat}}, \bibinfo {author} {\bibfnamefont {C.}~\bibnamefont {Pacilio}}, \ and\ \bibinfo {author} {\bibfnamefont {D.}~\bibnamefont {Gerosa}},\ }\href@noop {} {\  (\bibinfo {year} {2025})},\ \Eprint {http://arxiv.org/abs/2504.17021} {arXiv:2504.17021 [gr-qc]} \BibitemShut {NoStop}%
\bibitem [{\citenamefont {Carullo}(2024)}]{Carullo:2024smg}%
  \BibitemOpen
  \bibfield  {author} {\bibinfo {author} {\bibfnamefont {G.}~\bibnamefont {Carullo}},\ }\href {\doibase 10.1088/1475-7516/2024/10/061} {\bibfield  {journal} {\bibinfo  {journal} {JCAP}\ }\textbf {\bibinfo {volume} {10}},\ \bibinfo {pages} {061} (\bibinfo {year} {2024})},\ \Eprint {http://arxiv.org/abs/2406.19442} {arXiv:2406.19442 [gr-qc]} \BibitemShut {NoStop}%
\bibitem [{\citenamefont {Price}\ \emph {et~al.}(2016)\citenamefont {Price}, \citenamefont {Nampalliwar},\ and\ \citenamefont {Khanna}}]{Price:2015gia}%
  \BibitemOpen
  \bibfield  {author} {\bibinfo {author} {\bibfnamefont {R.~H.}\ \bibnamefont {Price}}, \bibinfo {author} {\bibfnamefont {S.}~\bibnamefont {Nampalliwar}}, \ and\ \bibinfo {author} {\bibfnamefont {G.}~\bibnamefont {Khanna}},\ }\href {\doibase 10.1103/PhysRevD.93.044060} {\bibfield  {journal} {\bibinfo  {journal} {Phys. Rev. D}\ }\textbf {\bibinfo {volume} {93}},\ \bibinfo {pages} {044060} (\bibinfo {year} {2016})},\ \Eprint {http://arxiv.org/abs/1508.04797} {arXiv:1508.04797 [gr-qc]} \BibitemShut {NoStop}%
\bibitem [{\citenamefont {Lagos}\ and\ \citenamefont {Hui}(2023)}]{Lagos:2022otp}%
  \BibitemOpen
  \bibfield  {author} {\bibinfo {author} {\bibfnamefont {M.}~\bibnamefont {Lagos}}\ and\ \bibinfo {author} {\bibfnamefont {L.}~\bibnamefont {Hui}},\ }\href {\doibase 10.1103/PhysRevD.107.044040} {\bibfield  {journal} {\bibinfo  {journal} {Phys. Rev. D}\ }\textbf {\bibinfo {volume} {107}},\ \bibinfo {pages} {044040} (\bibinfo {year} {2023})},\ \Eprint {http://arxiv.org/abs/2208.07379} {arXiv:2208.07379 [gr-qc]} \BibitemShut {NoStop}%
\bibitem [{\citenamefont {Oshita}\ \emph {et~al.}(2025)\citenamefont {Oshita}, \citenamefont {Ma}, \citenamefont {Chen},\ and\ \citenamefont {Yang}}]{Oshita:2025qmn}%
  \BibitemOpen
  \bibfield  {author} {\bibinfo {author} {\bibfnamefont {N.}~\bibnamefont {Oshita}}, \bibinfo {author} {\bibfnamefont {S.}~\bibnamefont {Ma}}, \bibinfo {author} {\bibfnamefont {Y.}~\bibnamefont {Chen}}, \ and\ \bibinfo {author} {\bibfnamefont {H.}~\bibnamefont {Yang}},\ }\href@noop {} {\  (\bibinfo {year} {2025})},\ \Eprint {http://arxiv.org/abs/2509.09165} {arXiv:2509.09165 [gr-qc]} \BibitemShut {NoStop}%
\bibitem [{\citenamefont {Lu}\ \emph {et~al.}(2025)\citenamefont {Lu}, \citenamefont {Ma}, \citenamefont {Piccinni}, \citenamefont {Chen},\ and\ \citenamefont {Sun}}]{Lu:2025vol}%
  \BibitemOpen
  \bibfield  {author} {\bibinfo {author} {\bibfnamefont {N.}~\bibnamefont {Lu}}, \bibinfo {author} {\bibfnamefont {S.}~\bibnamefont {Ma}}, \bibinfo {author} {\bibfnamefont {O.~J.}\ \bibnamefont {Piccinni}}, \bibinfo {author} {\bibfnamefont {Y.}~\bibnamefont {Chen}}, \ and\ \bibinfo {author} {\bibfnamefont {L.}~\bibnamefont {Sun}},\ }\href@noop {} {\  (\bibinfo {year} {2025})},\ \Eprint {http://arxiv.org/abs/2510.01001} {arXiv:2510.01001 [gr-qc]} \BibitemShut {NoStop}%
\bibitem [{\citenamefont {Sberna}\ \emph {et~al.}(2022)\citenamefont {Sberna}, \citenamefont {Bosch}, \citenamefont {East}, \citenamefont {Green},\ and\ \citenamefont {Lehner}}]{Sberna:2021eui}%
  \BibitemOpen
  \bibfield  {author} {\bibinfo {author} {\bibfnamefont {L.}~\bibnamefont {Sberna}}, \bibinfo {author} {\bibfnamefont {P.}~\bibnamefont {Bosch}}, \bibinfo {author} {\bibfnamefont {W.~E.}\ \bibnamefont {East}}, \bibinfo {author} {\bibfnamefont {S.~R.}\ \bibnamefont {Green}}, \ and\ \bibinfo {author} {\bibfnamefont {L.}~\bibnamefont {Lehner}},\ }\href {\doibase 10.1103/PhysRevD.105.064046} {\bibfield  {journal} {\bibinfo  {journal} {Phys. Rev. D}\ }\textbf {\bibinfo {volume} {105}},\ \bibinfo {pages} {064046} (\bibinfo {year} {2022})},\ \Eprint {http://arxiv.org/abs/2112.11168} {arXiv:2112.11168 [gr-qc]} \BibitemShut {NoStop}%
\bibitem [{\citenamefont {Redondo-Yuste}\ \emph {et~al.}(2024{\natexlab{b}})\citenamefont {Redondo-Yuste}, \citenamefont {Pere{\~n}iguez},\ and\ \citenamefont {Cardoso}}]{Redondo-Yuste:2023ipg}%
  \BibitemOpen
  \bibfield  {author} {\bibinfo {author} {\bibfnamefont {J.}~\bibnamefont {Redondo-Yuste}}, \bibinfo {author} {\bibfnamefont {D.}~\bibnamefont {Pere{\~n}iguez}}, \ and\ \bibinfo {author} {\bibfnamefont {V.}~\bibnamefont {Cardoso}},\ }\href {\doibase 10.1103/PhysRevD.109.044048} {\bibfield  {journal} {\bibinfo  {journal} {Phys. Rev. D}\ }\textbf {\bibinfo {volume} {109}},\ \bibinfo {pages} {044048} (\bibinfo {year} {2024}{\natexlab{b}})},\ \Eprint {http://arxiv.org/abs/2312.04633} {arXiv:2312.04633 [gr-qc]} \BibitemShut {NoStop}%
\bibitem [{\citenamefont {May}\ \emph {et~al.}(2024)\citenamefont {May}, \citenamefont {Ma}, \citenamefont {Ripley},\ and\ \citenamefont {East}}]{May:2024rrg}%
  \BibitemOpen
  \bibfield  {author} {\bibinfo {author} {\bibfnamefont {T.}~\bibnamefont {May}}, \bibinfo {author} {\bibfnamefont {S.}~\bibnamefont {Ma}}, \bibinfo {author} {\bibfnamefont {J.~L.}\ \bibnamefont {Ripley}}, \ and\ \bibinfo {author} {\bibfnamefont {W.~E.}\ \bibnamefont {East}},\ }\href {\doibase 10.1103/PhysRevD.110.084034} {\bibfield  {journal} {\bibinfo  {journal} {Phys. Rev. D}\ }\textbf {\bibinfo {volume} {110}},\ \bibinfo {pages} {084034} (\bibinfo {year} {2024})},\ \Eprint {http://arxiv.org/abs/2405.18303} {arXiv:2405.18303 [gr-qc]} \BibitemShut {NoStop}%
\bibitem [{\citenamefont {Capuano}\ \emph {et~al.}(2024)\citenamefont {Capuano}, \citenamefont {Santoni},\ and\ \citenamefont {Barausse}}]{Capuano:2024qhv}%
  \BibitemOpen
  \bibfield  {author} {\bibinfo {author} {\bibfnamefont {L.}~\bibnamefont {Capuano}}, \bibinfo {author} {\bibfnamefont {L.}~\bibnamefont {Santoni}}, \ and\ \bibinfo {author} {\bibfnamefont {E.}~\bibnamefont {Barausse}},\ }\href {\doibase 10.1103/PhysRevD.110.084081} {\bibfield  {journal} {\bibinfo  {journal} {Phys. Rev. D}\ }\textbf {\bibinfo {volume} {110}},\ \bibinfo {pages} {084081} (\bibinfo {year} {2024})},\ \Eprint {http://arxiv.org/abs/2407.06009} {arXiv:2407.06009 [gr-qc]} \BibitemShut {NoStop}%
\bibitem [{\citenamefont {Buonanno}\ and\ \citenamefont {Damour}(1999)}]{Buonanno:1999cu}%
  \BibitemOpen
  \bibfield  {author} {\bibinfo {author} {\bibfnamefont {A.}~\bibnamefont {Buonanno}}\ and\ \bibinfo {author} {\bibfnamefont {T.}~\bibnamefont {Damour}},\ }\href {\doibase 10.1103/PhysRevD.59.084006} {\bibfield  {journal} {\bibinfo  {journal} {Phys. Rev. D}\ }\textbf {\bibinfo {volume} {59}},\ \bibinfo {pages} {084006} (\bibinfo {year} {1999})},\ \Eprint {http://arxiv.org/abs/gr-qc/9811091} {arXiv:gr-qc/9811091 [gr-qc]} \BibitemShut {NoStop}%
\bibitem [{\citenamefont {Buonanno}\ and\ \citenamefont {Damour}(2000)}]{Buonanno:2000ef}%
  \BibitemOpen
  \bibfield  {author} {\bibinfo {author} {\bibfnamefont {A.}~\bibnamefont {Buonanno}}\ and\ \bibinfo {author} {\bibfnamefont {T.}~\bibnamefont {Damour}},\ }\href {\doibase 10.1103/PhysRevD.62.064015} {\bibfield  {journal} {\bibinfo  {journal} {Phys. Rev. D}\ }\textbf {\bibinfo {volume} {62}},\ \bibinfo {pages} {064015} (\bibinfo {year} {2000})},\ \Eprint {http://arxiv.org/abs/gr-qc/0001013} {arXiv:gr-qc/0001013 [gr-qc]} \BibitemShut {NoStop}%
\bibitem [{\citenamefont {Damour}(2001)}]{Damour:2001tt}%
  \BibitemOpen
  \bibfield  {author} {\bibinfo {author} {\bibfnamefont {T.}~\bibnamefont {Damour}},\ }\href {\doibase 10.1103/PhysRevD.64.124013} {\bibfield  {journal} {\bibinfo  {journal} {Phys. Rev. D}\ }\textbf {\bibinfo {volume} {64}},\ \bibinfo {pages} {124013} (\bibinfo {year} {2001})},\ \Eprint {http://arxiv.org/abs/gr-qc/0103018} {arXiv:gr-qc/0103018 [gr-qc]} \BibitemShut {NoStop}%
\bibitem [{\citenamefont {Damour}\ and\ \citenamefont {Nagar}(2014)}]{Damour:2014yha}%
  \BibitemOpen
  \bibfield  {author} {\bibinfo {author} {\bibfnamefont {T.}~\bibnamefont {Damour}}\ and\ \bibinfo {author} {\bibfnamefont {A.}~\bibnamefont {Nagar}},\ }\href {\doibase 10.1103/PhysRevD.90.024054} {\bibfield  {journal} {\bibinfo  {journal} {Phys. Rev. D}\ }\textbf {\bibinfo {volume} {90}},\ \bibinfo {pages} {024054} (\bibinfo {year} {2014})},\ \Eprint {http://arxiv.org/abs/1406.0401} {arXiv:1406.0401 [gr-qc]} \BibitemShut {NoStop}%
\bibitem [{\citenamefont {Del~Pozzo}\ and\ \citenamefont {Nagar}(2017)}]{DelPozzo:2017rka}%
  \BibitemOpen
  \bibfield  {author} {\bibinfo {author} {\bibfnamefont {W.}~\bibnamefont {Del~Pozzo}}\ and\ \bibinfo {author} {\bibfnamefont {A.}~\bibnamefont {Nagar}},\ }\href {\doibase 10.1103/PhysRevD.95.124034} {\bibfield  {journal} {\bibinfo  {journal} {Phys. Rev. D}\ }\textbf {\bibinfo {volume} {95}},\ \bibinfo {pages} {124034} (\bibinfo {year} {2017})},\ \Eprint {http://arxiv.org/abs/1606.03952} {arXiv:1606.03952 [gr-qc]} \BibitemShut {NoStop}%
\bibitem [{\citenamefont {Bohé}\ and\ \citenamefont {et~al.}(2017)}]{Bohe:2017bkm}%
  \BibitemOpen
  \bibfield  {author} {\bibinfo {author} {\bibfnamefont {A.}~\bibnamefont {Bohé}}\ and\ \bibinfo {author} {\bibnamefont {et~al.}},\ }\href {\doibase 10.1103/PhysRevD.95.044028} {\bibfield  {journal} {\bibinfo  {journal} {Phys. Rev. D}\ }\textbf {\bibinfo {volume} {95}},\ \bibinfo {pages} {044028} (\bibinfo {year} {2017})},\ \Eprint {http://arxiv.org/abs/1611.03703} {arXiv:1611.03703 [gr-qc]} \BibitemShut {NoStop}%
\bibitem [{\citenamefont {Cotesta}\ \emph {et~al.}(2018)\citenamefont {Cotesta}, \citenamefont {Buonanno}, \citenamefont {Bohé}, \citenamefont {Taracchini}, \citenamefont {Hinder},\ and\ \citenamefont {Ossokine}}]{Cotesta:2018fzk}%
  \BibitemOpen
  \bibfield  {author} {\bibinfo {author} {\bibfnamefont {R.}~\bibnamefont {Cotesta}}, \bibinfo {author} {\bibfnamefont {A.}~\bibnamefont {Buonanno}}, \bibinfo {author} {\bibfnamefont {A.}~\bibnamefont {Bohé}}, \bibinfo {author} {\bibfnamefont {A.}~\bibnamefont {Taracchini}}, \bibinfo {author} {\bibfnamefont {I.}~\bibnamefont {Hinder}}, \ and\ \bibinfo {author} {\bibfnamefont {S.}~\bibnamefont {Ossokine}},\ }\href {\doibase 10.1103/PhysRevD.98.084028} {\bibfield  {journal} {\bibinfo  {journal} {Phys. Rev. D}\ }\textbf {\bibinfo {volume} {98}},\ \bibinfo {pages} {084028} (\bibinfo {year} {2018})},\ \Eprint {http://arxiv.org/abs/1803.10701} {arXiv:1803.10701 [gr-qc]} \BibitemShut {NoStop}%
\bibitem [{\citenamefont {Nagar}\ \emph {et~al.}(2020{\natexlab{a}})\citenamefont {Nagar}, \citenamefont {Pratten}, \citenamefont {Riemenschneider},\ and\ \citenamefont {Gamba}}]{Nagar:2020eul}%
  \BibitemOpen
  \bibfield  {author} {\bibinfo {author} {\bibfnamefont {A.}~\bibnamefont {Nagar}}, \bibinfo {author} {\bibfnamefont {G.}~\bibnamefont {Pratten}}, \bibinfo {author} {\bibfnamefont {G.}~\bibnamefont {Riemenschneider}}, \ and\ \bibinfo {author} {\bibfnamefont {R.}~\bibnamefont {Gamba}},\ }\href {\doibase 10.1103/PhysRevD.101.024041} {\bibfield  {journal} {\bibinfo  {journal} {Phys. Rev. D}\ }\textbf {\bibinfo {volume} {101}},\ \bibinfo {pages} {024041} (\bibinfo {year} {2020}{\natexlab{a}})},\ \Eprint {http://arxiv.org/abs/1904.09550} {arXiv:1904.09550 [gr-qc]} \BibitemShut {NoStop}%
\bibitem [{\citenamefont {Nagar}\ \emph {et~al.}(2020{\natexlab{b}})\citenamefont {Nagar}, \citenamefont {Riemenschneider}, \citenamefont {Pratten}, \citenamefont {Rettegno},\ and\ \citenamefont {Messina}}]{Nagar:2020xsk}%
  \BibitemOpen
  \bibfield  {author} {\bibinfo {author} {\bibfnamefont {A.}~\bibnamefont {Nagar}}, \bibinfo {author} {\bibfnamefont {G.}~\bibnamefont {Riemenschneider}}, \bibinfo {author} {\bibfnamefont {G.}~\bibnamefont {Pratten}}, \bibinfo {author} {\bibfnamefont {P.}~\bibnamefont {Rettegno}}, \ and\ \bibinfo {author} {\bibfnamefont {F.}~\bibnamefont {Messina}},\ }\href {\doibase 10.1103/PhysRevD.102.024077} {\bibfield  {journal} {\bibinfo  {journal} {Phys. Rev. D}\ }\textbf {\bibinfo {volume} {102}},\ \bibinfo {pages} {024077} (\bibinfo {year} {2020}{\natexlab{b}})},\ \Eprint {http://arxiv.org/abs/2001.09082} {arXiv:2001.09082 [gr-qc]} \BibitemShut {NoStop}%
\bibitem [{\citenamefont {Estell{\'e}s}\ \emph {et~al.}(2022)\citenamefont {Estell{\'e}s}, \citenamefont {Husa}, \citenamefont {Colleoni}, \citenamefont {Keitel}, \citenamefont {Mateu-Lucena}, \citenamefont {Garc{\'\i}a-Quir{\'o}s}, \citenamefont {Ramos-Buades},\ and\ \citenamefont {Borchers}}]{Estelles:2020twz}%
  \BibitemOpen
  \bibfield  {author} {\bibinfo {author} {\bibfnamefont {H.}~\bibnamefont {Estell{\'e}s}}, \bibinfo {author} {\bibfnamefont {S.}~\bibnamefont {Husa}}, \bibinfo {author} {\bibfnamefont {M.}~\bibnamefont {Colleoni}}, \bibinfo {author} {\bibfnamefont {D.}~\bibnamefont {Keitel}}, \bibinfo {author} {\bibfnamefont {M.}~\bibnamefont {Mateu-Lucena}}, \bibinfo {author} {\bibfnamefont {C.}~\bibnamefont {Garc{\'\i}a-Quir{\'o}s}}, \bibinfo {author} {\bibfnamefont {A.}~\bibnamefont {Ramos-Buades}}, \ and\ \bibinfo {author} {\bibfnamefont {A.}~\bibnamefont {Borchers}},\ }\href {\doibase 10.1103/PhysRevD.105.084039} {\bibfield  {journal} {\bibinfo  {journal} {Phys. Rev. D}\ }\textbf {\bibinfo {volume} {105}},\ \bibinfo {pages} {084039} (\bibinfo {year} {2022})},\ \Eprint {http://arxiv.org/abs/2012.11923} {arXiv:2012.11923 [gr-qc]} \BibitemShut {NoStop}%
\bibitem [{\citenamefont {Pompili}\ and\ \citenamefont {et~al.}(2023)}]{Pompili:2023tna}%
  \BibitemOpen
  \bibfield  {author} {\bibinfo {author} {\bibfnamefont {L.}~\bibnamefont {Pompili}}\ and\ \bibinfo {author} {\bibnamefont {et~al.}},\ }\href {\doibase 10.1103/PhysRevD.108.124035} {\bibfield  {journal} {\bibinfo  {journal} {Phys. Rev. D}\ }\textbf {\bibinfo {volume} {108}},\ \bibinfo {pages} {124035} (\bibinfo {year} {2023})},\ \Eprint {http://arxiv.org/abs/2303.18039} {arXiv:2303.18039 [gr-qc]} \BibitemShut {NoStop}%
\bibitem [{\citenamefont {Isi}\ and\ \citenamefont {Farr}(2021)}]{Isi:2021iql}%
  \BibitemOpen
  \bibfield  {author} {\bibinfo {author} {\bibfnamefont {M.}~\bibnamefont {Isi}}\ and\ \bibinfo {author} {\bibfnamefont {W.~M.}\ \bibnamefont {Farr}},\ }\href@noop {} {\  (\bibinfo {year} {2021})},\ \Eprint {http://arxiv.org/abs/2107.05609} {arXiv:2107.05609 [gr-qc]} \BibitemShut {NoStop}%
\bibitem [{\citenamefont {Siegel}\ \emph {et~al.}(2024)\citenamefont {Siegel}, \citenamefont {Isi},\ and\ \citenamefont {Farr}}]{Siegel:2024jqd}%
  \BibitemOpen
  \bibfield  {author} {\bibinfo {author} {\bibfnamefont {H.}~\bibnamefont {Siegel}}, \bibinfo {author} {\bibfnamefont {M.}~\bibnamefont {Isi}}, \ and\ \bibinfo {author} {\bibfnamefont {W.~M.}\ \bibnamefont {Farr}},\ }\href@noop {} {\  (\bibinfo {year} {2024})},\ \Eprint {http://arxiv.org/abs/2410.02704} {arXiv:2410.02704 [gr-qc]} \BibitemShut {NoStop}%
\bibitem [{\citenamefont {London}(2020)}]{London:2018gaq}%
  \BibitemOpen
  \bibfield  {author} {\bibinfo {author} {\bibfnamefont {L.~T.}\ \bibnamefont {London}},\ }\href {\doibase 10.1103/PhysRevD.102.084052} {\bibfield  {journal} {\bibinfo  {journal} {Phys. Rev. D}\ }\textbf {\bibinfo {volume} {102}},\ \bibinfo {pages} {084052} (\bibinfo {year} {2020})},\ \Eprint {http://arxiv.org/abs/1801.08208} {arXiv:1801.08208 [gr-qc]} \BibitemShut {NoStop}%
\bibitem [{\citenamefont {Baibhav}\ and\ \citenamefont {Berti}(2019)}]{Baibhav:2018rfk}%
  \BibitemOpen
  \bibfield  {author} {\bibinfo {author} {\bibfnamefont {V.}~\bibnamefont {Baibhav}}\ and\ \bibinfo {author} {\bibfnamefont {E.}~\bibnamefont {Berti}},\ }\href {\doibase 10.1103/PhysRevD.99.024005} {\bibfield  {journal} {\bibinfo  {journal} {Phys. Rev. D}\ }\textbf {\bibinfo {volume} {99}},\ \bibinfo {pages} {024005} (\bibinfo {year} {2019})},\ \Eprint {http://arxiv.org/abs/1809.03500} {arXiv:1809.03500 [gr-qc]} \BibitemShut {NoStop}%
\bibitem [{\citenamefont {Baibhav}\ \emph {et~al.}(2020)\citenamefont {Baibhav}, \citenamefont {Berti},\ and\ \citenamefont {Cardoso}}]{Baibhav:2020tma}%
  \BibitemOpen
  \bibfield  {author} {\bibinfo {author} {\bibfnamefont {V.}~\bibnamefont {Baibhav}}, \bibinfo {author} {\bibfnamefont {E.}~\bibnamefont {Berti}}, \ and\ \bibinfo {author} {\bibfnamefont {V.}~\bibnamefont {Cardoso}},\ }\href {\doibase 10.1103/PhysRevD.101.084053} {\bibfield  {journal} {\bibinfo  {journal} {Phys. Rev. D}\ }\textbf {\bibinfo {volume} {101}},\ \bibinfo {pages} {084053} (\bibinfo {year} {2020})},\ \Eprint {http://arxiv.org/abs/2001.10011} {arXiv:2001.10011 [gr-qc]} \BibitemShut {NoStop}%
\bibitem [{\citenamefont {Capuano}\ \emph {et~al.}(2025)\citenamefont {Capuano}, \citenamefont {Vaglio}, \citenamefont {Chandramouli}, \citenamefont {Pitte}, \citenamefont {Kuntz},\ and\ \citenamefont {Barausse}}]{Capuano:2025kkl}%
  \BibitemOpen
  \bibfield  {author} {\bibinfo {author} {\bibfnamefont {L.}~\bibnamefont {Capuano}}, \bibinfo {author} {\bibfnamefont {M.}~\bibnamefont {Vaglio}}, \bibinfo {author} {\bibfnamefont {R.~S.}\ \bibnamefont {Chandramouli}}, \bibinfo {author} {\bibfnamefont {C.~L.}\ \bibnamefont {Pitte}}, \bibinfo {author} {\bibfnamefont {A.}~\bibnamefont {Kuntz}}, \ and\ \bibinfo {author} {\bibfnamefont {E.}~\bibnamefont {Barausse}},\ }\href@noop {} {\  (\bibinfo {year} {2025})},\ \Eprint {http://arxiv.org/abs/2506.21181} {arXiv:2506.21181 [gr-qc]} \BibitemShut {NoStop}%
\bibitem [{\citenamefont {Crescimbeni}(2025)}]{crescimbeni2025interpolating}%
  \BibitemOpen
  \bibfield  {author} {\bibinfo {author} {\bibfnamefont {F.}~\bibnamefont {Crescimbeni}},\ }\href@noop {} {\enquote {\bibinfo {title} {Interpolating function of ringdown starting time},}\ }\bibinfo {howpublished} {\url{https://github.com/francesco-crescimbeni/Interpolating-function-of-ringdown-starting-time}} (\bibinfo {year} {2025}),\ \bibinfo {note} {gitHub repository}\BibitemShut {NoStop}%
\bibitem [{\citenamefont {Press}\ and\ \citenamefont {Teukolsky}(1973)}]{Press:1973zz}%
  \BibitemOpen
  \bibfield  {author} {\bibinfo {author} {\bibfnamefont {W.~H.}\ \bibnamefont {Press}}\ and\ \bibinfo {author} {\bibfnamefont {S.~A.}\ \bibnamefont {Teukolsky}},\ }\href {\doibase 10.1086/152445} {\bibfield  {journal} {\bibinfo  {journal} {Astrophys. J.}\ }\textbf {\bibinfo {volume} {185}},\ \bibinfo {pages} {649} (\bibinfo {year} {1973})}\BibitemShut {NoStop}%
\bibitem [{\citenamefont {Berti}\ \emph {et~al.}(2006{\natexlab{b}})\citenamefont {Berti}, \citenamefont {Cardoso},\ and\ \citenamefont {Casals}}]{Berti:2005gp}%
  \BibitemOpen
  \bibfield  {author} {\bibinfo {author} {\bibfnamefont {E.}~\bibnamefont {Berti}}, \bibinfo {author} {\bibfnamefont {V.}~\bibnamefont {Cardoso}}, \ and\ \bibinfo {author} {\bibfnamefont {M.}~\bibnamefont {Casals}},\ }\href {\doibase 10.1103/PhysRevD.73.109902} {\bibfield  {journal} {\bibinfo  {journal} {Phys. Rev. D}\ }\textbf {\bibinfo {volume} {73}},\ \bibinfo {pages} {024013} (\bibinfo {year} {2006}{\natexlab{b}})},\ \bibinfo {note} {[Erratum: Phys.Rev.D 73, 109902 (2006)]},\ \Eprint {http://arxiv.org/abs/gr-qc/0511111} {arXiv:gr-qc/0511111} \BibitemShut {NoStop}%
\bibitem [{\citenamefont {Berti}\ and\ \citenamefont {Klein}(2014)}]{Berti:2014fga}%
  \BibitemOpen
  \bibfield  {author} {\bibinfo {author} {\bibfnamefont {E.}~\bibnamefont {Berti}}\ and\ \bibinfo {author} {\bibfnamefont {A.}~\bibnamefont {Klein}},\ }\href {\doibase 10.1103/PhysRevD.90.064012} {\bibfield  {journal} {\bibinfo  {journal} {Phys. Rev. D}\ }\textbf {\bibinfo {volume} {90}},\ \bibinfo {pages} {064012} (\bibinfo {year} {2014})},\ \Eprint {http://arxiv.org/abs/1408.1860} {arXiv:1408.1860 [gr-qc]} \BibitemShut {NoStop}%
\bibitem [{\citenamefont {Berti}\ \emph {et~al.}(2007{\natexlab{b}})\citenamefont {Berti}, \citenamefont {Cardoso}, \citenamefont {Gonzalez}, \citenamefont {Sperhake}, \citenamefont {Hannam}, \citenamefont {Husa},\ and\ \citenamefont {Bruegmann}}]{Berti:2007fi}%
  \BibitemOpen
  \bibfield  {author} {\bibinfo {author} {\bibfnamefont {E.}~\bibnamefont {Berti}}, \bibinfo {author} {\bibfnamefont {V.}~\bibnamefont {Cardoso}}, \bibinfo {author} {\bibfnamefont {J.~A.}\ \bibnamefont {Gonzalez}}, \bibinfo {author} {\bibfnamefont {U.}~\bibnamefont {Sperhake}}, \bibinfo {author} {\bibfnamefont {M.}~\bibnamefont {Hannam}}, \bibinfo {author} {\bibfnamefont {S.}~\bibnamefont {Husa}}, \ and\ \bibinfo {author} {\bibfnamefont {B.}~\bibnamefont {Bruegmann}},\ }\href {\doibase 10.1103/PhysRevD.76.064034} {\bibfield  {journal} {\bibinfo  {journal} {Phys. Rev. D}\ }\textbf {\bibinfo {volume} {76}},\ \bibinfo {pages} {064034} (\bibinfo {year} {2007}{\natexlab{b}})},\ \Eprint {http://arxiv.org/abs/gr-qc/0703053} {arXiv:gr-qc/0703053} \BibitemShut {NoStop}%
\bibitem [{\citenamefont {Blanchet}(2014)}]{Blanchet:2013haa}%
  \BibitemOpen
  \bibfield  {author} {\bibinfo {author} {\bibfnamefont {L.}~\bibnamefont {Blanchet}},\ }\href {\doibase 10.12942/lrr-2014-2} {\bibfield  {journal} {\bibinfo  {journal} {Living Rev. Rel.}\ }\textbf {\bibinfo {volume} {17}},\ \bibinfo {pages} {2} (\bibinfo {year} {2014})},\ \Eprint {http://arxiv.org/abs/1310.1528} {arXiv:1310.1528 [gr-qc]} \BibitemShut {NoStop}%
\bibitem [{\citenamefont {{LIGO Scientific Collaboration}}(2018)}]{LIGOT1800044}%
  \BibitemOpen
  \bibfield  {author} {\bibinfo {author} {\bibnamefont {{LIGO Scientific Collaboration}}},\ }\href@noop {} {\enquote {\bibinfo {title} {Instrument science white paper 2018},}\ }\bibinfo {howpublished} {LIGO Document T1800044} (\bibinfo {year} {2018}),\ \bibinfo {note} {\url{https://dcc.ligo.org/LIGO-T1800044/public}}\BibitemShut {NoStop}%
\bibitem [{iff()}]{ifft}%
  \BibitemOpen
  \href@noop {} {}\bibinfo {howpublished} {\url{https://numpy.org/doc/stable/reference/routines.fft.html}}\BibitemShut {NoStop}%
\bibitem [{\citenamefont {Abbott}\ \emph {et~al.}(2020)\citenamefont {Abbott} \emph {et~al.}}]{LIGOScientific:2019hgc}%
  \BibitemOpen
  \bibfield  {author} {\bibinfo {author} {\bibfnamefont {B.~P.}\ \bibnamefont {Abbott}} \emph {et~al.} (\bibinfo {collaboration} {LIGO Scientific, Virgo}),\ }\href {\doibase 10.1088/1361-6382/ab685e} {\bibfield  {journal} {\bibinfo  {journal} {Class. Quant. Grav.}\ }\textbf {\bibinfo {volume} {37}},\ \bibinfo {pages} {055002} (\bibinfo {year} {2020})},\ \Eprint {http://arxiv.org/abs/1908.11170} {arXiv:1908.11170 [gr-qc]} \BibitemShut {NoStop}%
\bibitem [{\citenamefont {Carullo}\ \emph {et~al.}(2018)\citenamefont {Carullo} \emph {et~al.}}]{Carullo:2018sfu}%
  \BibitemOpen
  \bibfield  {author} {\bibinfo {author} {\bibfnamefont {G.}~\bibnamefont {Carullo}} \emph {et~al.},\ }\href {\doibase 10.1103/PhysRevD.98.104020} {\bibfield  {journal} {\bibinfo  {journal} {Phys. Rev. D}\ }\textbf {\bibinfo {volume} {98}},\ \bibinfo {pages} {104020} (\bibinfo {year} {2018})},\ \Eprint {http://arxiv.org/abs/1805.04760} {arXiv:1805.04760 [gr-qc]} \BibitemShut {NoStop}%
\bibitem [{\citenamefont {Boyle}\ \emph {et~al.}(2019)\citenamefont {Boyle} \emph {et~al.}}]{Boyle:2019kee}%
  \BibitemOpen
  \bibfield  {author} {\bibinfo {author} {\bibfnamefont {M.}~\bibnamefont {Boyle}} \emph {et~al.},\ }\href {\doibase 10.1088/1361-6382/ab34e2} {\bibfield  {journal} {\bibinfo  {journal} {Class. Quant. Grav.}\ }\textbf {\bibinfo {volume} {36}},\ \bibinfo {pages} {195006} (\bibinfo {year} {2019})},\ \Eprint {http://arxiv.org/abs/1904.04831} {arXiv:1904.04831 [gr-qc]} \BibitemShut {NoStop}%
\bibitem [{\citenamefont {Scheel}\ \emph {et~al.}(2025)\citenamefont {Scheel} \emph {et~al.}}]{Scheel:2025jct}%
  \BibitemOpen
  \bibfield  {author} {\bibinfo {author} {\bibfnamefont {M.~A.}\ \bibnamefont {Scheel}} \emph {et~al.},\ }\href@noop {} {\  (\bibinfo {year} {2025})},\ \Eprint {http://arxiv.org/abs/2505.13378} {arXiv:2505.13378 [gr-qc]} \BibitemShut {NoStop}%
\bibitem [{\citenamefont {Boyle}\ \emph {et~al.}(2025)\citenamefont {Boyle}, \citenamefont {Mitman}, \citenamefont {Scheel},\ and\ \citenamefont {Stein}}]{SXSPackage_v2025.0.15}%
  \BibitemOpen
  \bibfield  {author} {\bibinfo {author} {\bibfnamefont {M.}~\bibnamefont {Boyle}}, \bibinfo {author} {\bibfnamefont {K.}~\bibnamefont {Mitman}}, \bibinfo {author} {\bibfnamefont {M.}~\bibnamefont {Scheel}}, \ and\ \bibinfo {author} {\bibfnamefont {L.}~\bibnamefont {Stein}},\ }\href {\doibase 10.5281/ZENODO.15412737} {\enquote {\bibinfo {title} {The sxs package},}\ } (\bibinfo {year} {2025})\BibitemShut {NoStop}%
\bibitem [{\citenamefont {Carullo}\ \emph {et~al.}(2023{\natexlab{a}})\citenamefont {Carullo}, \citenamefont {De~Amicis},\ and\ \citenamefont {Redondo-Yuste}}]{carullo_gregorio_2023_8284026}%
  \BibitemOpen
  \bibfield  {author} {\bibinfo {author} {\bibfnamefont {G.}~\bibnamefont {Carullo}}, \bibinfo {author} {\bibfnamefont {M.}~\bibnamefont {De~Amicis}}, \ and\ \bibinfo {author} {\bibfnamefont {J.}~\bibnamefont {Redondo-Yuste}},\ }\href {\doibase 10.5281/zenodo.8284026} {\enquote {\bibinfo {title} {bayring},}\ }\bibinfo {howpublished} {\href{https://github.com/GCArullo/bayRing}{github.com/GCArullo/bayRing}} (\bibinfo {year} {2023}{\natexlab{a}})\BibitemShut {NoStop}%
\bibitem [{\citenamefont {Carullo}\ \emph {et~al.}(2023{\natexlab{b}})\citenamefont {Carullo}, \citenamefont {Del~Pozzo},\ and\ \citenamefont {Veitch}}]{pyRing}%
  \BibitemOpen
  \bibfield  {author} {\bibinfo {author} {\bibfnamefont {G.}~\bibnamefont {Carullo}}, \bibinfo {author} {\bibfnamefont {W.}~\bibnamefont {Del~Pozzo}}, \ and\ \bibinfo {author} {\bibfnamefont {J.}~\bibnamefont {Veitch}},\ }\href {\doibase 10.5281/zenodo.8165508} {\enquote {\bibinfo {title} {\texttt{pyRing}: a time-domain ringdown analysis python package},}\ }\bibinfo {howpublished} {\href{https://git.ligo.org/lscsoft/pyring}{git.ligo.org/lscsoft/pyring}} (\bibinfo {year} {2023}{\natexlab{b}})\BibitemShut {NoStop}%
\bibitem [{\citenamefont {Stein}(2019)}]{Stein:2019mop}%
  \BibitemOpen
  \bibfield  {author} {\bibinfo {author} {\bibfnamefont {L.~C.}\ \bibnamefont {Stein}},\ }\href {\doibase 10.21105/joss.01683} {\bibfield  {journal} {\bibinfo  {journal} {J. Open Source Softw.}\ }\textbf {\bibinfo {volume} {4}},\ \bibinfo {pages} {1683} (\bibinfo {year} {2019})},\ \Eprint {http://arxiv.org/abs/1908.10377} {arXiv:1908.10377 [gr-qc]} \BibitemShut {NoStop}%
\bibitem [{\citenamefont {Jani}\ \emph {et~al.}(2016)\citenamefont {Jani}, \citenamefont {Healy}, \citenamefont {Clark}, \citenamefont {London}, \citenamefont {Laguna},\ and\ \citenamefont {Shoemaker}}]{Jani:2016wkt}%
  \BibitemOpen
  \bibfield  {author} {\bibinfo {author} {\bibfnamefont {K.}~\bibnamefont {Jani}}, \bibinfo {author} {\bibfnamefont {J.}~\bibnamefont {Healy}}, \bibinfo {author} {\bibfnamefont {J.~A.}\ \bibnamefont {Clark}}, \bibinfo {author} {\bibfnamefont {L.}~\bibnamefont {London}}, \bibinfo {author} {\bibfnamefont {P.}~\bibnamefont {Laguna}}, \ and\ \bibinfo {author} {\bibfnamefont {D.}~\bibnamefont {Shoemaker}},\ }\href {\doibase 10.1088/0264-9381/33/20/204001} {\bibfield  {journal} {\bibinfo  {journal} {Class. Quant. Grav.}\ }\textbf {\bibinfo {volume} {33}},\ \bibinfo {pages} {204001} (\bibinfo {year} {2016})},\ \Eprint {http://arxiv.org/abs/1605.03204} {arXiv:1605.03204 [gr-qc]} \BibitemShut {NoStop}%
\bibitem [{\citenamefont {Gennari}\ \emph {et~al.}(2024)\citenamefont {Gennari}, \citenamefont {Carullo},\ and\ \citenamefont {Del~Pozzo}}]{Gennari:2023gmx}%
  \BibitemOpen
  \bibfield  {author} {\bibinfo {author} {\bibfnamefont {V.}~\bibnamefont {Gennari}}, \bibinfo {author} {\bibfnamefont {G.}~\bibnamefont {Carullo}}, \ and\ \bibinfo {author} {\bibfnamefont {W.}~\bibnamefont {Del~Pozzo}},\ }\href {\doibase 10.1140/epjc/s10052-024-12550-x} {\bibfield  {journal} {\bibinfo  {journal} {Eur. Phys. J. C}\ }\textbf {\bibinfo {volume} {84}},\ \bibinfo {pages} {233} (\bibinfo {year} {2024})},\ \Eprint {http://arxiv.org/abs/2312.12515} {arXiv:2312.12515 [gr-qc]} \BibitemShut {NoStop}%
\bibitem [{\citenamefont {Finch}\ and\ \citenamefont {Moore}(2021)}]{Finch:2021iip}%
  \BibitemOpen
  \bibfield  {author} {\bibinfo {author} {\bibfnamefont {E.}~\bibnamefont {Finch}}\ and\ \bibinfo {author} {\bibfnamefont {C.~J.}\ \bibnamefont {Moore}},\ }\href {\doibase 10.1103/PhysRevD.103.084048} {\bibfield  {journal} {\bibinfo  {journal} {Phys. Rev. D}\ }\textbf {\bibinfo {volume} {103}},\ \bibinfo {pages} {084048} (\bibinfo {year} {2021})},\ \Eprint {http://arxiv.org/abs/2102.07794} {arXiv:2102.07794 [gr-qc]} \BibitemShut {NoStop}%
\bibitem [{\citenamefont {Zhu}\ \emph {et~al.}(2025)\citenamefont {Zhu} \emph {et~al.}}]{Zhu:2023fnf}%
  \BibitemOpen
  \bibfield  {author} {\bibinfo {author} {\bibfnamefont {H.}~\bibnamefont {Zhu}} \emph {et~al.},\ }\href {\doibase 10.1103/PhysRevD.111.064052} {\bibfield  {journal} {\bibinfo  {journal} {Phys. Rev. D}\ }\textbf {\bibinfo {volume} {111}},\ \bibinfo {pages} {064052} (\bibinfo {year} {2025})},\ \Eprint {http://arxiv.org/abs/2312.08588} {arXiv:2312.08588 [gr-qc]} \BibitemShut {NoStop}%
\bibitem [{\citenamefont {Flanagan}\ and\ \citenamefont {Hughes}(1998)}]{Flanagan:1997kp}%
  \BibitemOpen
  \bibfield  {author} {\bibinfo {author} {\bibfnamefont {E.~E.}\ \bibnamefont {Flanagan}}\ and\ \bibinfo {author} {\bibfnamefont {S.~A.}\ \bibnamefont {Hughes}},\ }\href {\doibase 10.1103/PhysRevD.57.4566} {\bibfield  {journal} {\bibinfo  {journal} {Phys. Rev. D}\ }\textbf {\bibinfo {volume} {57}},\ \bibinfo {pages} {4566} (\bibinfo {year} {1998})},\ \Eprint {http://arxiv.org/abs/gr-qc/9710129} {arXiv:gr-qc/9710129} \BibitemShut {NoStop}%
\bibitem [{\citenamefont {Lindblom}\ \emph {et~al.}(2008)\citenamefont {Lindblom}, \citenamefont {Owen},\ and\ \citenamefont {Brown}}]{Lindblom:2008cm}%
  \BibitemOpen
  \bibfield  {author} {\bibinfo {author} {\bibfnamefont {L.}~\bibnamefont {Lindblom}}, \bibinfo {author} {\bibfnamefont {B.~J.}\ \bibnamefont {Owen}}, \ and\ \bibinfo {author} {\bibfnamefont {D.~A.}\ \bibnamefont {Brown}},\ }\href {\doibase 10.1103/PhysRevD.78.124020} {\bibfield  {journal} {\bibinfo  {journal} {Phys. Rev. D}\ }\textbf {\bibinfo {volume} {78}},\ \bibinfo {pages} {124020} (\bibinfo {year} {2008})},\ \Eprint {http://arxiv.org/abs/0809.3844} {arXiv:0809.3844 [gr-qc]} \BibitemShut {NoStop}%
\bibitem [{\citenamefont {Chandramouli}\ \emph {et~al.}(2025)\citenamefont {Chandramouli}, \citenamefont {Prokup}, \citenamefont {Berti},\ and\ \citenamefont {Yunes}}]{Chandramouli:2024vhw}%
  \BibitemOpen
  \bibfield  {author} {\bibinfo {author} {\bibfnamefont {R.~S.}\ \bibnamefont {Chandramouli}}, \bibinfo {author} {\bibfnamefont {K.}~\bibnamefont {Prokup}}, \bibinfo {author} {\bibfnamefont {E.}~\bibnamefont {Berti}}, \ and\ \bibinfo {author} {\bibfnamefont {N.}~\bibnamefont {Yunes}},\ }\href {\doibase 10.1103/PhysRevD.111.044026} {\bibfield  {journal} {\bibinfo  {journal} {Phys. Rev. D}\ }\textbf {\bibinfo {volume} {111}},\ \bibinfo {pages} {044026} (\bibinfo {year} {2025})},\ \Eprint {http://arxiv.org/abs/2410.06254} {arXiv:2410.06254 [gr-qc]} \BibitemShut {NoStop}%
\bibitem [{\citenamefont {Thompson}\ \emph {et~al.}(2025)\citenamefont {Thompson}, \citenamefont {Hoy}, \citenamefont {Fauchon-Jones},\ and\ \citenamefont {Hannam}}]{Thompson:2025hhc}%
  \BibitemOpen
  \bibfield  {author} {\bibinfo {author} {\bibfnamefont {J.~E.}\ \bibnamefont {Thompson}}, \bibinfo {author} {\bibfnamefont {C.}~\bibnamefont {Hoy}}, \bibinfo {author} {\bibfnamefont {E.}~\bibnamefont {Fauchon-Jones}}, \ and\ \bibinfo {author} {\bibfnamefont {M.}~\bibnamefont {Hannam}},\ }\href {\doibase 10.1103/ddz7-x9zz} {\bibfield  {journal} {\bibinfo  {journal} {Phys. Rev. D}\ }\textbf {\bibinfo {volume} {112}},\ \bibinfo {pages} {064011} (\bibinfo {year} {2025})},\ \Eprint {http://arxiv.org/abs/2506.10530} {arXiv:2506.10530 [gr-qc]} \BibitemShut {NoStop}%
\bibitem [{\citenamefont {Bhagwat}\ \emph {et~al.}(2023)\citenamefont {Bhagwat}, \citenamefont {Pacilio}, \citenamefont {Pani},\ and\ \citenamefont {Mapelli}}]{Bhagwat:2023jwv}%
  \BibitemOpen
  \bibfield  {author} {\bibinfo {author} {\bibfnamefont {S.}~\bibnamefont {Bhagwat}}, \bibinfo {author} {\bibfnamefont {C.}~\bibnamefont {Pacilio}}, \bibinfo {author} {\bibfnamefont {P.}~\bibnamefont {Pani}}, \ and\ \bibinfo {author} {\bibfnamefont {M.}~\bibnamefont {Mapelli}},\ }\href {\doibase 10.1103/PhysRevD.108.043019} {\bibfield  {journal} {\bibinfo  {journal} {Phys. Rev. D}\ }\textbf {\bibinfo {volume} {108}},\ \bibinfo {pages} {043019} (\bibinfo {year} {2023})},\ \Eprint {http://arxiv.org/abs/2304.02283} {arXiv:2304.02283 [gr-qc]} \BibitemShut {NoStop}%
\bibitem [{\citenamefont {Berti}\ \emph {et~al.}(2016)\citenamefont {Berti}, \citenamefont {Sesana}, \citenamefont {Barausse}, \citenamefont {Cardoso},\ and\ \citenamefont {Belczynski}}]{Berti:2016lat}%
  \BibitemOpen
  \bibfield  {author} {\bibinfo {author} {\bibfnamefont {E.}~\bibnamefont {Berti}}, \bibinfo {author} {\bibfnamefont {A.}~\bibnamefont {Sesana}}, \bibinfo {author} {\bibfnamefont {E.}~\bibnamefont {Barausse}}, \bibinfo {author} {\bibfnamefont {V.}~\bibnamefont {Cardoso}}, \ and\ \bibinfo {author} {\bibfnamefont {K.}~\bibnamefont {Belczynski}},\ }\href {\doibase 10.1103/PhysRevLett.117.101102} {\bibfield  {journal} {\bibinfo  {journal} {Phys. Rev. Lett.}\ }\textbf {\bibinfo {volume} {117}},\ \bibinfo {pages} {101102} (\bibinfo {year} {2016})},\ \Eprint {http://arxiv.org/abs/1605.09286} {arXiv:1605.09286 [gr-qc]} \BibitemShut {NoStop}%
\bibitem [{\citenamefont {Bhagwat}\ \emph {et~al.}(2022)\citenamefont {Bhagwat}, \citenamefont {Pacilio}, \citenamefont {Barausse},\ and\ \citenamefont {Pani}}]{Bhagwat:2021kwv}%
  \BibitemOpen
  \bibfield  {author} {\bibinfo {author} {\bibfnamefont {S.}~\bibnamefont {Bhagwat}}, \bibinfo {author} {\bibfnamefont {C.}~\bibnamefont {Pacilio}}, \bibinfo {author} {\bibfnamefont {E.}~\bibnamefont {Barausse}}, \ and\ \bibinfo {author} {\bibfnamefont {P.}~\bibnamefont {Pani}},\ }\href {\doibase 10.1103/PhysRevD.105.124063} {\bibfield  {journal} {\bibinfo  {journal} {Phys. Rev. D}\ }\textbf {\bibinfo {volume} {105}},\ \bibinfo {pages} {124063} (\bibinfo {year} {2022})},\ \Eprint {http://arxiv.org/abs/2201.00023} {arXiv:2201.00023 [gr-qc]} \BibitemShut {NoStop}%
\bibitem [{\citenamefont {Mitman}\ \emph {et~al.}(2021)\citenamefont {Mitman} \emph {et~al.}}]{Mitman:2020bjf}%
  \BibitemOpen
  \bibfield  {author} {\bibinfo {author} {\bibfnamefont {K.}~\bibnamefont {Mitman}} \emph {et~al.},\ }\href {\doibase 10.1103/PhysRevD.103.024031} {\bibfield  {journal} {\bibinfo  {journal} {Phys. Rev. D}\ }\textbf {\bibinfo {volume} {103}},\ \bibinfo {pages} {024031} (\bibinfo {year} {2021})},\ \Eprint {http://arxiv.org/abs/2011.01309} {arXiv:2011.01309 [gr-qc]} \BibitemShut {NoStop}%
\bibitem [{\citenamefont {Blackman}\ \emph {et~al.}(2015)\citenamefont {Blackman}, \citenamefont {Field}, \citenamefont {Galley}, \citenamefont {Szilagyi}, \citenamefont {Scheel},\ and\ \citenamefont {Tiglio}}]{Blackman:2015pia}%
  \BibitemOpen
  \bibfield  {author} {\bibinfo {author} {\bibfnamefont {J.}~\bibnamefont {Blackman}}, \bibinfo {author} {\bibfnamefont {S.~E.}\ \bibnamefont {Field}}, \bibinfo {author} {\bibfnamefont {C.~R.}\ \bibnamefont {Galley}}, \bibinfo {author} {\bibfnamefont {B.}~\bibnamefont {Szilagyi}}, \bibinfo {author} {\bibfnamefont {M.~A.}\ \bibnamefont {Scheel}}, \ and\ \bibinfo {author} {\bibfnamefont {M.}~\bibnamefont {Tiglio}},\ }\href {\doibase 10.1103/PhysRevLett.115.121102} {\bibfield  {journal} {\bibinfo  {journal} {Phys. Rev. Lett.}\ }\textbf {\bibinfo {volume} {115}},\ \bibinfo {pages} {121102} (\bibinfo {year} {2015})},\ \Eprint {http://arxiv.org/abs/1508.07253} {arXiv:1508.07253 [gr-qc]} \BibitemShut {NoStop}%
\bibitem [{\citenamefont {Khan}\ \emph {et~al.}(2016)\citenamefont {Khan}, \citenamefont {Husa}, \citenamefont {Hannam}, \citenamefont {Ohme}, \citenamefont {P\"urrer}, \citenamefont {Jim\'enez~Forteza},\ and\ \citenamefont {Boh\'e}}]{Khan:2015jqa}%
  \BibitemOpen
  \bibfield  {author} {\bibinfo {author} {\bibfnamefont {S.}~\bibnamefont {Khan}}, \bibinfo {author} {\bibfnamefont {S.}~\bibnamefont {Husa}}, \bibinfo {author} {\bibfnamefont {M.}~\bibnamefont {Hannam}}, \bibinfo {author} {\bibfnamefont {F.}~\bibnamefont {Ohme}}, \bibinfo {author} {\bibfnamefont {M.}~\bibnamefont {P\"urrer}}, \bibinfo {author} {\bibfnamefont {X.}~\bibnamefont {Jim\'enez~Forteza}}, \ and\ \bibinfo {author} {\bibfnamefont {A.}~\bibnamefont {Boh\'e}},\ }\href {\doibase 10.1103/PhysRevD.93.044007} {\bibfield  {journal} {\bibinfo  {journal} {Phys. Rev. D}\ }\textbf {\bibinfo {volume} {93}},\ \bibinfo {pages} {044007} (\bibinfo {year} {2016})},\ \Eprint {http://arxiv.org/abs/1508.07253} {arXiv:1508.07253 [gr-qc]} \BibitemShut {NoStop}%
\bibitem [{\citenamefont {Pratten}\ \emph {et~al.}(2021)\citenamefont {Pratten} \emph {et~al.}}]{Pratten:2020ceb}%
  \BibitemOpen
  \bibfield  {author} {\bibinfo {author} {\bibfnamefont {G.}~\bibnamefont {Pratten}} \emph {et~al.},\ }\href {\doibase 10.1103/PhysRevD.103.104056} {\bibfield  {journal} {\bibinfo  {journal} {Phys. Rev. D}\ }\textbf {\bibinfo {volume} {103}},\ \bibinfo {pages} {104056} (\bibinfo {year} {2021})},\ \Eprint {http://arxiv.org/abs/2004.06503} {arXiv:2004.06503 [gr-qc]} \BibitemShut {NoStop}%
\bibitem [{\citenamefont {Biwer}\ \emph {et~al.}(2019)\citenamefont {Biwer}, \citenamefont {Capano}, \citenamefont {De}, \citenamefont {Cabero}, \citenamefont {Brown}, \citenamefont {Nitz},\ and\ \citenamefont {Raymond}}]{Biwer:2018osg}%
  \BibitemOpen
  \bibfield  {author} {\bibinfo {author} {\bibfnamefont {C.~M.}\ \bibnamefont {Biwer}}, \bibinfo {author} {\bibfnamefont {C.~D.}\ \bibnamefont {Capano}}, \bibinfo {author} {\bibfnamefont {S.}~\bibnamefont {De}}, \bibinfo {author} {\bibfnamefont {M.}~\bibnamefont {Cabero}}, \bibinfo {author} {\bibfnamefont {D.~A.}\ \bibnamefont {Brown}}, \bibinfo {author} {\bibfnamefont {A.~H.}\ \bibnamefont {Nitz}}, \ and\ \bibinfo {author} {\bibfnamefont {V.}~\bibnamefont {Raymond}},\ }\href {\doibase 10.1088/1538-3873/aaef0b} {\bibfield  {journal} {\bibinfo  {journal} {Publ. Astron. Soc. Pac.}\ }\textbf {\bibinfo {volume} {131}},\ \bibinfo {pages} {024503} (\bibinfo {year} {2019})},\ \Eprint {http://arxiv.org/abs/1807.10312} {arXiv:1807.10312 [astro-ph.IM]} \BibitemShut {NoStop}%
\bibitem [{\citenamefont {PyCBC}(2025)}]{pycbc_waveform}%
  \BibitemOpen
  \bibfield  {author} {\bibinfo {author} {\bibnamefont {PyCBC}},\ }\href {https://pycbc.org/pycbc/latest/html/waveform.html#calculating-the-match-between-waveforms} {\enquote {\bibinfo {title} {Calculating the match between waveforms},}\ } (\bibinfo {year} {2025}),\ \bibinfo {note} {accessed: 2025-02-09}\BibitemShut {NoStop}%
\bibitem [{\citenamefont {P{\"u}rrer}\ and\ \citenamefont {Haster}(2020)}]{Purrer:2019jcp}%
  \BibitemOpen
  \bibfield  {author} {\bibinfo {author} {\bibfnamefont {M.}~\bibnamefont {P{\"u}rrer}}\ and\ \bibinfo {author} {\bibfnamefont {C.-J.}\ \bibnamefont {Haster}},\ }\href {\doibase 10.1103/PhysRevResearch.2.023151} {\bibfield  {journal} {\bibinfo  {journal} {Phys. Rev. Res.}\ }\textbf {\bibinfo {volume} {2}},\ \bibinfo {pages} {023151} (\bibinfo {year} {2020})},\ \Eprint {http://arxiv.org/abs/1912.10055} {arXiv:1912.10055 [gr-qc]} \BibitemShut {NoStop}%
\end{thebibliography}%
\let\addcontentsline\oldaddcontentsline

\end{document}
